\documentclass[twocolumn,english,aps,prb,twocolum,superscriptaddress,bibnotes,amsmath,amssymb,floatfix]{revtex4-2}
\pdfoutput=1
\usepackage{amsfonts}
\usepackage{amsmath}
\usepackage{amssymb}
\usepackage[colorlinks=true,citecolor=blue,linkcolor=blue,breaklinks=true]{hyperref}
\usepackage{graphicx}

\usepackage{soul}
\usepackage{url}
\usepackage{dcolumn}% Align table columns on decimal point
\usepackage{bm}% bold math
\usepackage{stmaryrd}
\usepackage{changes}
\usepackage{multirow}
\usepackage{rotating}
\usepackage{titlesec}
\usepackage{float}
\usepackage{tabularx}
\usepackage{textcomp}
\usepackage[flushleft]{threeparttable}
\usepackage{booktabs}
\usepackage{dcolumn}% Align table columns on decimal point

\usepackage{titlesec}
\renewcommand{\thesection}{\arabic{section}}
\titleformat{\section}
{\normalfont\large\bfseries}{\thesection.}{1em}{}
\usepackage{afterpage}
\usepackage{soul}

\newcommand{\figref}[1]{\mbox{Fig.~\ref{#1}}}
\newcommand{\figsref}[1]{\mbox{Figs.~\ref{#1}}}

\newcommand{\secref}[1]{\mbox{Sec.~\ref{#1}}}

\begin{document}
	\title{Compact superconducting vacuum-gap capacitors with low microwave loss and high mechanical coherence for scalable quantum circuits}

\author{Amir Youssefi}\thanks{These authors contributed equally.}
\affiliation{EDWATEC SA, EPFL Innovation Park, Lausanne, Switzerland.}
\affiliation{Institute of Physics, Swiss Federal Institute of Technology Lausanne (EPFL), Lausanne, Switzerland.}
\affiliation{Institute of Electrical and Micro Engineering, Swiss Federal Institute of Technology Lausanne (EPFL), Lausanne, Switzerland.}
\author{Mahdi Chegnizadeh}\thanks{These authors contributed equally.}
\author{Marco Scigliuzzo}
\author{Tobias~J.~Kippenberg}
\email[]{tobias.kippenberg@epfl.ch}
\affiliation{Institute of Physics, Swiss Federal Institute of Technology Lausanne (EPFL), Lausanne, Switzerland.}
\affiliation{Institute of Electrical and Micro Engineering, Swiss Federal Institute of Technology Lausanne (EPFL), Lausanne, Switzerland.}

\begin{abstract}
		Vacuum-gap capacitors have recently attracted significant interest in superconducting circuit platforms due to their compact design and exceptionally low dielectric losses in the microwave regime. Their intrinsic ability to support mechanical vibrational modes makes them well-suited for circuit optomechanics. However, precise control over the gap size and the realization of high-coherence mechanical modes remain longstanding challenges. Here, we present a detailed and scalable fabrication process for vacuum-gap capacitors that support ultra-high-coherence mechanical motion, exhibit low microwave loss, and occupy a significantly smaller footprint compared to conventional planar geometries. By employing a planarized SiO$_2$ sacrificial layer, we achieve vacuum gaps on the order of 150~nm. Using this platform, we have recently demonstrated ground-state cooling and motion squeezing of a mechanical oscillator with a quality factor of 40 million—a 100-fold improvement compared to prior works—as well as a single-photon optomechanical coupling rate of approximately 15Hz~\cite{youssefi2023squeezed}. Additional achievements include the realization of an optomechanical topological lattice with 24 sites~\cite{youssefi2022topological} and the observation of quantum collective dynamics in a mechanical hexamer~\cite{chegnizadeh2024quantum}. Collectively, these results underscore the potential of vacuum-gap capacitors as a platform for coupling superconducting qubits to mechanical systems, enabling quantum storage, and probing gravitational effects in quantum mechanics.
\end{abstract}
\maketitle

\textbf{}

	\section{Introduction}
Over the past two decades, quantum control of mechanical systems has been firmly established, following the quantum control of individual atoms \cite{phillips1998nobel} and ions \cite{leibfried2003quantum} in the first wave of development, and superconducting circuits \cite{o2010quantum} in the second wave. This progress has been particularly catalyzed by cavity optomechanics~\cite{RMP_optomechanics}, which utilizes radiation-pressure coupling between mechanical oscillators and electromagnetic cavities. More recently, these advancements have been extended to coupling mechanical systems with superconducting qubits via quantum acoustics~\cite{o2010quantum,chu2017quantum}. These developments have paved the way for quantum optomechanics~\cite{teufel2009nanomechanical}, enabling breakthroughs such as the cooling of low-frequency mechanical oscillators to their quantum ground state~\cite{teufel2011sideband,chan2011laser}—unattainable with passive cooling alone. Other milestones include the generation of entanglement between electromagnetic fields and mechanical oscillators~\cite{palomaki2013entangling,barzanjeh2019stationary}, entanglement among macroscopic mechanical oscillators~\cite{ockeloen2018stabilized,kotler2021direct}, observation of quantum sideband asymmetry in micromechanical oscillators~\cite{weinstein2014observation,youssefi2023squeezed}, realization of back-action-evading measurements of mechanical motion~\cite{hertzberg2010back,shomroni2019optical}, ponderomotive squeezing of light~\cite{purdy2013strong}, quantum-coherent coupling of light and mechanical oscillators~\cite{verhagen2012quantum}, and real-time quantum feedback control of mechanical oscillators~\cite{wilson2015measurement}. Furthermore, quantum optomechanics has spurred novel quantum technological innovations, such as interfaces for converting microwave to optical fields with minimal added noise~\cite{andrews2014bidirectional, delaney2022superconducting} and have been used to amplify microwave signals~\cite{massel2011microwave}.

A particularly promising platform for the quantum control of mechanical oscillators is circuit optomechanics~\cite{rocheleau2010preparation,teufel2011sideband}, where a mechanically compliant vacuum-gap capacitor is shunted with an inductor to form a microwave resonator. These capacitors were first introduced in the field of circuit quantum electrodynamics (cQED) to reduce losses in microwave resonators by eliminating the lossy dielectric layer typical of capacitors and increasing the participation ratio of the electric field in vacuum~\cite{cicak2010low}. Over the years, such circuits have enabled remarkable achievements, including mechanical ground-state cooling~\cite{teufel2011sideband}, even below the backaction limit~\cite{clark2017sideband}, mechanical squeezing~\cite{wollman2015quantum,pirkkalainen2015squeezing,lecocq2015quantum}, entanglement~\cite{kotler2021direct,ockeloen2018stabilized,palomaki2013entangling,mercier2021quantum}, non-classical state storage~\cite{reed2017faithful,palomaki2013coherent}, and non-reciprocal circuits~\cite{Bernier2017,de2019realization,barzanjeh2017mechanical}. However, despite these experimental advances, the design and fabrication processes for vacuum-gap capacitors have not kept pace. In particular, circuit optomechanics remains hindered by limited mechanical quality factors, predominantly dictated by the fabrication methods employed for vacuum-gap capacitors, as well as variations in microwave and mechanical properties. These limitations pose significant challenges to scaling up and realizing large-scale lattices.

\begin{figure*}[t]
	\centering 
	\includegraphics[width=\textwidth]{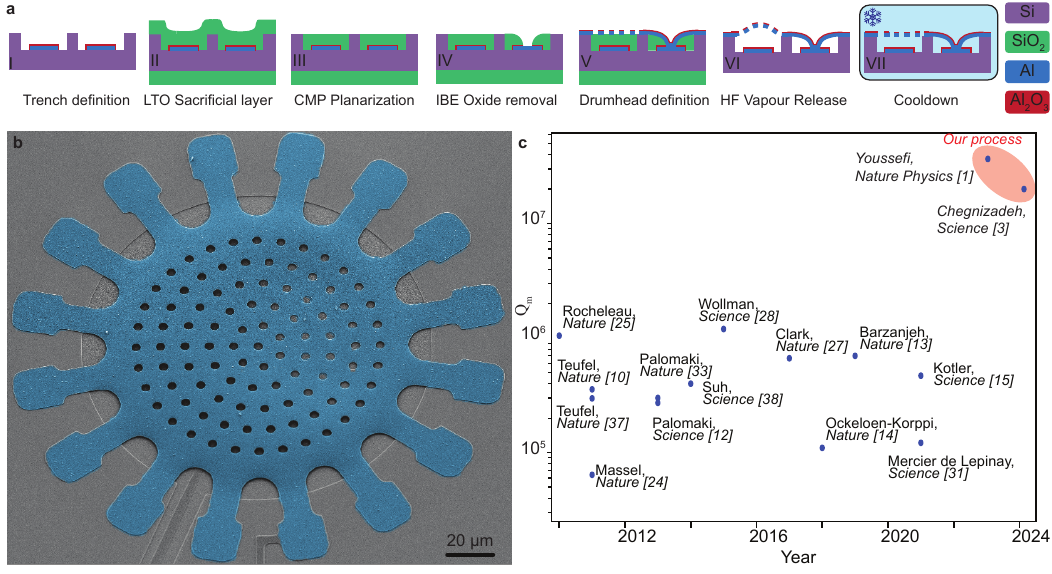} 
	\caption[Overview of the fabrication technique for the next-generation circuit optomechanics]{\textbf{Overview of the fabrication technique for the next-generation circuit optomechanical platform.} \textbf{a}, The main steps of the process consists of etching a trench in the substrate followed by deposition of a sacrificial layer, planarization, top layer definition, release, and finally cool down. Due to the compressive stresses, the top plate will buckle up after the release. However, the drumhead shrinks and flattens at cryogenic temperatures, resulting in a controllable gap size. \textbf{b}, A drumhead parallel plate capacitor after releasing the top layer. \textbf{c} An overview of experimental realizations of circuit optomechanics with vacuum gap capacitors since 2010. The shaded red area highlights the results obtained with our platform, effectively boosting the mechanical coherence by almost two orders of magnitudes.
	}
	\label{fig:fab_OV_PF} 
\end{figure*}

In this work, we present a comprehensive account of the detailed fabrication process for low-loss vacuum gap capacitors, tailored for circuit quantum optomechanics and circuit quantum electrodynamics applications, and outline the challenges we encountered. We introduce the ``flat-geometry" vacuum gap capacitor, developed to address the limitations of the conventional platform~\cite{teufel2011circuit,suh2014mechanically,pirkkalainen2015squeezing,toth2017dissipative}, and describe the full optimization process. As shown in \figref{fig:fab_OV_PF}a, the flat geometry offers significant advantages, enabling precise control of both the gap and the mechanical frequency through the lithographic process. In \figref{fig:fab_OV_PF}b, we provide a SEM micrograph of the top electrode of the vacuum gap capacitor. Furthermore, \figref{fig:fab_OV_PF}c illustrates the performance of our fabrication method compared to other approaches reported in the literature, excluding metalized membranes that feature very large footprint~\cite{liu2025degeneracy,seis2022ground} (see supplementary information). Finally, we demonstrate the application of this device in a circuit optomechanical platform, achieving accurate target frequencies and quality factors as high as 40 million. This platform has facilitated the realization of topological lattices in optomechanical circuits~\cite{youssefi2022topological}, the preparation of squeezed mechanical states and the observation of their decoherence in real time~\cite{youssefi2023squeezed}, and, most recently, collective ground-state cooling~\cite{chegnizadeh2024quantum}.

\section{Flat geometry vacuum gap capacitors}
The conventional fabrication process of vacuum gap relies on deposition and following lithography definition of a sacrificial layer covering the bottom layer. Several sacrificial materials on different substrates such as Si$_3$N$_4$ on sapphire~\cite{teufel2011circuit}, polymer on Si~\cite{suh2014mechanically}, SiO$_2$ on quartz~\cite{pirkkalainen2015squeezing}, and a-Si on sapphire~\cite{toth2017dissipative} have been tested. In these platforms, due to the deposition-induced compressive stress in the superconducting thin film, the drumhead capacitor buckles up after the release at room temperature, which increases the gap size between two plates up to a few micrometers. Cooling down such devices induces tensile stress in the thin film metal due to the significant difference in thermal expansion rates between the thin film and the substrate. Under tensile stress, the drumhead shrinks, resulting in an approximately 50~nm gap size that is neither predictable nor reproducible. This prevents the precise control of the microwave and mechanical properties of the system at low temperatures and reduces the reproducibility of the design given the high probability of deformations and collapses after the release~\cite{toth2018dissipation}. In practice, any non-uniformity of the stress distribution in the drumhead after release or asymmetric buckling results in an uncertainty in the final gap size at cryogenic temperatures. This can be in the order of tens of nano-meters, hence limiting the frequency fluctuation in the microwave LC resonator in the order of $\mathcal{O}(10\%)$. 

The key idea to overcome the existing challenges, granting better reproducibility and longer mechanical coherence time, is the flat geometry of the vibrating plate.
A tensioned vibrating plate results in lower mechanical losses~\cite{bereyhi2019clamp} and prevents thermal-induced deformations affecting the capacitor's gap size. In our fabrication process (\figref{fig:fab_OV_PF}a), we first define a trench in a silicon substrate by dry etching. Next, we deposit and pattern the bottom plate of the capacitor inside the trench. The trench is then covered by a thick SiO$_2$ sacrificial layer, which inherits the same topography of the layer underneath. To remove this topography and obtain a flat surface, we use chemical mechanical polishing (CMP) to planarize the SiO$_2$ surface. We then etch back the sacrificial layer down to the substrate layer and deposit the top Al plate of the capacitor. The sacrificial layer will be removed by HF vapor isotropic etching to suspend the structure. After the release process, the drumhead may buckle up (depending on the deposition-induced stress of the thin film) due to the compressive stress; however, at cryogenic temperature the high tensile stress ensures the flatness of the top plate. This will guarantee the gap size to be precisely defined by the trench's depth and the bottom plate's thickness. Furthermore, the top plate's flat geometry significantly reduces the drumhead resonator's mechanical dissipation. Such advance can be implemented using different materials for substrate, superconducting metal, and sacrificial layer. However, process compatibility of materials and their resilience against different etching steps conduct us to choose a specific set of materials for this process.

To minimize dielectric loss of the superconducting circuits, substrates with low bulk tangent loss, such as intrinsic silicon and sapphire are usually preferred. We find that sacrificial layers present insufficient adhesion to the substrate, preventing full planarization (details are provided in SI). In addition, micro-structuring sapphire is challenging since lacking of established processes to etch and manipulate this material. For our process we determine that silicon wafers suit best our process. In particular, we use high-resistivity ($> 20$~k$\mathbf{\mathrm{\Omega}}$cm), low-bow ($< 20 \mu$m), low total thickness variation (TTV$< 5 \mu$m), and float-zone intrinsic silicon wafer with 10~cm diameter and 523~$\mu$m thickness supplied from Topsil$^\text{®}$. Importantly wafer's flatness, uniformity, and bow play an essential role in the CMP planarization step.

High coherence superconducting circuits are traditionally realized in aluminum~\cite{burnett2019decoherence}, niobium~\cite{gordon2022environmental,kono2024mechanically} and more recently tantalum~\cite{place2021new,crowley2023disentangling}. With the aim of integrating vacuum gap capacitor in optomechanical system, we decide to use aluminum due to its lighter density which provides larger optomechanical coupling (see SI for more details). This makes future integration of our circuits with conventional superconducting qubits more straightforward.

For the sacrificial layer, amorphous silicon (a-Si), silicon nitride (Si$_3$N$_4$), silicon oxide (SiO$_2$), and polymer photoresists are four candidates which were used in the previous generation of circuit optomechanical devices. Each one needs different isotropic etching for the release process. For example, a-Si can be removed by XeF$_2$, which is an exothermic gas etching. Si$_3$N$_4$ can be etched by SF$_6$ plasma, and polymer resists by oxygen plasma. We decided to use SiO$_2$ as our sacrificial layer which can be removed with HF vapor, enabling us to release high aspect ratio structures by avoiding plasma or wet etching, therefore increasing the yield and successful release rate of the process. In addition, it has infinite selectivity to aluminum and silicon. More details are given in the release section (Sec.~\ref{Sec:fab_release}).

\section{Vacuum gap fabrication process}

\subsection{Etching trenches in silicon}
We use optical lithography to transfer patterns on the photoresist. It is performed by direct mask-less optical lithography (Heidelberg$^\text{®}$ MLA 150). We spin coat a 1~$\mu$m thick AZ$^\text{®}$ ECI 3007 positive photoresist after HMDS surface preparation. All photoresist coating and developing steps are processed using automatic coater/developer (Süss$^\text{®}$ ACS200 GEN3). The exposure dose and depth of focus vary based on the tool and need to be calibrated by dose tests, but are typically set to $\sim~150$ mJ/cm$^2$ and 0, respectively. After the exposure, the resist is developed, and the wafer is rinsed in a spin dryer to clean any unwanted contamination. To remove residual photoresist on the surface of exposed areas, we conduct a short (10-20 seconds) oxygen plasma descum at 200~Watts and 200~sccm (Tepla$^\text{®}$ GiGAbatch). After the descum, the wafer is ready for the etching step.

After lithography, we use deep reactive ion etching (DRIE) to etch the trenches in the silicon substrate. We use $\rm{C_4F_8}$ chemistry plasma as etchant (Alcatel$^\text{®}$ AMS200) with a typical etch rate of $\sim13$~nm/sec and selectivity of Si:PR$\sim10:1$. Due to the small fluctuation of the etch rate in the machine, we use test wafers with a similar pattern to calculate the etch rate by removing the resist and measuring the trench depth using a mechanical profilometer. In addition, we set the total etching $\sim30$~nm deeper than the target capacitor gap size plus the thickness of the bottom electrode to compensate for potential non-uniformity in the CMP planarization step among different chips on a wafer. The excess depth after CMP can be etched back by IBE in the following steps to reach the desired gap size. The roughness of the silicon inside the trenches is measured $R_\mathrm{a} \approx 1.5$~nm with a trench depth uniformity of $\sim 1\%$.

After each etching step, the photoresist is stripped first using UFT remover 1165 wet process, followed by rinse and drying, and then 3 minutes 200~Watt and  200~sccm Oxygen plasma (Tepla$^\text{®}$ GiGAbatch). In the fabrication steps where the wafer contains uncovered thin-film aluminum, it is recommended to reduce either the power or exposure time of the oxygen plasma to avoid additional oxidation and local heating of the metal, specifically for the vibrating top plate.

\begin{figure}
	\includegraphics[width=\columnwidth]{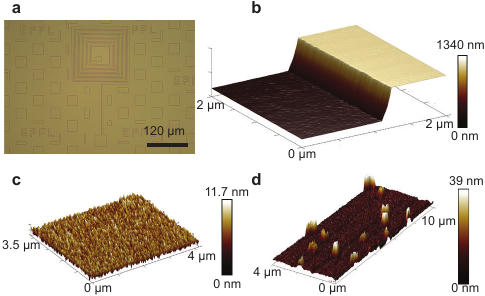} 
	\caption[Si etching]{\textbf{Etching a trench in silicon.} \textbf{a}, Optical microscope image of trenches. \textbf{b}, Atomic Force Microscopy (AFM) of a test trench etched in Si with DRIE. \textbf{c}, AFM of the Si surface inside the trench. The average roughness is $R_\mathrm{a} = 1.5$~nm. \textbf{d}, An example of a trench etch when the resist descum was not enough. After the etch, the residual photoresist in the trenches results in big hillocks of Si.}
	\label{fig:fab_trench} 
\end{figure}

\subsection{Bottom layer}
The first electrode of the vacuum gap capacitor is placed within the silicon trenches. We also add a spiral inductor that will form an LC resonator. For microwave circuits, the metal-substrate interface has a major effect on the total loss of the superconducting circuit~\cite{murray2021material}. While the electric field in the vacuum gap capacitor is mainly stored in the space between the two electrodes, we still clean the wafer after trench etching with Piranha and dip it into HF (1\% diluted) for a few minutes to remove the native silicon oxide and minimize as much as possible dielectric loss. Then we rinse and dry the wafer and immediately transfer it to the deposition tool (less than 3 minutes) and pump down the chamber to avoid regrowth of the native oxide.

Deposition of the bottom aluminum layer can be done by either sputtering or electron beam evaporation. However, we find that evaporated films have better thickness uniformity and thickness control compared to sputtered ones. We typically choose 100~nm thickness (Alliance-Concept EVA 760) for the bottom layer, deposited with 0.5~nm/sec rate. 

\begin{figure}
	\centering 
	\includegraphics[width=\columnwidth]{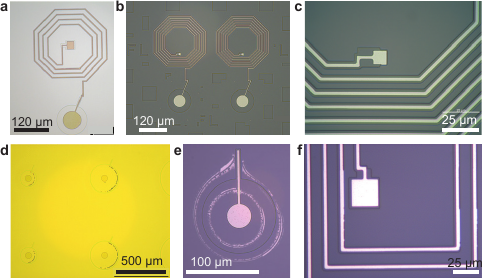} 
	\caption[Bottom Al layer patterning]{\textbf{Bottom Aluminum layer patterning.} \textbf{a}, Micrograph of a successfully patterned photoresist inside the trench to etch Al. \textbf{b} and \textbf{c}, Bottom layer circuits after a successful aluminum etching. \textbf{d} (\textbf{e}), Effect of shallow resist on development (etching). The areas close to edges are under exposed.  \textbf{f}, When the metal wires pattern are too close to the trench edges, the trench's edge prevents a proper exposure in corners.}
	\label{fig:fab_Al_bottom} 
\end{figure}

After deposition of the aluminum layer, we repeat the lithography step to pattern the bottom circuit. However, to reduce the topography thickness variation after spin coating deriving from the trenches, we increase the thickness of the resist to 1.2~$\mu$m. In addition, the metal outside of the design area and close to the wafer edge is removed to improve uniformity of the CMP step.

We use wet etching to remove aluminum using the following chemistry at 35°C: $\rm{H_3PO_4}\; 85\% + \rm{CH_3COOH }\; 100\% + \rm{HNO_3 }\; 70\%$ 83:5.5:5.5. Although we measure an etching rate of 2.2~nm/s, we keep the wafer in the solution 5 additional seconds after the main pattern appears to ensure etching of small area. Importantly, the wet etching has infinite selectivity to silicon, maintaining edge sharpness of the trenches at which the mechanical drum is clamped. 

\subsection{Sacrificial layer}

\begin{figure}
	\centering 
	\includegraphics[width=\columnwidth]{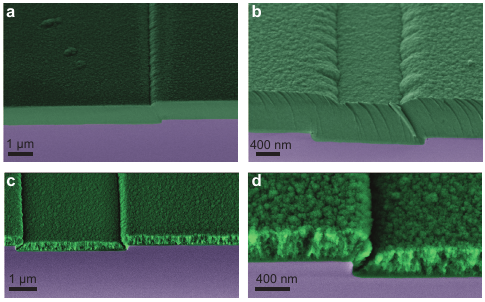} 
	\caption[SiO$_2$ sacrificial layer]{\textbf{SiO$_2$ sacrificial layer deposition.} \textbf{a} and \textbf{b}, False color SEM micrograph of the cross section of low temperature oxide low pressure chemical vapor deposition (LTO LPCVD) of SiO$_2$ sacrificial layer covering the trenches. The step coverage and gap filling is perfect, the porosity is low, and the oxide layer is dense. \textbf{c} and \textbf{d}, Plasma enhanced chemical vapor deposition (PECVD) of oxide at ${200^ \circ }{\rm{C}}$. The oxide layer is porous and does not show a good step coverage, forming void areas at the corners of the trench.  }
	\label{fig:fab_sacrificial} 
\end{figure}

The SiO$_2$ sacrificial layer in our process must satisfy several conditions: 1- It should be grown at low temperatures (below the melting point of aluminum at 660°C) to minimize damage to the aluminum. 2- It should provide good step coverage. 3- It should not be porous, which is important to maintain a flat surface after planarization. 4- It should have high adhesion to the substrate to prevent delamination or dishing when subjected to significant mechanical shear stress during CMP polishing.

PECVD and LTO deposition can both be operated at low temperatures, 100-250°C and 300-450°C, respectively. In \figsref{fig:fab_sacrificial}c and d, we report SEM image of cross-section of the PECVD silicon oxide deposited at 200\,$^{\circ}$C.  The mechanical softness of the oxide is responsible for delamination in CMP, and the porosity produced a large roughness of the aluminum top layer, resulting in lower quality suspended aluminum film. In contrast, in \figref{fig:fab_sacrificial}a and b we display an SEM image of LTO-deposited silicon oxide. Such layer presents a higher density and better adhesion to the substrate. Such property are reflected in the CMP step, when the etch/polish rate of LTO-grown oxide is measured $\sim30\%$ lower than PECVD oxide with same polishing parameters.  
We find the thickness of the sacrificial layer that optimize the topography removal is around 6 times larger than the maximum topography of the wafer, i.e. the trench depth. For example for a 300~nm trench we deposit 2~$\mu$m oxide layer. This guarantees enough room to run CMP which simultaneously etches and planarizes the surface. We observe less than 0.5\% wafer-scale non-uniformity for 3~$\mu$m depositions.

The aluminum film, beneath the sacrificial layer, occasionally displays small and sparse holes (less than 1~$\mu$m diameter) that we attribute to variation in the precursors' concentration in the LTO chamber, due to previous usage.  Nevertheless, we did not observe any sizable impact on the circuits due to this effect. 

CMP planarization simultaneously etches the oxide layer and smooths the edges and reduces the topography. For the oxide we deposit, we measure an etching rate between 100 and 300~nm/minute that varies based on pressure, slurry rate and concentration, and rotation speed. A list of the optimal parameters we obtain in our fabrication methods is reported in Table~\ref{tab:CMP}.

\begin{table}
	\caption[CMP optimized parameters]{Optimized parameters for one cycle of CMP.}
	\label{tab:CMP}
	\centering
	\resizebox{\columnwidth}{!}{
		\begin{tabular}{ccccc}
			\toprule
			Step & Head speed (rpm) & Pad speed (rpm) & Pressure (Bar)& Time (s) \\ 
			\midrule
			Preparation	& 40	& 60 	& 0.2 	& 15 \\
			Polishing	& 78	& 85 	& 0.4 	& 120 \\
			Cleaning	& 100	& 100 	& 0.25 	& 30  \\
			\bottomrule
	\end{tabular}}
	%\resizebox{\columnwidth}{!}{
		\begin{tabular}{cc}
			Slurry &7A5 Gal (pure) or 30N50 (1:1 diluted)\\
			Slurry flow & 1/10\\
			Back pressure & 0.25 Bar \\
			\bottomrule 
		\end{tabular}%}
\end{table}

\begin{figure}
	\centering 
	\includegraphics[width=\columnwidth]{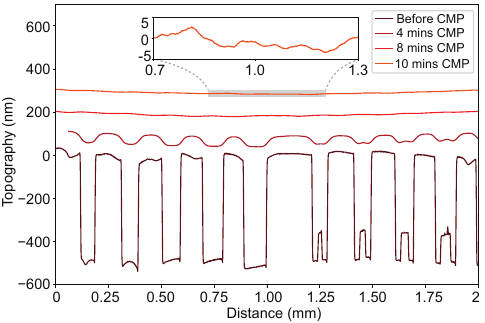} 
	\caption[Topography planarization in CMP]{\textbf{Topography planarization in CMP.}  CMP enables us to reduce the surface topography from $\sim500$~nm (the trench depth) to below 10 nm. The figure shows the effect of successive CMP runs on the topography measured by a mechanical profilometer. The final global curve in the topography shows the wafer bow. The inset shows magnified final topography. Adapted from~\cite{youssefi2022topological,youssefi2023squeezed}.}
	\label{fig:fab_CMP_plot} 
\end{figure}

We find that there is a trade-off between uniformity and residual topography after polishing: longer times reduce the topography (that finally saturates by dishing effect) but increases the thickness non-uniformity at wafer scale. In order to minimize dishing and delamination effect and increase the polishing uniformity, we fill all the empty areas of the wafer between circuits with \textit{dummy patterns}. These patterns are squares with the size of $60 \mu$m with double of this size spacing (see SI for more details). In \figref{fig:fab_CMP_plot}, we plot the mechanical profilometer (KLA$^\text{®}$ Tencor D600) traces  across trenches after each of these cycles (with arbitrary translation to make the different profiles comparable).  After 10 minutes, topography is removed to below 10\,nm, as it is shown in the inset (see SI for more details).

\begin{figure}
	\centering 
	\includegraphics[width=\columnwidth]{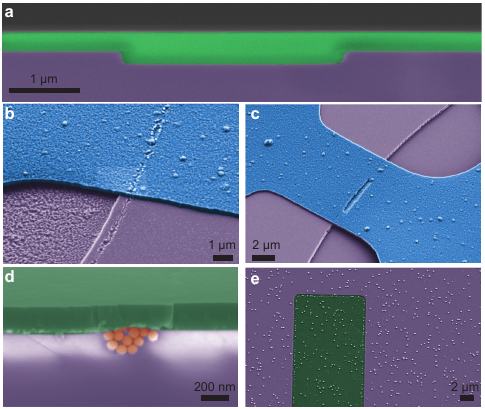} 
	\caption[CMP planarization]{\textbf{CMP planarization.} \textbf{a,} A cross section SEM showing successful CMP planarization of oxide sacrificial layer covering a trench. The remaining oxide will be removed by IBE to prevent oxide delamination. \textbf{b} and \textbf{c}, SEM of released devices where the CMP reached too close to the substrate surface, resulting in delamination of the sacrificial layer, creating voids at the edges of the trench and creaks on sidewalls of the trenches. \textbf{d} and \textbf{e}, SEM showing the slurry nano-particles after CMP (with PECVD oxide sacrificial layer for these samples). The slurry particles should be cleaned before IBE step using post CMP cleaner mentioned in the text.}
	\label{fig:fab_CMP_SEM} 
\end{figure}

The thickness of the residual sacrificial layer (inside or outside of the trenches) is measured with an optical spectroscopic reflectometer (Nanospec$^\text{®}$ AFT-6100 or FilMetrics$^\text{®}$ F54-XY). To measure the thickness inside the trenches, we always locate a "test trench" with dimension $\sim0.5\times0.5$~mm$^2$ (larger than the waist diameter of the optical beam) on every chip to be able to individually measure chips and extract a wafer map of the residual thickness. We aim to have few hundreds of nanometers in thickness for the residual layer above the substrate, as visible in the false color SEM micrograph reported in \figref{fig:fab_CMP_SEM}a. Failing to stop the polishing before this threshold results in large shear stress during CMP that peels off the sacrificial layer from the trench and creates voids around its edge (see Figs.~\ref{fig:fab_CMP_SEM}b-d). Residual slurry particles (see Figs.~\ref{fig:fab_CMP_SEM}e and f)  are removed by a post-CMP cleaning tool (GnP$^\text{®}$ Cleaner 428) immediately after the CMP before the wafer dries out.

\begin{figure}
	\centering 
	\includegraphics[width=\columnwidth]{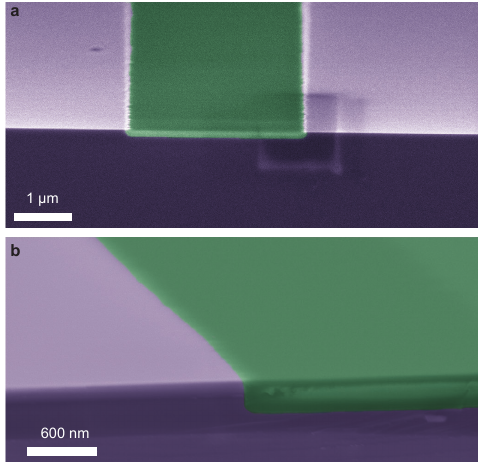} 
	\caption[IBE etch-back]{\textbf{IBE etch-back.} \textbf{a} and \textbf{b}, SEM images showing trenches after planarization and IBE etch-back. The oxide-silicon border is dense and smooth, making a perfect condition for top-layer deposition.}
	\label{fig:fab_CMP_IBE} 
\end{figure}

The remaining SiO$_2$ on the Si surface is etched by argon milling (Veeco$^\text{®}$ Nexus IBE350) with a slow etch rate of 35~nm/minute and 1:1 etch selectivity for Si:SiO$_2$, which increases the controllability of the residual layer thickness and avoid increasing topography when the etching transitions from  SiO$_2$ to Si.
To further improve uniformity, the wafer rotates during etching at 10~rpm. In addition, to avoid redepositions, the wafer is tilted by 45$^\circ$ during etching. Post-etching surface roughness measurements on Si yields to $R_\mathrm{a} \simeq 0.75$~nm demonstrating negligible surface damage. 
We target an over-etch of $\sim20$~nm to make sure all SiO$_2$ is removed from surface. See \figref{fig:fab_CMP_IBE} to see the oxide in the trench after IBE. 

\begin{figure}
	\centering 
	\includegraphics[width=\columnwidth]{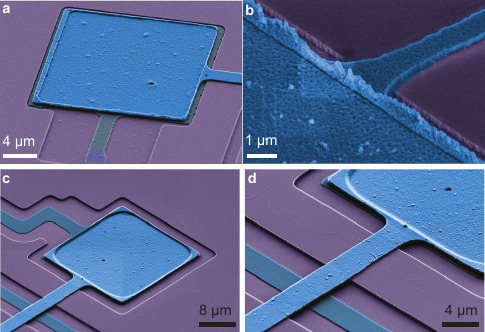} 
	\caption[Problem of galvanic connection with sharp edges.]{\textbf{Problem of galvanic connection with sharp edges.} \textbf{a}, The galvanic connection on SiO$_2$ openings with sharp edges. Due to the local thickness decrease of aluminum layer on the sharp edges, these circuits show frequency shift when the intra-cavity photon number -in other words, circulating current- is high. \textbf{b}, In addition, the sharp edges of the opening may result in accumulation of aluminum during the top layer deposition. \textbf{c} and \textbf{d}, SEM of a galvanic connection with resist reflow process after the release showing smooth transition of the top layer to the lower level.}
	\label{fig:fab_opening_bad} 
\end{figure}

To have electrical access to the bottom electrode, we open a via through the silicon oxide. We define a square pad of $\sim20 \mu\rm{m}\times20\mu\rm{m}$ (in the trench) galvanically connected to the bottom plate of the capacitor. With the same lithography procedure described in the beginning, we pattern a smaller rectangle on the resist on top of the SiO$_2$ layer which covers the bottom connection pad. 
To have a smooth metal coverage, we reflow the resist by heating up the wafer to 180 Celsius for 30 seconds using a standard hot plate. After the reflow, we do the standard descum to remove resist residues. Afterward, we use DRIE plasma etching (SPTS$^\text{®}$ APS) with $\rm{CHF_3}$ chemistry which offers 1:1 selectivity for $\rm{SiO_2}$:photoresist and transfer the photoresist pattern into the oxide. Then the resist is removed by the standard procedure discussed earlier. Avoiding reflow produces thin aluminum contact ($<50$~nm) on the edges of the galvanic connection as displayed in Figs.~\ref{fig:fab_opening_bad}a and b. In this case, we observe strong high-power nonlinearities in the microwave response, that we attribute to high local current densities in the connection region. The galvanic connections after the reflow step, reported in Figs.~\ref{fig:fab_opening_bad}c~and~d, do not show non-idealities.

\subsection{Top aluminum layer deposition}
\label{sec:top_al}
We tested two different electron beam evaporators (\textit{Eva}: Alliance-Concept$^\text{®}$ EVA 760, and \textit{Plassys}: Plassys$^\text{®}$ MEB550SL3) for the deposition of the top aluminum layer. Eva is a standard electron beam evaporator with 450~mm working distance. A 200~nm aluminum film grown with 0.5~nm/s deposition rate in Eva results in $\sim 50$~MPa compressive stress, that produces a buckling of the drumhead in a dome shape after the release at room temperature. This effect has the advantage of improving the release yield since the gap size will increase more than $\sim1 \mu$m, and HF vapor can penetrate easier to remove the remaining sacrificial layer. 
With this method, however, a thin layer of native aluminum oxide remains in the top layer to the bottom layer connection (through the pad that was explained in the previous section), making it a resistive connection.

\begin{figure*}[t] 
	\centering 
	\includegraphics[width=\textwidth]{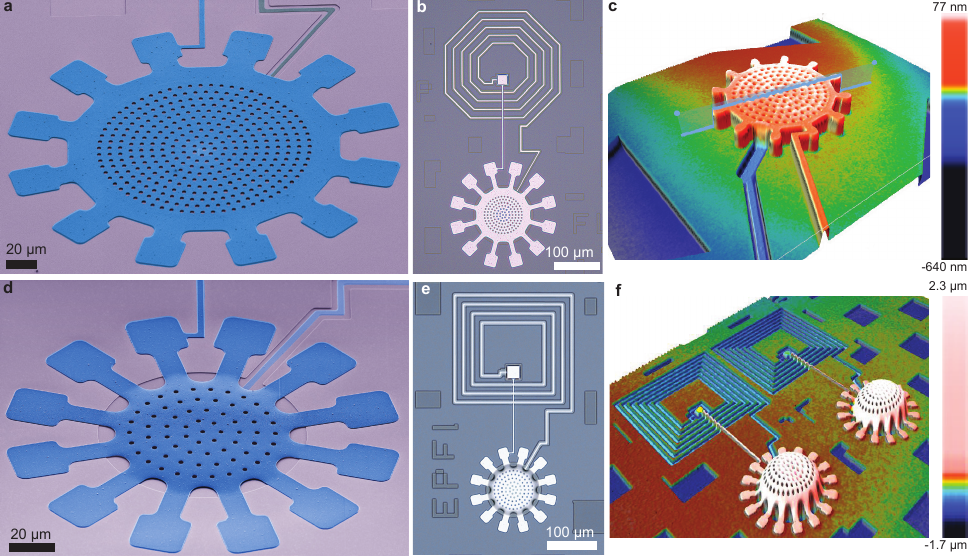} 
	\caption[Released vacuum gap capacitors]{\textbf{Released devices.} \textbf{a} to \textbf{c}, SEM image, microscope image, and optical profilometry of released drumhead capacitors for the case of near-zero or tensile stress (Plassys deposition), receptively. \textbf{d} to \textbf{f}, SEM image, microscope image, and optical profilometry of released drumhead capacitors for the case of compressive stress in the top aluminum layer (using Eva evaporation machine). Drums with compressive stress buckle up to $\sim2\mu$m and form a dome shape visible under the microscope. However the gap size at cryogenic temperatures goes to the designed value due to the temperature induced tensile stress and consequently flatness of the drum.}
	\label{fig:fab_full_release} 
\end{figure*}
To overcome this drawback, we argon-ion mill and deposit the metal without breaking the vacuum in Plassys. This can help to significantly reduce the microwave cavity heating when sending high power to the device \cite{youssefi2023squeezed}. An optimized evaporation on 200~nm Al in Plassys gives minor tensile stress in the film. Finally, an Al$_2$O$_3$ layer is grown on the drumhead resonator by injecting 10\,mBar of 99.99\% pure oxygen in Plassys oxidation chamber. Additional deposition parameters are discussed in supplementary information (SI).

After the deposition, we use the standard lithography technique to pattern the top layer and etch it using the same wet process which has been mentioned earlier.

\subsection{Release}\label{Sec:fab_release}
We dice the wafer into chips before the release (see details on dicing in SI). The last step of the fabrication process is releasing the vacuum gap capacitors by removing the $\rm{SiO_2}$ sacrificial layer. This is done by vapor phase Hydrofluoric acid etching (SPTS$^\text{®}$ uEtch). 
Reaction with the sacrificial SiO$_2$ on the wafer surface (in the presence of ethanol as the catalyst) produces silicon tetrafluoride (SiF$_4$) gas and water vapor:
\begin{equation}
	\rm{SiO_2} + \rm{4HF}^{-2} + \rm{4C_2H_5OH}^{+2} \longrightarrow \rm{SiF_4} + \rm{2H_2O} + \rm{4C_2H_5OH}
\end{equation}

\begin{figure}[h!] 
	\centering 
	\includegraphics[width=\columnwidth]{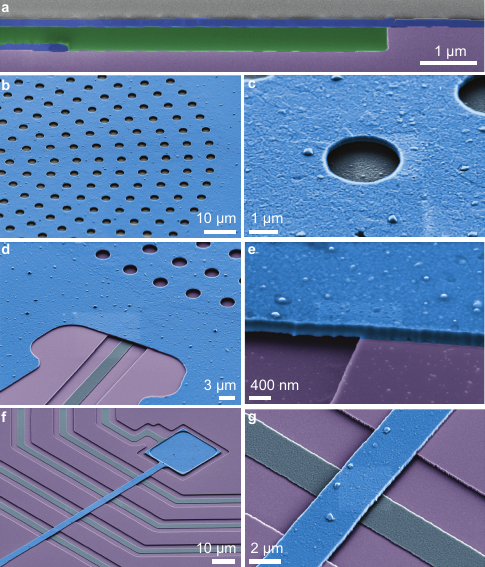} 
	\caption[Elements of the final device]{\textbf{Elements of the final device.} \textbf{a}, SEM micrograph of a focused ion beam cross-section of a capacitor before removing the SiO$_2$ sacrificial layer (Pt is used as the focused ion beam protective layer). The flatness of the top layer is visible in the image indicating a successful CMP planarization. \textbf{b}, SEM image of the perforated released drumhead. The holes facilitate the release process. \textbf{c}, Magnified SEM of one release hole. The bottom Al layer is visible from the hole. The small particles seen on the aluminum surface are aluminum hillocks, a well-know accumulation of aluminum in evaporation technique. The size and distribution of hillocks depends on the evaporation rate and the pressure of the chamber. \textbf{d}, The bottom wire going under a drum. \textbf{e}, A magnified SEM of a drumhead clamp. \textbf{f} and \textbf{g}, The spiral inductor air bridges and the galvanic connection.}
	\label{fig:fab_final_device} 
\end{figure}

Although the liquid HF attacks aluminum, we observe that the vapor HF does not deteriorate or degrade Al films. 
We use a recipe with 125~Torr pressure and $\sim 100$~nm/minute etch rate for 900~seconds etching time in every cycle. Since the vapor HF needs to penetrate horizontally between the top Al layer and the trench bottom surface, the total number of cycles needed should be calculated based on the maximum lateral distance between two penetration windows for the gas etchant, considering the pattern of Al covering the trenches. We multiply this number by a factor of four to prevent any residual oxide and to ensure the whole structure is released. 
The etching process is liquid-free, which is crucial to release structures with high aspect ratios, in our case $\sim100$~$\mu$m big drums suspended by $\sim200$~nm gap above another metallic layer. Any liquid  formation will result in the collapse and sticking of two capacitor plates. Examples of successfully released devices are shown in Figs.~\ref{fig:fab_full_release}~and~\ref{fig:fab_final_device}.

To facilitate the release, specifically for big drums, we perforate drums by defining small holes with 1.8~$\mu$m diameter and distance of $\sim10 \mu$m as shown in Fig.~\ref{fig:fab_final_device}. We successfully release vacuum gaps with the gap size down to 75~nm, but the success rate of the release was not high for such a low gap size (see SI). For gaps above 150~nm, we find almost 100\% yield of release. The success rate depends on the trench size, the thickness of the top plate, and the room temperature stress. Smaller drums have a lower risk of collapse and can tolerate smaller gap sizes. We usually use 150-200~nm thick Al top layer. Attempts to release thin top layers (50~nm) were not successful (see SI). 

\section{Measurement and results}
To determine the mechanical and microwave properties of the vacuum gap capacitor, we shunt it by a meander inductor and thermally anchor the sample to the mixing chamber of a dilution refrigerator at 10\,mK. To determine the mechanical frequency of the vacuum gap capacitor we use optomechanically induced transparency (OMIT)~\cite{weis2010optomechanically}.

\begin{figure}
	\includegraphics[width=\columnwidth]{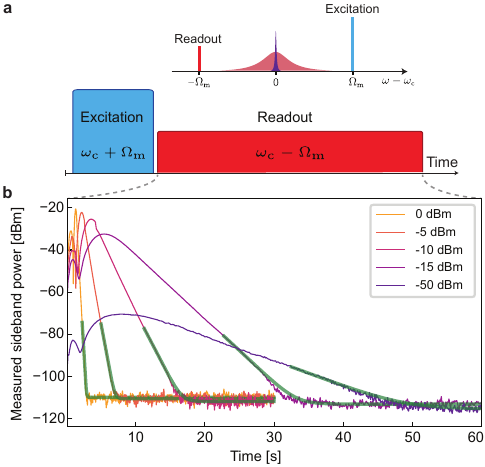}
	\caption{\textbf{Ringdown measurement of a high-$Q_\mathrm{m}$ electromechanical system.} \textbf{a}, Pulse and frequency scheme of the ringdown measurement. A strong blue detuned pulse is exciting the mechanical oscillator through optomechanical parametric instability. A red-detuned readout pulse generates an optomechanical sideband on resonance. \textbf{b}, Example of ringdown traces measured for different readout powers. The initial nonlinear behavior in the ringdown trace may be due to the energy exchange between different mechanical modes of the drumhead at high amplitude vibrations. For the exponential fitting (shown by green solid lines), we only use the low-power linear part of the ring down. The trace corresponding to pump power of -50~dBm shows $Q_\mathrm{m}=4\times10^{7}$. Adapted from \cite{youssefi2023squeezed}.}
	\label{fig:char_ringdown}
\end{figure}

The bare mechanical damping rates can be either characterized by extracting the linewidth of the OMIT features while pumping the microwave resonance with low powers corresponding to optomechanical cooperativities below one ($\mathcal{C}\ll 1$) or conducting a ring-down experiment. Since the presented fabrication process results in sub-Hertz damping rates of mechanical modes, the OMIT measurement requires low frequency resolution and typically low resolution bandwidth. Therefore, the ringdown measurement is more suitable for such characterizations. For higher red-detuned probe powers, the effective mechanical damping rate is $\Gamma_\mathrm{tot} = \Gamma_\mathrm{m} (1+\mathcal{C})$. Sweeping the power of the red probe enables us to directly measure $\Gamma_\mathrm{m}$ (Figs.~\ref{fig:char_ringdown}c and d).

We studied 16 separate electromechanical LC circuits fabricated on a 9.5~mm$\times$6.5~mm chip (See SI for the chip design). The frequencies of microwave and mechanical resonances are multiplexed in the chip in the range of 5-7 GHz and 1.5-2.5 MHz, respectively. This is done by changing the trench radius for mechanical frequency tuning (from 60~$\mu$m to 100~$\mu$m), and the capacitor bottom plate radius for microwave frequency tuning. All 16 LC circuits were magnetically coupled to a micro-strip waveguide.

\begin{figure}[h!]
	\includegraphics[width = \columnwidth]{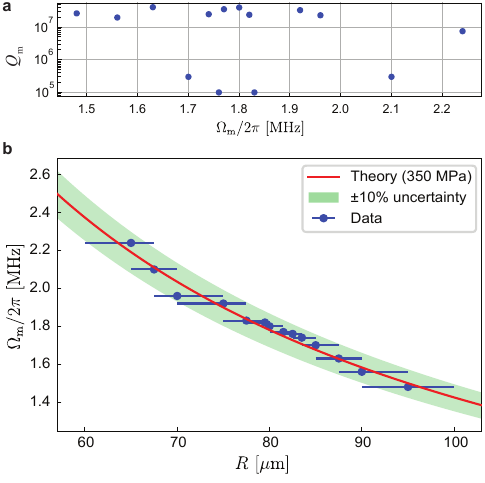} \caption{\textbf{Quality factor distribution and Extraction of the aluminum film stress at low temperatures.} \textbf{a}, Measured mechanical quality factors versus the mechanical frequencies in a chip with 14 separate electromechanical LC circuits. Outlier quality factors may be caused by post-fabrication device contamination. \textbf{b}, Measured mechanical frequencies on a chip with 16 separate electromechanical LC resonators. In the chip layout the trench radius of drums is swept from 60~$\mu$m to 100~$\mu$m. Since two LC resonances were inaccessible due to frequency overlap with the JTWPA stopband, we measured 14 resonances and considered an error bar showing the uncertainty in the trench radii. The red line shows the theoretical curve with $\sigma_\text{Al}= 350$~MPa and the green shade shows theory bounds corresponding to $350 \text{MPa}\times(1\pm10\%)$ uncertainty.}
	\label{fig:des_stress}
\end{figure}

We provide the measured quality factors and mechanical frequencies for a chip with 16 independent electromechanical LC resonators in a chip - we did not observe two LC resonators, due to overlapping their frequencies with the stop-band of a Josephson traveling wave parametric amplifier~\cite{macklin2015near} used in the measurement chain. More than 50\% of the resonators exhibit above $20\times10^6$ mechanical quality factor (Fig.~\ref{fig:des_stress}a), which demonstrates a high yield in our new fabrication process. 

\begin{figure*}[h] 
	\centering 
	\includegraphics{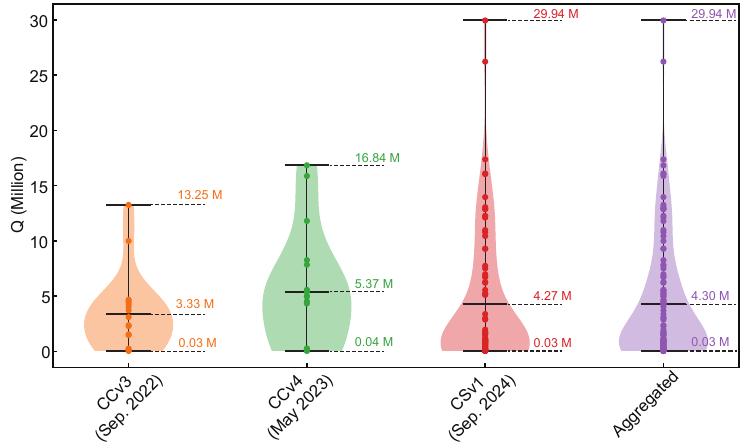}
	\caption{\textbf{Batch-to-batch mechanical quality factors}. 
		The mechanical quality factors for three different batches CCv3, CCv4, and CSv1, which are fabricated on September 2022, May 2023, and September 2024, respectively. The transparent region is the estimated density of quality factors, the top line show the maximum quality factor, the middle line shows the average quality factor, and the bottom line shows the minimum quality factor. All the drums have 70~$\mu $m radius (average mechanical frequency 2~MHz).
	}
	\label{fig:Qs_col} 
\end{figure*}

In Fig.~\ref{fig:des_stress}a, we reported the mechanical quality factors of 14 devices fabricated on a single chip. In \figref{fig:Qs_col}, we present data from a much broader dataset comprising 119 mechanical oscillators measured over a span of three years, encompassing three different fabrication batches across six wafers. The fabrication of batches CCv3, CCv4, and CSv1 was completed in September 2022, May 2023, and September 2024, respectively. The average mechanical quality factor across 119 devices is remarkably high, at 4.3~million. These results demonstrate the high yield and consistency of our fabrication process across batches and over time.

\begin{figure}
	\centering 
	\includegraphics[width=\columnwidth]{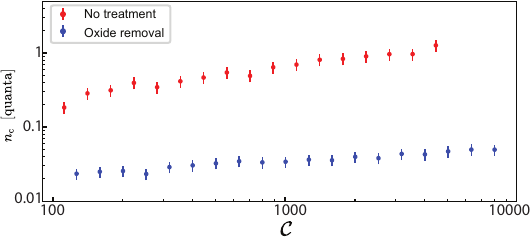} 
	\caption{\textbf{Microwave cavity heating treatment.} Heating of the microwave cavity (in units of quanta) as a function of the red pump cooperativity for a device without ion milling treatment (red circles) and with milling treatment (blue circles). Adapted from~\cite{youssefi2023squeezed}. }
	\label{fig:heating} 
\end{figure}

A persistent challenge in circuit optomechanical platforms has been the heating of the microwave cavity caused mainly by high circulating currents in the circuit~\cite{teufel2011sideband}, which limits the minimum occupation of the mechanical oscillator using sideband cooling. As discussed in \secref{sec:top_al}, argon milling of the bottom-layer aluminum prior to making the galvanic connection removes the resistive aluminum oxide layer, thereby reducing microwave loss. Figure \ref{fig:heating} shows the microwave cavity heating with and without ion milling treatment, demonstrating that this method reduces heating by an order of magnitude. With this improvement, the microwave resonator quality factor is limited by dielectric loss of its capacitive element. By measuring the internal microwave quality factor as a function of the intracavity photon number, in \cite{chegnizadeh2024quantum} we report internal quality factors up to $10^5$ for photon number $n\sim\mathcal{O}(100)$ going up to $2\times10^6$ for photon number $n\sim\mathcal{O}(10^7)$ for a resonance frequency  $\sim\mathcal{O}(5\,\rm{GHz})$. The reported microwave quality factor is compatible with the large participation ratio of the electric field in the native aluminum oxide present on the capacitor leads and other vacuum gap capacitors have shown similar performance in compact capacitor for resonator \cite{boussaha2020development} and for transmon qubit \cite{zemlicka2023compact}.

The fundamental mechanical frequency of an ideal and fully clamped drum is given by:
\begin{equation}
	\Omega_{\mathrm{m}} = \frac{\alpha_{0,1}}{R} \sqrt{\frac{\sigma_\mathrm{m}}{\rho}},
	\label{eq:omega_m}
\end{equation}
where $\alpha_{0,1}$ is the first root of the zeroth order Bessel function ($J_0$), and  $\sigma_\mathrm{m}$ and $\rho$ are the mechanical stress and density of the material, respectively, in our case aluminum with $\rho_\mathrm{Al} = 2700 \ \mathrm{ kg}/\mathrm{m}^3$. 

To experimentally extract the value of the mechanical stress in drumheads at low temperature, we plot mechanical frequencies versus trench radius and fit the theoretically expected frequencies to extract the stress as $\sigma_\text{Al}= 350 (\pm 10\%)$~MPa, as shown in Fig.~\ref{fig:des_stress}b. Tapering the clamping points will increase the local stress proportional to the tapering ratio~\cite{bereyhi2019clamp}. We swept the clamp ratio of drums with 50~$\mu$m radius and observed broken legs for the ones with more than $\sim1$~GPa stress. This can indicate an upper limit for the yield stress of the aluminum thin films at low temperatures which has not been reported in the literature. Detailed discussion on the aluminum thin film stress at low temperatures is provided in SI.

\begin{figure*}[t] 
	\centering 
	\includegraphics[width=\columnwidth]{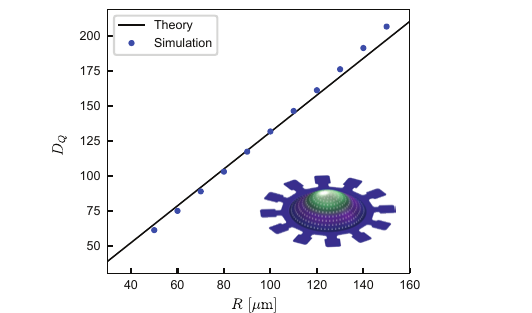} 
	\caption{\textbf{Dilution factor estimation}. Mechanical dilution factor as a function of the drum resonator radius, obtained through finite element method simulations in COMSOL. The inset illustrates the simulated drum model in COMSOL. Adapted from~\cite{youssefi2023squeezed}.}
	\label{fig:dilution}
\end{figure*}

The quality factor of a nano-mechanical oscillator is expressed as $Q_{\rm m} = Q_0 \times D_Q$, where $Q_0$ represents the material quality factor, and $D_Q$ is the loss dilution factor~\cite{fedorov2020mechanical,fedorov2019generalized}. The dilution factor $D_Q$ depends on the oscillator's geometry and material properties such as stress and Young's modulus. Using finite element simulations of our drum resonator in COMSOL (\figref{fig:dilution}), we estimate the dilution factor in our device to be $D_Q \approx 100$. Additionally, the aluminum quality factor at 10~mK is estimated to be $Q_0 \approx 4 \times 10^5$.

\begin{figure*}[t] 
	\centering 
	\includegraphics[width=\textwidth]{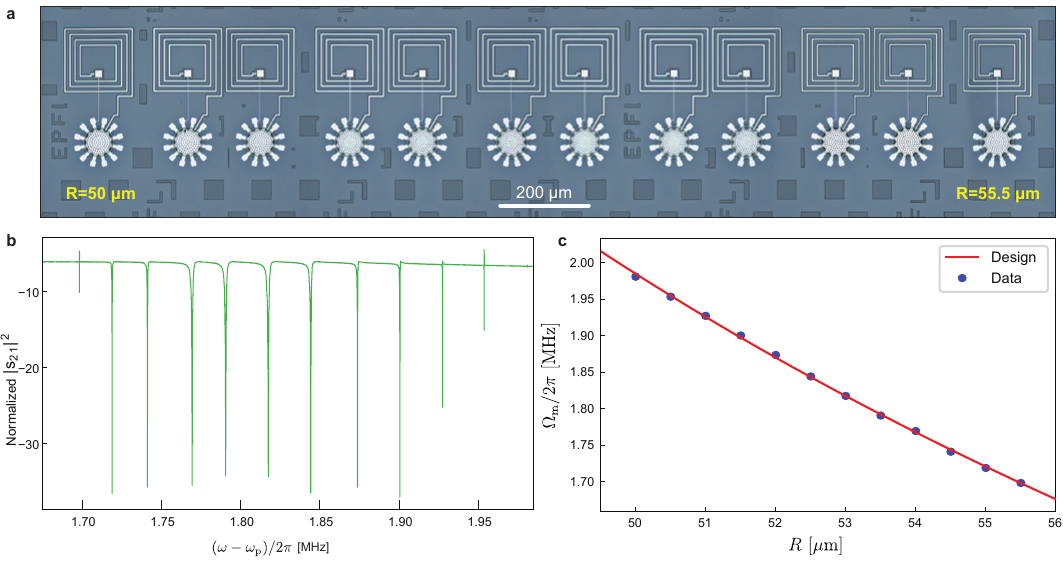} 
	\caption[Released vacuum gap capacitors]{\textbf{Low-disorder optomechanical arrays.}  \textbf{a}, a 12-site electromechanical array realizing SSH model. All sites are identical, but with slightly different trench radius (between 50~$\mu$m to 55.5~$\mu$m with 0.5~$\mu$m increment) resulting in mechanical frequency. \textbf{b}, OMIT response showing all 12 mechanical modes. The normalized scattering parameter is shown versus the pump-probe frequency detuning. \textbf{c}, Measured mechanical frequencies versus the designed trench radius (dots). The red line shows theoretical fit with $1/R$ scaling rule.}
	\label{fig:OMIT}
\end{figure*}

Optomechanical building blocks can be coupled with each other to form arrays and lattices~\cite{heinrich2011collective,xuereb2012strong,ludwig2013quantum,raeisi2020quench,peano2015topological,zangeneh2020topological}. Namely, such multimode systems can implement Su-Schrieffer-Heeger (SSH) arrays~\cite{asboth2016short,ozawa2019topological}, as a fundamental topological model. Such optomechanical arrays can exhibit non-trivial topological properties, localized edge states, and be exploited to directly measure hybridized microwave modeshapes to fully reconstruct the system's Hamiltonian~\cite{youssefi2022topological}. To demonstrate the scalability of our process, we fabricated a 12-site 1D array of identical electromechanical LC circuits with alternating mutual inductance realizing SSH model (Fig.~\ref{fig:OMIT}a). We slightly sweep the trench radius of drumhead resonators in the array (by 500~nm) to shift their frequencies and be able to identify them in the OMIT response (Fig.~\ref{fig:OMIT}b). As shown in Fig.~\ref{fig:OMIT}c, the measured mechanical frequencies from OMIT experiment perfectly follow the theoretical fit. The standard deviation of mechanical frequency disorder from the theoretical fit is less than 0.2\%, which shows a perfect control and high reproducibility in the process.

\section{Discussion and Conclusion}

In summary, we report the nano-fabrication technique for ultra-coherent and reproducible superconducting circuit optomechanics. Such systems have shown millisecond scale quantum decoherence~\cite{youssefi2023squeezed}, were used to make large-scale optomechanical lattices~\cite{youssefi2022topological}, and can be used to demonstrate quantum collective dynamics of mechanical oscillators~\cite{chegnizadeh2024quantum}. 

Having such controllable vacuum gap capacitor allows to realize resonators with precise resonance frequencies~\cite{chegnizadeh2024quantum}, on par with other platform as kinetic inductance resonators~\cite{jouanny2024band} or Josephson junction-based circuitry~\cite{osman2023mitigation}.  

The presented ultra-coherent electromechanical system can be exploited in quantum sensing applications~\cite{li2021cavity}. The expected on-resonance force sensitivity of our device can be estimated as $\sqrt{S_{FF}} = \sqrt{2 k_\mathrm{B}Tm_\mathrm{eff}\Gamma_\mathrm{m}} \simeq 200\times 10^{-21} \left[ \frac{\text{N}}{\sqrt{\mathrm{Hz}}} \right ]$, which is considerably low compared with several other optomechanical platforms thanks to the high mechanical quality factor and the low operating temperature of the device ($T=10$~mK and $m_\mathrm{eff}\simeq2$~ng).
Moreover, such high-Q electromechanical system may benefit the implementation of qubit-mechanics interfaces~\cite{reed2017faithful}, generation of mechanical non-classical states~\cite{gely2021phonon}, realization of long life-time memories for quantum computation and communication~\cite{wallucks2020quantum,pechal2018superconducting},  and it may set the stage to perform fundamental tests of quantum mechanics in macroscopic scales such as quantum gravity tests~\cite{liu2021gravitational,gely2021superconducting}, high fidelity Bell tests~\cite{marinkovic2018optomechanical,hong2017hanbury}, quantum teleportation~\cite{fiaschi2021optomechanical}, or even the search for Dark matter~\cite{carney2021mechanical,manley2021searching}.

\newpage
\section*{Acknowledgement}
This work was supported by the Swiss National Science Foundation (SNSF) under grant No. 10002358 (CoolMe). M.S. acknowledges support from the EPFL Center for Quantum Science and Engineering postdoctoral fellowship. All devices were fabricated in the Center of MicroNanoTechnology (CMi) at EPFL. 

\section*{Data availability}
The data that support the findings of this article are openly available~\cite{chegnizadeh_2025_15413211}.
%\bibliography{References}
%\bibliography{References}

%\bibliography{References}

\section*{References}
\begin{thebibliography}{85}%
	\makeatletter
	\providecommand \@ifxundefined [1]{%
		\@ifx{#1\undefined}
	}%
	\providecommand \@ifnum [1]{%
		\ifnum #1\expandafter \@firstoftwo
		\else \expandafter \@secondoftwo
		\fi
	}%
	\providecommand \@ifx [1]{%
		\ifx #1\expandafter \@firstoftwo
		\else \expandafter \@secondoftwo
		\fi
	}%
	\providecommand \natexlab [1]{#1}%
	\providecommand \enquote  [1]{``#1''}%
	\providecommand \bibnamefont  [1]{#1}%
	\providecommand \bibfnamefont [1]{#1}%
	\providecommand \citenamefont [1]{#1}%
	\providecommand \href@noop [0]{\@secondoftwo}%
	\providecommand \href [0]{\begingroup \@sanitize@url \@href}%
	\providecommand \@href[1]{\@@startlink{#1}\@@href}%
	\providecommand \@@href[1]{\endgroup#1\@@endlink}%
	\providecommand \@sanitize@url [0]{\catcode `\\12\catcode `\$12\catcode
		`\&12\catcode `\#12\catcode `\^12\catcode `\_12\catcode `\%12\relax}%
	\providecommand \@@startlink[1]{}%
	\providecommand \@@endlink[0]{}%
	\providecommand \url  [0]{\begingroup\@sanitize@url \@url }%
	\providecommand \@url [1]{\endgroup\@href {#1}{\urlprefix }}%
	\providecommand \urlprefix  [0]{URL }%
	\providecommand \Eprint [0]{\href }%
	\providecommand \doibase [0]{https://doi.org/}%
	\providecommand \selectlanguage [0]{\@gobble}%
	\providecommand \bibinfo  [0]{\@secondoftwo}%
	\providecommand \bibfield  [0]{\@secondoftwo}%
	\providecommand \translation [1]{[#1]}%
	\providecommand \BibitemOpen [0]{}%
	\providecommand \bibitemStop [0]{}%
	\providecommand \bibitemNoStop [0]{.\EOS\space}%
	\providecommand \EOS [0]{\spacefactor3000\relax}%
	\providecommand \BibitemShut  [1]{\csname bibitem#1\endcsname}%
	\let\auto@bib@innerbib\@empty
	%</preamble>
	\bibitem [{\citenamefont {Youssefi}\ \emph {et~al.}(2023)\citenamefont
		{Youssefi}, \citenamefont {Kono}, \citenamefont {Chegnizadeh},\ and\
		\citenamefont {Kippenberg}}]{youssefi2023squeezed}%
	\BibitemOpen
	\bibfield  {author} {\bibinfo {author} {\bibfnamefont {A.}~\bibnamefont
			{Youssefi}}, \bibinfo {author} {\bibfnamefont {S.}~\bibnamefont {Kono}},
		\bibinfo {author} {\bibfnamefont {M.}~\bibnamefont {Chegnizadeh}},\ and\
		\bibinfo {author} {\bibfnamefont {T.~J.}\ \bibnamefont {Kippenberg}},\
	}\bibfield  {title} {\bibinfo {title} {A squeezed mechanical oscillator with
			millisecond quantum decoherence},\ }\href
	{https://www.nature.com/articles/s41567-023-02135-y} {\bibfield  {journal}
		{\bibinfo  {journal} {Nature Physics}\ ,\ \bibinfo {pages} {1}} (\bibinfo
		{year} {2023})}\BibitemShut {NoStop}%
	\bibitem [{\citenamefont {Youssefi}\ \emph {et~al.}(2022)\citenamefont
		{Youssefi}, \citenamefont {Kono}, \citenamefont {Bancora}, \citenamefont
		{Chegnizadeh}, \citenamefont {Pan}, \citenamefont {Vovk},\ and\ \citenamefont
		{Kippenberg}}]{youssefi2022topological}%
	\BibitemOpen
	\bibfield  {author} {\bibinfo {author} {\bibfnamefont {A.}~\bibnamefont
			{Youssefi}}, \bibinfo {author} {\bibfnamefont {S.}~\bibnamefont {Kono}},
		\bibinfo {author} {\bibfnamefont {A.}~\bibnamefont {Bancora}}, \bibinfo
		{author} {\bibfnamefont {M.}~\bibnamefont {Chegnizadeh}}, \bibinfo {author}
		{\bibfnamefont {J.}~\bibnamefont {Pan}}, \bibinfo {author} {\bibfnamefont
			{T.}~\bibnamefont {Vovk}},\ and\ \bibinfo {author} {\bibfnamefont {T.~J.}\
			\bibnamefont {Kippenberg}},\ }\bibfield  {title} {\bibinfo {title}
		{Topological lattices realized in superconducting circuit optomechanics},\
	}\href {https://www.nature.com/articles/s41586-022-05367-9} {\bibfield
		{journal} {\bibinfo  {journal} {Nature}\ }\textbf {\bibinfo {volume} {612}},\
		\bibinfo {pages} {666} (\bibinfo {year} {2022})}\BibitemShut {NoStop}%
	\bibitem [{\citenamefont {Chegnizadeh}\ \emph {et~al.}(2024)\citenamefont
		{Chegnizadeh}, \citenamefont {Scigliuzzo}, \citenamefont {Youssefi},
		\citenamefont {Kono}, \citenamefont {Guzovskii},\ and\ \citenamefont
		{Kippenberg}}]{chegnizadeh2024quantum}%
	\BibitemOpen
	\bibfield  {author} {\bibinfo {author} {\bibfnamefont {M.}~\bibnamefont
			{Chegnizadeh}}, \bibinfo {author} {\bibfnamefont {M.}~\bibnamefont
			{Scigliuzzo}}, \bibinfo {author} {\bibfnamefont {A.}~\bibnamefont
			{Youssefi}}, \bibinfo {author} {\bibfnamefont {S.}~\bibnamefont {Kono}},
		\bibinfo {author} {\bibfnamefont {E.}~\bibnamefont {Guzovskii}},\ and\
		\bibinfo {author} {\bibfnamefont {T.~J.}\ \bibnamefont {Kippenberg}},\
	}\bibfield  {title} {\bibinfo {title} {Quantum collective motion of
			macroscopic mechanical oscillators},\ }\href@noop {} {\bibfield  {journal}
		{\bibinfo  {journal} {Science}\ }\textbf {\bibinfo {volume} {386}},\ \bibinfo
		{pages} {1383} (\bibinfo {year} {2024})}\BibitemShut {NoStop}%
	\bibitem [{\citenamefont {Phillips}(1998)}]{phillips1998nobel}%
	\BibitemOpen
	\bibfield  {author} {\bibinfo {author} {\bibfnamefont {W.~D.}\ \bibnamefont
			{Phillips}},\ }\bibfield  {title} {\bibinfo {title} {Nobel lecture: Laser
			cooling and trapping of neutral atoms},\ }\href@noop {} {\bibfield  {journal}
		{\bibinfo  {journal} {Reviews of Modern Physics}\ }\textbf {\bibinfo {volume}
			{70}},\ \bibinfo {pages} {721} (\bibinfo {year} {1998})}\BibitemShut
	{NoStop}%
	\bibitem [{\citenamefont {Leibfried}\ \emph {et~al.}(2003)\citenamefont
		{Leibfried}, \citenamefont {Blatt}, \citenamefont {Monroe},\ and\
		\citenamefont {Wineland}}]{leibfried2003quantum}%
	\BibitemOpen
	\bibfield  {author} {\bibinfo {author} {\bibfnamefont {D.}~\bibnamefont
			{Leibfried}}, \bibinfo {author} {\bibfnamefont {R.}~\bibnamefont {Blatt}},
		\bibinfo {author} {\bibfnamefont {C.}~\bibnamefont {Monroe}},\ and\ \bibinfo
		{author} {\bibfnamefont {D.}~\bibnamefont {Wineland}},\ }\bibfield  {title}
	{\bibinfo {title} {Quantum dynamics of single trapped ions},\ }\href
	{https://journals.aps.org/rmp/abstract/10.1103/RevModPhys.75.281} {\bibfield
		{journal} {\bibinfo  {journal} {Reviews of Modern Physics}\ }\textbf
		{\bibinfo {volume} {75}},\ \bibinfo {pages} {281} (\bibinfo {year}
		{2003})}\BibitemShut {NoStop}%
	\bibitem [{\citenamefont {O’Connell}\ \emph {et~al.}(2010)\citenamefont
		{O’Connell}, \citenamefont {Hofheinz}, \citenamefont {Ansmann},
		\citenamefont {Bialczak}, \citenamefont {Lenander}, \citenamefont {Lucero},
		\citenamefont {Neeley}, \citenamefont {Sank}, \citenamefont {Wang},
		\citenamefont {Weides} \emph {et~al.}}]{o2010quantum}%
	\BibitemOpen
	\bibfield  {author} {\bibinfo {author} {\bibfnamefont {A.~D.}\ \bibnamefont
			{O’Connell}}, \bibinfo {author} {\bibfnamefont {M.}~\bibnamefont
			{Hofheinz}}, \bibinfo {author} {\bibfnamefont {M.}~\bibnamefont {Ansmann}},
		\bibinfo {author} {\bibfnamefont {R.~C.}\ \bibnamefont {Bialczak}}, \bibinfo
		{author} {\bibfnamefont {M.}~\bibnamefont {Lenander}}, \bibinfo {author}
		{\bibfnamefont {E.}~\bibnamefont {Lucero}}, \bibinfo {author} {\bibfnamefont
			{M.}~\bibnamefont {Neeley}}, \bibinfo {author} {\bibfnamefont
			{D.}~\bibnamefont {Sank}}, \bibinfo {author} {\bibfnamefont {H.}~\bibnamefont
			{Wang}}, \bibinfo {author} {\bibfnamefont {M.}~\bibnamefont {Weides}}, \emph
		{et~al.},\ }\bibfield  {title} {\bibinfo {title} {Quantum ground state and
			single-phonon control of a mechanical resonator},\ }\href@noop {} {\bibfield
		{journal} {\bibinfo  {journal} {Nature}\ }\textbf {\bibinfo {volume} {464}},\
		\bibinfo {pages} {697} (\bibinfo {year} {2010})}\BibitemShut {NoStop}%
	\bibitem [{\citenamefont {Aspelmeyer}\ \emph {et~al.}(2014)\citenamefont
		{Aspelmeyer}, \citenamefont {Kippenberg},\ and\ \citenamefont
		{Marquardt}}]{RMP_optomechanics}%
	\BibitemOpen
	\bibfield  {author} {\bibinfo {author} {\bibfnamefont {M.}~\bibnamefont
			{Aspelmeyer}}, \bibinfo {author} {\bibfnamefont {T.~J.}\ \bibnamefont
			{Kippenberg}},\ and\ \bibinfo {author} {\bibfnamefont {F.}~\bibnamefont
			{Marquardt}},\ }\bibfield  {title} {\bibinfo {title} {Cavity optomechanics},\
	}\href {https://journals.aps.org/rmp/abstract/10.1103/RevModPhys.86.1391}
	{\bibfield  {journal} {\bibinfo  {journal} {Reviews of Modern Physics}\
		}\textbf {\bibinfo {volume} {86}},\ \bibinfo {pages} {1391} (\bibinfo {year}
		{2014})}\BibitemShut {NoStop}%
	\bibitem [{\citenamefont {Chu}\ \emph {et~al.}(2017)\citenamefont {Chu},
		\citenamefont {Kharel}, \citenamefont {Renninger}, \citenamefont {Burkhart},
		\citenamefont {Frunzio}, \citenamefont {Rakich},\ and\ \citenamefont
		{Schoelkopf}}]{chu2017quantum}%
	\BibitemOpen
	\bibfield  {author} {\bibinfo {author} {\bibfnamefont {Y.}~\bibnamefont
			{Chu}}, \bibinfo {author} {\bibfnamefont {P.}~\bibnamefont {Kharel}},
		\bibinfo {author} {\bibfnamefont {W.~H.}\ \bibnamefont {Renninger}}, \bibinfo
		{author} {\bibfnamefont {L.~D.}\ \bibnamefont {Burkhart}}, \bibinfo {author}
		{\bibfnamefont {L.}~\bibnamefont {Frunzio}}, \bibinfo {author} {\bibfnamefont
			{P.~T.}\ \bibnamefont {Rakich}},\ and\ \bibinfo {author} {\bibfnamefont
			{R.~J.}\ \bibnamefont {Schoelkopf}},\ }\bibfield  {title} {\bibinfo {title}
		{Quantum acoustics with superconducting qubits},\ }\href@noop {} {\bibfield
		{journal} {\bibinfo  {journal} {Science}\ }\textbf {\bibinfo {volume}
			{358}},\ \bibinfo {pages} {199} (\bibinfo {year} {2017})}\BibitemShut
	{NoStop}%
	\bibitem [{\citenamefont {Teufel}\ \emph {et~al.}(2009)\citenamefont {Teufel},
		\citenamefont {Donner}, \citenamefont {Castellanos-Beltran}, \citenamefont
		{Harlow},\ and\ \citenamefont {Lehnert}}]{teufel2009nanomechanical}%
	\BibitemOpen
	\bibfield  {author} {\bibinfo {author} {\bibfnamefont {J.~D.}\ \bibnamefont
			{Teufel}}, \bibinfo {author} {\bibfnamefont {T.}~\bibnamefont {Donner}},
		\bibinfo {author} {\bibfnamefont {M.}~\bibnamefont {Castellanos-Beltran}},
		\bibinfo {author} {\bibfnamefont {J.~W.}\ \bibnamefont {Harlow}},\ and\
		\bibinfo {author} {\bibfnamefont {K.~W.}\ \bibnamefont {Lehnert}},\
	}\bibfield  {title} {\bibinfo {title} {Nanomechanical motion measured with an
			imprecision below that at the standard quantum limit},\ }\href
	{https://doi.org/https://doi.org/10.1038/nnano.2009.343} {\bibfield
		{journal} {\bibinfo  {journal} {Nature nanotechnology}\ }\textbf {\bibinfo
			{volume} {4}},\ \bibinfo {pages} {820} (\bibinfo {year} {2009})}\BibitemShut
	{NoStop}%
	\bibitem [{\citenamefont {Teufel}\ \emph
		{et~al.}(2011{\natexlab{a}})\citenamefont {Teufel}, \citenamefont {Donner},
		\citenamefont {Li}, \citenamefont {Harlow}, \citenamefont {Allman},
		\citenamefont {Cicak}, \citenamefont {Sirois}, \citenamefont {Whittaker},
		\citenamefont {Lehnert},\ and\ \citenamefont
		{Simmonds}}]{teufel2011sideband}%
	\BibitemOpen
	\bibfield  {author} {\bibinfo {author} {\bibfnamefont {J.~D.}\ \bibnamefont
			{Teufel}}, \bibinfo {author} {\bibfnamefont {T.}~\bibnamefont {Donner}},
		\bibinfo {author} {\bibfnamefont {D.}~\bibnamefont {Li}}, \bibinfo {author}
		{\bibfnamefont {J.~W.}\ \bibnamefont {Harlow}}, \bibinfo {author}
		{\bibfnamefont {M.}~\bibnamefont {Allman}}, \bibinfo {author} {\bibfnamefont
			{K.}~\bibnamefont {Cicak}}, \bibinfo {author} {\bibfnamefont {A.~J.}\
			\bibnamefont {Sirois}}, \bibinfo {author} {\bibfnamefont {J.~D.}\
			\bibnamefont {Whittaker}}, \bibinfo {author} {\bibfnamefont {K.~W.}\
			\bibnamefont {Lehnert}},\ and\ \bibinfo {author} {\bibfnamefont {R.~W.}\
			\bibnamefont {Simmonds}},\ }\bibfield  {title} {\bibinfo {title} {Sideband
			cooling of micromechanical motion to the quantum ground state},\ }\href
	{https://doi.org/https://doi.org/10.1038/nature10261} {\bibfield  {journal}
		{\bibinfo  {journal} {Nature}\ }\textbf {\bibinfo {volume} {475}},\ \bibinfo
		{pages} {359} (\bibinfo {year} {2011}{\natexlab{a}})}\BibitemShut {NoStop}%
	\bibitem [{\citenamefont {Chan}\ \emph {et~al.}(2011)\citenamefont {Chan},
		\citenamefont {Alegre}, \citenamefont {Safavi-Naeini}, \citenamefont {Hill},
		\citenamefont {Krause}, \citenamefont {Gr{\"o}blacher}, \citenamefont
		{Aspelmeyer},\ and\ \citenamefont {Painter}}]{chan2011laser}%
	\BibitemOpen
	\bibfield  {author} {\bibinfo {author} {\bibfnamefont {J.}~\bibnamefont
			{Chan}}, \bibinfo {author} {\bibfnamefont {T.~M.}\ \bibnamefont {Alegre}},
		\bibinfo {author} {\bibfnamefont {A.~H.}\ \bibnamefont {Safavi-Naeini}},
		\bibinfo {author} {\bibfnamefont {J.~T.}\ \bibnamefont {Hill}}, \bibinfo
		{author} {\bibfnamefont {A.}~\bibnamefont {Krause}}, \bibinfo {author}
		{\bibfnamefont {S.}~\bibnamefont {Gr{\"o}blacher}}, \bibinfo {author}
		{\bibfnamefont {M.}~\bibnamefont {Aspelmeyer}},\ and\ \bibinfo {author}
		{\bibfnamefont {O.}~\bibnamefont {Painter}},\ }\bibfield  {title} {\bibinfo
		{title} {Laser cooling of a nanomechanical oscillator into its quantum ground
			state},\ }\href {https://www.nature.com/articles/nature10461} {\bibfield
		{journal} {\bibinfo  {journal} {Nature}\ }\textbf {\bibinfo {volume} {478}},\
		\bibinfo {pages} {89} (\bibinfo {year} {2011})}\BibitemShut {NoStop}%
	\bibitem [{\citenamefont {Palomaki}\ \emph
		{et~al.}(2013{\natexlab{a}})\citenamefont {Palomaki}, \citenamefont {Teufel},
		\citenamefont {Simmonds},\ and\ \citenamefont
		{Lehnert}}]{palomaki2013entangling}%
	\BibitemOpen
	\bibfield  {author} {\bibinfo {author} {\bibfnamefont {T.}~\bibnamefont
			{Palomaki}}, \bibinfo {author} {\bibfnamefont {J.}~\bibnamefont {Teufel}},
		\bibinfo {author} {\bibfnamefont {R.}~\bibnamefont {Simmonds}},\ and\
		\bibinfo {author} {\bibfnamefont {K.~W.}\ \bibnamefont {Lehnert}},\
	}\bibfield  {title} {\bibinfo {title} {Entangling mechanical motion with
			microwave fields},\ }\href
	{https://www.science.org/doi/full/10.1126/science.1244563} {\bibfield
		{journal} {\bibinfo  {journal} {Science}\ }\textbf {\bibinfo {volume}
			{342}},\ \bibinfo {pages} {710} (\bibinfo {year}
		{2013}{\natexlab{a}})}\BibitemShut {NoStop}%
	\bibitem [{\citenamefont {Barzanjeh}\ \emph {et~al.}(2019)\citenamefont
		{Barzanjeh}, \citenamefont {Redchenko}, \citenamefont {Peruzzo},
		\citenamefont {Wulf}, \citenamefont {Lewis}, \citenamefont {Arnold},\ and\
		\citenamefont {Fink}}]{barzanjeh2019stationary}%
	\BibitemOpen
	\bibfield  {author} {\bibinfo {author} {\bibfnamefont {S.}~\bibnamefont
			{Barzanjeh}}, \bibinfo {author} {\bibfnamefont {E.}~\bibnamefont
			{Redchenko}}, \bibinfo {author} {\bibfnamefont {M.}~\bibnamefont {Peruzzo}},
		\bibinfo {author} {\bibfnamefont {M.}~\bibnamefont {Wulf}}, \bibinfo {author}
		{\bibfnamefont {D.}~\bibnamefont {Lewis}}, \bibinfo {author} {\bibfnamefont
			{G.}~\bibnamefont {Arnold}},\ and\ \bibinfo {author} {\bibfnamefont {J.~M.}\
			\bibnamefont {Fink}},\ }\bibfield  {title} {\bibinfo {title} {Stationary
			entangled radiation from micromechanical motion},\ }\href
	{https://doi.org/https://doi.org/10.1038/s41586-019-1320-2} {\bibfield
		{journal} {\bibinfo  {journal} {Nature}\ }\textbf {\bibinfo {volume} {570}},\
		\bibinfo {pages} {480} (\bibinfo {year} {2019})}\BibitemShut {NoStop}%
	\bibitem [{\citenamefont {Ockeloen-Korppi}\ \emph {et~al.}(2018)\citenamefont
		{Ockeloen-Korppi}, \citenamefont {Damsk{\"a}gg}, \citenamefont
		{Pirkkalainen}, \citenamefont {Asjad}, \citenamefont {Clerk} \emph
		{et~al.}}]{ockeloen2018stabilized}%
	\BibitemOpen
	\bibfield  {author} {\bibinfo {author} {\bibfnamefont {C.}~\bibnamefont
			{Ockeloen-Korppi}}, \bibinfo {author} {\bibfnamefont {E.}~\bibnamefont
			{Damsk{\"a}gg}}, \bibinfo {author} {\bibfnamefont {J.-M.}\ \bibnamefont
			{Pirkkalainen}}, \bibinfo {author} {\bibfnamefont {M.}~\bibnamefont {Asjad}},
		\bibinfo {author} {\bibfnamefont {A.}~\bibnamefont {Clerk}}, \emph {et~al.},\
	}\bibfield  {title} {\bibinfo {title} {Stabilized entanglement of massive
			mechanical oscillators},\ }\href
	{https://www.nature.com/articles/s41586-018-0038-x} {\bibfield  {journal}
		{\bibinfo  {journal} {Nature}\ }\textbf {\bibinfo {volume} {556}},\ \bibinfo
		{pages} {478} (\bibinfo {year} {2018})}\BibitemShut {NoStop}%
	\bibitem [{\citenamefont {Kotler}\ \emph {et~al.}(2021)\citenamefont {Kotler},
		\citenamefont {Peterson}, \citenamefont {Shojaee}, \citenamefont {Lecocq},
		\citenamefont {Cicak}, \citenamefont {Kwiatkowski}, \citenamefont {Geller},
		\citenamefont {Glancy}, \citenamefont {Knill}, \citenamefont {Simmonds} \emph
		{et~al.}}]{kotler2021direct}%
	\BibitemOpen
	\bibfield  {author} {\bibinfo {author} {\bibfnamefont {S.}~\bibnamefont
			{Kotler}}, \bibinfo {author} {\bibfnamefont {G.~A.}\ \bibnamefont
			{Peterson}}, \bibinfo {author} {\bibfnamefont {E.}~\bibnamefont {Shojaee}},
		\bibinfo {author} {\bibfnamefont {F.}~\bibnamefont {Lecocq}}, \bibinfo
		{author} {\bibfnamefont {K.}~\bibnamefont {Cicak}}, \bibinfo {author}
		{\bibfnamefont {A.}~\bibnamefont {Kwiatkowski}}, \bibinfo {author}
		{\bibfnamefont {S.}~\bibnamefont {Geller}}, \bibinfo {author} {\bibfnamefont
			{S.}~\bibnamefont {Glancy}}, \bibinfo {author} {\bibfnamefont
			{E.}~\bibnamefont {Knill}}, \bibinfo {author} {\bibfnamefont {R.~W.}\
			\bibnamefont {Simmonds}}, \emph {et~al.},\ }\bibfield  {title} {\bibinfo
		{title} {Direct observation of deterministic macroscopic entanglement},\
	}\href {https://science.sciencemag.org/content/372/6542/622} {\bibfield
		{journal} {\bibinfo  {journal} {Science}\ }\textbf {\bibinfo {volume}
			{372}},\ \bibinfo {pages} {622} (\bibinfo {year} {2021})}\BibitemShut
	{NoStop}%
	\bibitem [{\citenamefont {Weinstein}\ \emph {et~al.}(2014)\citenamefont
		{Weinstein}, \citenamefont {Lei}, \citenamefont {Wollman}, \citenamefont
		{Suh}, \citenamefont {Metelmann}, \citenamefont {Clerk},\ and\ \citenamefont
		{Schwab}}]{weinstein2014observation}%
	\BibitemOpen
	\bibfield  {author} {\bibinfo {author} {\bibfnamefont {A.}~\bibnamefont
			{Weinstein}}, \bibinfo {author} {\bibfnamefont {C.}~\bibnamefont {Lei}},
		\bibinfo {author} {\bibfnamefont {E.}~\bibnamefont {Wollman}}, \bibinfo
		{author} {\bibfnamefont {J.}~\bibnamefont {Suh}}, \bibinfo {author}
		{\bibfnamefont {A.}~\bibnamefont {Metelmann}}, \bibinfo {author}
		{\bibfnamefont {A.}~\bibnamefont {Clerk}},\ and\ \bibinfo {author}
		{\bibfnamefont {K.}~\bibnamefont {Schwab}},\ }\bibfield  {title} {\bibinfo
		{title} {Observation and interpretation of motional sideband asymmetry in a
			quantum electromechanical device},\ }\href
	{https://journals.aps.org/prx/abstract/10.1103/PhysRevX.4.041003} {\bibfield
		{journal} {\bibinfo  {journal} {Physical Review X}\ }\textbf {\bibinfo
			{volume} {4}},\ \bibinfo {pages} {041003} (\bibinfo {year}
		{2014})}\BibitemShut {NoStop}%
	\bibitem [{\citenamefont {Hertzberg}\ \emph {et~al.}(2010)\citenamefont
		{Hertzberg}, \citenamefont {Rocheleau}, \citenamefont {Ndukum}, \citenamefont
		{Savva}, \citenamefont {Clerk},\ and\ \citenamefont
		{Schwab}}]{hertzberg2010back}%
	\BibitemOpen
	\bibfield  {author} {\bibinfo {author} {\bibfnamefont {J.}~\bibnamefont
			{Hertzberg}}, \bibinfo {author} {\bibfnamefont {T.}~\bibnamefont
			{Rocheleau}}, \bibinfo {author} {\bibfnamefont {T.}~\bibnamefont {Ndukum}},
		\bibinfo {author} {\bibfnamefont {M.}~\bibnamefont {Savva}}, \bibinfo
		{author} {\bibfnamefont {A.~A.}\ \bibnamefont {Clerk}},\ and\ \bibinfo
		{author} {\bibfnamefont {K.}~\bibnamefont {Schwab}},\ }\bibfield  {title}
	{\bibinfo {title} {Back-action-evading measurements of nanomechanical
			motion},\ }\href@noop {} {\bibfield  {journal} {\bibinfo  {journal} {Nature
				Physics}\ }\textbf {\bibinfo {volume} {6}},\ \bibinfo {pages} {213} (\bibinfo
		{year} {2010})}\BibitemShut {NoStop}%
	\bibitem [{\citenamefont {Shomroni}\ \emph {et~al.}(2019)\citenamefont
		{Shomroni}, \citenamefont {Qiu}, \citenamefont {Malz}, \citenamefont
		{Nunnenkamp},\ and\ \citenamefont {Kippenberg}}]{shomroni2019optical}%
	\BibitemOpen
	\bibfield  {author} {\bibinfo {author} {\bibfnamefont {I.}~\bibnamefont
			{Shomroni}}, \bibinfo {author} {\bibfnamefont {L.}~\bibnamefont {Qiu}},
		\bibinfo {author} {\bibfnamefont {D.}~\bibnamefont {Malz}}, \bibinfo {author}
		{\bibfnamefont {A.}~\bibnamefont {Nunnenkamp}},\ and\ \bibinfo {author}
		{\bibfnamefont {T.~J.}\ \bibnamefont {Kippenberg}},\ }\bibfield  {title}
	{\bibinfo {title} {Optical backaction-evading measurement of a mechanical
			oscillator},\ }\href {https://www.nature.com/articles/s41467-019-10024-3}
	{\bibfield  {journal} {\bibinfo  {journal} {Nature communications}\ }\textbf
		{\bibinfo {volume} {10}},\ \bibinfo {pages} {2086} (\bibinfo {year}
		{2019})}\BibitemShut {NoStop}%
	\bibitem [{\citenamefont {Purdy}\ \emph {et~al.}(2013)\citenamefont {Purdy},
		\citenamefont {Yu}, \citenamefont {Peterson}, \citenamefont {Kampel},\ and\
		\citenamefont {Regal}}]{purdy2013strong}%
	\BibitemOpen
	\bibfield  {author} {\bibinfo {author} {\bibfnamefont {T.~P.}\ \bibnamefont
			{Purdy}}, \bibinfo {author} {\bibfnamefont {P.-L.}\ \bibnamefont {Yu}},
		\bibinfo {author} {\bibfnamefont {R.}~\bibnamefont {Peterson}}, \bibinfo
		{author} {\bibfnamefont {N.}~\bibnamefont {Kampel}},\ and\ \bibinfo {author}
		{\bibfnamefont {C.}~\bibnamefont {Regal}},\ }\bibfield  {title} {\bibinfo
		{title} {Strong optomechanical squeezing of light},\ }\href
	{https://journals.aps.org/prx/abstract/10.1103/PhysRevX.3.031012} {\bibfield
		{journal} {\bibinfo  {journal} {Physical Review X}\ }\textbf {\bibinfo
			{volume} {3}},\ \bibinfo {pages} {031012} (\bibinfo {year}
		{2013})}\BibitemShut {NoStop}%
	\bibitem [{\citenamefont {Verhagen}\ \emph {et~al.}(2012)\citenamefont
		{Verhagen}, \citenamefont {Del{\'e}glise}, \citenamefont {Weis},
		\citenamefont {Schliesser},\ and\ \citenamefont
		{Kippenberg}}]{verhagen2012quantum}%
	\BibitemOpen
	\bibfield  {author} {\bibinfo {author} {\bibfnamefont {E.}~\bibnamefont
			{Verhagen}}, \bibinfo {author} {\bibfnamefont {S.}~\bibnamefont
			{Del{\'e}glise}}, \bibinfo {author} {\bibfnamefont {S.}~\bibnamefont {Weis}},
		\bibinfo {author} {\bibfnamefont {A.}~\bibnamefont {Schliesser}},\ and\
		\bibinfo {author} {\bibfnamefont {T.~J.}\ \bibnamefont {Kippenberg}},\
	}\bibfield  {title} {\bibinfo {title} {Quantum-coherent coupling of a
			mechanical oscillator to an optical cavity mode},\ }\href@noop {} {\bibfield
		{journal} {\bibinfo  {journal} {Nature}\ }\textbf {\bibinfo {volume} {482}},\
		\bibinfo {pages} {63} (\bibinfo {year} {2012})}\BibitemShut {NoStop}%
	\bibitem [{\citenamefont {Wilson}\ \emph {et~al.}(2015)\citenamefont {Wilson},
		\citenamefont {Sudhir}, \citenamefont {Piro}, \citenamefont {Schilling},
		\citenamefont {Ghadimi},\ and\ \citenamefont
		{Kippenberg}}]{wilson2015measurement}%
	\BibitemOpen
	\bibfield  {author} {\bibinfo {author} {\bibfnamefont {D.~J.}\ \bibnamefont
			{Wilson}}, \bibinfo {author} {\bibfnamefont {V.}~\bibnamefont {Sudhir}},
		\bibinfo {author} {\bibfnamefont {N.}~\bibnamefont {Piro}}, \bibinfo {author}
		{\bibfnamefont {R.}~\bibnamefont {Schilling}}, \bibinfo {author}
		{\bibfnamefont {A.}~\bibnamefont {Ghadimi}},\ and\ \bibinfo {author}
		{\bibfnamefont {T.~J.}\ \bibnamefont {Kippenberg}},\ }\bibfield  {title}
	{\bibinfo {title} {Measurement-based control of a mechanical oscillator at
			its thermal decoherence rate},\ }\href
	{https://www.nature.com/articles/nature14672} {\bibfield  {journal} {\bibinfo
			{journal} {Nature}\ }\textbf {\bibinfo {volume} {524}},\ \bibinfo {pages}
		{325} (\bibinfo {year} {2015})}\BibitemShut {NoStop}%
	\bibitem [{\citenamefont {Andrews}\ \emph {et~al.}(2014)\citenamefont
		{Andrews}, \citenamefont {Peterson}, \citenamefont {Purdy}, \citenamefont
		{Cicak}, \citenamefont {Simmonds} \emph {et~al.}}]{andrews2014bidirectional}%
	\BibitemOpen
	\bibfield  {author} {\bibinfo {author} {\bibfnamefont {R.~W.}\ \bibnamefont
			{Andrews}}, \bibinfo {author} {\bibfnamefont {R.~W.}\ \bibnamefont
			{Peterson}}, \bibinfo {author} {\bibfnamefont {T.~P.}\ \bibnamefont {Purdy}},
		\bibinfo {author} {\bibfnamefont {K.}~\bibnamefont {Cicak}}, \bibinfo
		{author} {\bibfnamefont {R.~W.}\ \bibnamefont {Simmonds}}, \emph {et~al.},\
	}\bibfield  {title} {\bibinfo {title} {Bidirectional and efficient conversion
			between microwave and optical light},\ }\href
	{https://www.nature.com/articles/nphys2911} {\bibfield  {journal} {\bibinfo
			{journal} {Nature Physics}\ }\textbf {\bibinfo {volume} {10}},\ \bibinfo
		{pages} {321} (\bibinfo {year} {2014})}\BibitemShut {NoStop}%
	\bibitem [{\citenamefont {Delaney}\ \emph {et~al.}(2022)\citenamefont
		{Delaney}, \citenamefont {Urmey}, \citenamefont {Mittal}, \citenamefont
		{Brubaker}, \citenamefont {Kindem}, \citenamefont {Burns}, \citenamefont
		{Regal},\ and\ \citenamefont {Lehnert}}]{delaney2022superconducting}%
	\BibitemOpen
	\bibfield  {author} {\bibinfo {author} {\bibfnamefont {R.}~\bibnamefont
			{Delaney}}, \bibinfo {author} {\bibfnamefont {M.}~\bibnamefont {Urmey}},
		\bibinfo {author} {\bibfnamefont {S.}~\bibnamefont {Mittal}}, \bibinfo
		{author} {\bibfnamefont {B.}~\bibnamefont {Brubaker}}, \bibinfo {author}
		{\bibfnamefont {J.}~\bibnamefont {Kindem}}, \bibinfo {author} {\bibfnamefont
			{P.}~\bibnamefont {Burns}}, \bibinfo {author} {\bibfnamefont
			{C.}~\bibnamefont {Regal}},\ and\ \bibinfo {author} {\bibfnamefont
			{K.}~\bibnamefont {Lehnert}},\ }\bibfield  {title} {\bibinfo {title}
		{Superconducting-qubit readout via low-backaction electro-optic
			transduction},\ }\href {https://www.nature.com/articles/s41586-022-04720-2}
	{\bibfield  {journal} {\bibinfo  {journal} {Nature}\ }\textbf {\bibinfo
			{volume} {606}},\ \bibinfo {pages} {489} (\bibinfo {year}
		{2022})}\BibitemShut {NoStop}%
	\bibitem [{\citenamefont {Massel}\ \emph {et~al.}(2011)\citenamefont {Massel},
		\citenamefont {Heikkil{\"a}}, \citenamefont {Pirkkalainen}, \citenamefont
		{Cho}, \citenamefont {Saloniemi}, \citenamefont {Hakonen},\ and\
		\citenamefont {Sillanp{\"a}{\"a}}}]{massel2011microwave}%
	\BibitemOpen
	\bibfield  {author} {\bibinfo {author} {\bibfnamefont {F.}~\bibnamefont
			{Massel}}, \bibinfo {author} {\bibfnamefont {T.~T.}\ \bibnamefont
			{Heikkil{\"a}}}, \bibinfo {author} {\bibfnamefont {J.-M.}\ \bibnamefont
			{Pirkkalainen}}, \bibinfo {author} {\bibfnamefont {S.-U.}\ \bibnamefont
			{Cho}}, \bibinfo {author} {\bibfnamefont {H.}~\bibnamefont {Saloniemi}},
		\bibinfo {author} {\bibfnamefont {P.~J.}\ \bibnamefont {Hakonen}},\ and\
		\bibinfo {author} {\bibfnamefont {M.~A.}\ \bibnamefont {Sillanp{\"a}{\"a}}},\
	}\bibfield  {title} {\bibinfo {title} {Microwave amplification with
			nanomechanical resonators},\ }\href@noop {} {\bibfield  {journal} {\bibinfo
			{journal} {Nature}\ }\textbf {\bibinfo {volume} {480}},\ \bibinfo {pages}
		{351} (\bibinfo {year} {2011})}\BibitemShut {NoStop}%
	\bibitem [{\citenamefont {Rocheleau}\ \emph {et~al.}(2010)\citenamefont
		{Rocheleau}, \citenamefont {Ndukum}, \citenamefont {Macklin}, \citenamefont
		{Hertzberg}, \citenamefont {Clerk},\ and\ \citenamefont
		{Schwab}}]{rocheleau2010preparation}%
	\BibitemOpen
	\bibfield  {author} {\bibinfo {author} {\bibfnamefont {T.}~\bibnamefont
			{Rocheleau}}, \bibinfo {author} {\bibfnamefont {T.}~\bibnamefont {Ndukum}},
		\bibinfo {author} {\bibfnamefont {C.}~\bibnamefont {Macklin}}, \bibinfo
		{author} {\bibfnamefont {J.}~\bibnamefont {Hertzberg}}, \bibinfo {author}
		{\bibfnamefont {A.}~\bibnamefont {Clerk}},\ and\ \bibinfo {author}
		{\bibfnamefont {K.}~\bibnamefont {Schwab}},\ }\bibfield  {title} {\bibinfo
		{title} {Preparation and detection of a mechanical resonator near the ground
			state of motion},\ }\href@noop {} {\bibfield  {journal} {\bibinfo  {journal}
			{Nature}\ }\textbf {\bibinfo {volume} {463}},\ \bibinfo {pages} {72}
		(\bibinfo {year} {2010})}\BibitemShut {NoStop}%
	\bibitem [{\citenamefont {Cicak}\ \emph {et~al.}(2010)\citenamefont {Cicak},
		\citenamefont {Li}, \citenamefont {Strong}, \citenamefont {Allman},
		\citenamefont {Altomare}, \citenamefont {Sirois}, \citenamefont {Whittaker},
		\citenamefont {Teufel},\ and\ \citenamefont {Simmonds}}]{cicak2010low}%
	\BibitemOpen
	\bibfield  {author} {\bibinfo {author} {\bibfnamefont {K.}~\bibnamefont
			{Cicak}}, \bibinfo {author} {\bibfnamefont {D.}~\bibnamefont {Li}}, \bibinfo
		{author} {\bibfnamefont {J.~A.}\ \bibnamefont {Strong}}, \bibinfo {author}
		{\bibfnamefont {M.~S.}\ \bibnamefont {Allman}}, \bibinfo {author}
		{\bibfnamefont {F.}~\bibnamefont {Altomare}}, \bibinfo {author}
		{\bibfnamefont {A.~J.}\ \bibnamefont {Sirois}}, \bibinfo {author}
		{\bibfnamefont {J.~D.}\ \bibnamefont {Whittaker}}, \bibinfo {author}
		{\bibfnamefont {J.~D.}\ \bibnamefont {Teufel}},\ and\ \bibinfo {author}
		{\bibfnamefont {R.~W.}\ \bibnamefont {Simmonds}},\ }\bibfield  {title}
	{\bibinfo {title} {Low-loss superconducting resonant circuits using
			vacuum-gap-based microwave components},\ }\href
	{https://doi.org/https://doi.org/10.1063/1.3304168} {\bibfield  {journal}
		{\bibinfo  {journal} {Applied Physics Letters}\ }\textbf {\bibinfo {volume}
			{96}},\ \bibinfo {pages} {093502} (\bibinfo {year} {2010})}\BibitemShut
	{NoStop}%
	\bibitem [{\citenamefont {Clark}\ \emph {et~al.}(2017)\citenamefont {Clark},
		\citenamefont {Lecocq}, \citenamefont {Simmonds}, \citenamefont {Aumentado},\
		and\ \citenamefont {Teufel}}]{clark2017sideband}%
	\BibitemOpen
	\bibfield  {author} {\bibinfo {author} {\bibfnamefont {J.~B.}\ \bibnamefont
			{Clark}}, \bibinfo {author} {\bibfnamefont {F.}~\bibnamefont {Lecocq}},
		\bibinfo {author} {\bibfnamefont {R.~W.}\ \bibnamefont {Simmonds}}, \bibinfo
		{author} {\bibfnamefont {J.}~\bibnamefont {Aumentado}},\ and\ \bibinfo
		{author} {\bibfnamefont {J.~D.}\ \bibnamefont {Teufel}},\ }\bibfield  {title}
	{\bibinfo {title} {Sideband cooling beyond the quantum backaction limit with
			squeezed light},\ }\href {https://www.nature.com/articles/nature20604}
	{\bibfield  {journal} {\bibinfo  {journal} {Nature}\ }\textbf {\bibinfo
			{volume} {541}},\ \bibinfo {pages} {191} (\bibinfo {year}
		{2017})}\BibitemShut {NoStop}%
	\bibitem [{\citenamefont {Wollman}\ \emph {et~al.}(2015)\citenamefont
		{Wollman}, \citenamefont {Lei}, \citenamefont {Weinstein}, \citenamefont
		{Suh}, \citenamefont {Kronwald}, \citenamefont {Marquardt}, \citenamefont
		{Clerk},\ and\ \citenamefont {Schwab}}]{wollman2015quantum}%
	\BibitemOpen
	\bibfield  {author} {\bibinfo {author} {\bibfnamefont {E.~E.}\ \bibnamefont
			{Wollman}}, \bibinfo {author} {\bibfnamefont {C.}~\bibnamefont {Lei}},
		\bibinfo {author} {\bibfnamefont {A.}~\bibnamefont {Weinstein}}, \bibinfo
		{author} {\bibfnamefont {J.}~\bibnamefont {Suh}}, \bibinfo {author}
		{\bibfnamefont {A.}~\bibnamefont {Kronwald}}, \bibinfo {author}
		{\bibfnamefont {F.}~\bibnamefont {Marquardt}}, \bibinfo {author}
		{\bibfnamefont {A.~A.}\ \bibnamefont {Clerk}},\ and\ \bibinfo {author}
		{\bibfnamefont {K.}~\bibnamefont {Schwab}},\ }\bibfield  {title} {\bibinfo
		{title} {Quantum squeezing of motion in a mechanical resonator},\ }\href
	{https://www.science.org/doi/10.1126/science.aac5138} {\bibfield  {journal}
		{\bibinfo  {journal} {Science}\ }\textbf {\bibinfo {volume} {349}},\ \bibinfo
		{pages} {952} (\bibinfo {year} {2015})}\BibitemShut {NoStop}%
	\bibitem [{\citenamefont {Pirkkalainen}\ \emph {et~al.}(2015)\citenamefont
		{Pirkkalainen}, \citenamefont {Damsk{\"a}gg}, \citenamefont {Brandt},
		\citenamefont {Massel},\ and\ \citenamefont
		{Sillanp{\"a}{\"a}}}]{pirkkalainen2015squeezing}%
	\BibitemOpen
	\bibfield  {author} {\bibinfo {author} {\bibfnamefont {J.-M.}\ \bibnamefont
			{Pirkkalainen}}, \bibinfo {author} {\bibfnamefont {E.}~\bibnamefont
			{Damsk{\"a}gg}}, \bibinfo {author} {\bibfnamefont {M.}~\bibnamefont
			{Brandt}}, \bibinfo {author} {\bibfnamefont {F.}~\bibnamefont {Massel}},\
		and\ \bibinfo {author} {\bibfnamefont {M.~A.}\ \bibnamefont
			{Sillanp{\"a}{\"a}}},\ }\bibfield  {title} {\bibinfo {title} {Squeezing of
			quantum noise of motion in a micromechanical resonator},\ }\href
	{https://journals.aps.org/prl/abstract/10.1103/PhysRevLett.115.243601}
	{\bibfield  {journal} {\bibinfo  {journal} {Physical Review Letters}\
		}\textbf {\bibinfo {volume} {115}},\ \bibinfo {pages} {243601} (\bibinfo
		{year} {2015})}\BibitemShut {NoStop}%
	\bibitem [{\citenamefont {Lecocq}\ \emph {et~al.}(2015)\citenamefont {Lecocq},
		\citenamefont {Clark}, \citenamefont {Simmonds}, \citenamefont {Aumentado},\
		and\ \citenamefont {Teufel}}]{lecocq2015quantum}%
	\BibitemOpen
	\bibfield  {author} {\bibinfo {author} {\bibfnamefont {F.}~\bibnamefont
			{Lecocq}}, \bibinfo {author} {\bibfnamefont {J.~B.}\ \bibnamefont {Clark}},
		\bibinfo {author} {\bibfnamefont {R.~W.}\ \bibnamefont {Simmonds}}, \bibinfo
		{author} {\bibfnamefont {J.}~\bibnamefont {Aumentado}},\ and\ \bibinfo
		{author} {\bibfnamefont {J.~D.}\ \bibnamefont {Teufel}},\ }\bibfield  {title}
	{\bibinfo {title} {Quantum nondemolition measurement of a nonclassical state
			of a massive object},\ }\href
	{https://journals.aps.org/prx/abstract/10.1103/PhysRevX.5.041037} {\bibfield
		{journal} {\bibinfo  {journal} {Physical Review X}\ }\textbf {\bibinfo
			{volume} {5}},\ \bibinfo {pages} {041037} (\bibinfo {year}
		{2015})}\BibitemShut {NoStop}%
	\bibitem [{\citenamefont {Mercier~de L{\'e}pinay}\ \emph
		{et~al.}(2021)\citenamefont {Mercier~de L{\'e}pinay}, \citenamefont
		{Ockeloen-Korppi}, \citenamefont {Woolley},\ and\ \citenamefont
		{Sillanp{\"a}{\"a}}}]{mercier2021quantum}%
	\BibitemOpen
	\bibfield  {author} {\bibinfo {author} {\bibfnamefont {L.}~\bibnamefont
			{Mercier~de L{\'e}pinay}}, \bibinfo {author} {\bibfnamefont {C.~F.}\
			\bibnamefont {Ockeloen-Korppi}}, \bibinfo {author} {\bibfnamefont {M.~J.}\
			\bibnamefont {Woolley}},\ and\ \bibinfo {author} {\bibfnamefont {M.~A.}\
			\bibnamefont {Sillanp{\"a}{\"a}}},\ }\bibfield  {title} {\bibinfo {title}
		{Quantum mechanics--free subsystem with mechanical oscillators},\ }\href@noop
	{} {\bibfield  {journal} {\bibinfo  {journal} {Science}\ }\textbf {\bibinfo
			{volume} {372}},\ \bibinfo {pages} {625} (\bibinfo {year}
		{2021})}\BibitemShut {NoStop}%
	\bibitem [{\citenamefont {Reed}\ \emph {et~al.}(2017)\citenamefont {Reed},
		\citenamefont {Mayer}, \citenamefont {Teufel}, \citenamefont {Burkhart},
		\citenamefont {Pfaff}, \citenamefont {Reagor}, \citenamefont {Sletten},
		\citenamefont {Ma}, \citenamefont {Schoelkopf}, \citenamefont {Knill} \emph
		{et~al.}}]{reed2017faithful}%
	\BibitemOpen
	\bibfield  {author} {\bibinfo {author} {\bibfnamefont {A.}~\bibnamefont
			{Reed}}, \bibinfo {author} {\bibfnamefont {K.}~\bibnamefont {Mayer}},
		\bibinfo {author} {\bibfnamefont {J.}~\bibnamefont {Teufel}}, \bibinfo
		{author} {\bibfnamefont {L.}~\bibnamefont {Burkhart}}, \bibinfo {author}
		{\bibfnamefont {W.}~\bibnamefont {Pfaff}}, \bibinfo {author} {\bibfnamefont
			{M.}~\bibnamefont {Reagor}}, \bibinfo {author} {\bibfnamefont
			{L.}~\bibnamefont {Sletten}}, \bibinfo {author} {\bibfnamefont
			{X.}~\bibnamefont {Ma}}, \bibinfo {author} {\bibfnamefont {R.}~\bibnamefont
			{Schoelkopf}}, \bibinfo {author} {\bibfnamefont {E.}~\bibnamefont {Knill}},
		\emph {et~al.},\ }\bibfield  {title} {\bibinfo {title} {Faithful conversion
			of propagating quantum information to mechanical motion},\ }\href
	{https://www.nature.com/articles/nphys4251} {\bibfield  {journal} {\bibinfo
			{journal} {Nature Physics}\ }\textbf {\bibinfo {volume} {13}},\ \bibinfo
		{pages} {1163} (\bibinfo {year} {2017})}\BibitemShut {NoStop}%
	\bibitem [{\citenamefont {Palomaki}\ \emph
		{et~al.}(2013{\natexlab{b}})\citenamefont {Palomaki}, \citenamefont {Harlow},
		\citenamefont {Teufel}, \citenamefont {Simmonds},\ and\ \citenamefont
		{Lehnert}}]{palomaki2013coherent}%
	\BibitemOpen
	\bibfield  {author} {\bibinfo {author} {\bibfnamefont {T.}~\bibnamefont
			{Palomaki}}, \bibinfo {author} {\bibfnamefont {J.}~\bibnamefont {Harlow}},
		\bibinfo {author} {\bibfnamefont {J.}~\bibnamefont {Teufel}}, \bibinfo
		{author} {\bibfnamefont {R.}~\bibnamefont {Simmonds}},\ and\ \bibinfo
		{author} {\bibfnamefont {K.~W.}\ \bibnamefont {Lehnert}},\ }\bibfield
	{title} {\bibinfo {title} {Coherent state transfer between itinerant
			microwave fields and a mechanical oscillator},\ }\href
	{https://www.nature.com/articles/nature11915} {\bibfield  {journal} {\bibinfo
			{journal} {Nature}\ }\textbf {\bibinfo {volume} {495}},\ \bibinfo {pages}
		{210} (\bibinfo {year} {2013}{\natexlab{b}})}\BibitemShut {NoStop}%
	\bibitem [{\citenamefont {Bernier}\ \emph {et~al.}(2017)\citenamefont
		{Bernier}, \citenamefont {Tóth}, \citenamefont {Koottandavida},
		\citenamefont {Ioannou}, \citenamefont {Malz} \emph {et~al.}}]{Bernier2017}%
	\BibitemOpen
	\bibfield  {author} {\bibinfo {author} {\bibfnamefont {N.~R.}\ \bibnamefont
			{Bernier}}, \bibinfo {author} {\bibfnamefont {L.~D.}\ \bibnamefont {Tóth}},
		\bibinfo {author} {\bibfnamefont {A.}~\bibnamefont {Koottandavida}}, \bibinfo
		{author} {\bibfnamefont {M.~A.}\ \bibnamefont {Ioannou}}, \bibinfo {author}
		{\bibfnamefont {D.}~\bibnamefont {Malz}}, \emph {et~al.},\ }\bibfield
	{title} {\bibinfo {title} {Nonreciprocal reconfigurable microwave
			optomechanical circuit},\ }\bibfield  {journal} {\bibinfo  {journal} {Nature
			Communications}\ }\textbf {\bibinfo {volume} {8}},\ \href
	{https://doi.org/10.1038/s41467-017-00447-1} {10.1038/s41467-017-00447-1}
	(\bibinfo {year} {2017})\BibitemShut {NoStop}%
	\bibitem [{\citenamefont {de~L{\'e}pinay}\ \emph {et~al.}(2019)\citenamefont
		{de~L{\'e}pinay}, \citenamefont {Damsk{\"a}gg}, \citenamefont
		{Ockeloen-Korppi},\ and\ \citenamefont
		{Sillanp{\"a}{\"a}}}]{de2019realization}%
	\BibitemOpen
	\bibfield  {author} {\bibinfo {author} {\bibfnamefont {L.~M.}\ \bibnamefont
			{de~L{\'e}pinay}}, \bibinfo {author} {\bibfnamefont {E.}~\bibnamefont
			{Damsk{\"a}gg}}, \bibinfo {author} {\bibfnamefont {C.~F.}\ \bibnamefont
			{Ockeloen-Korppi}},\ and\ \bibinfo {author} {\bibfnamefont {M.~A.}\
			\bibnamefont {Sillanp{\"a}{\"a}}},\ }\bibfield  {title} {\bibinfo {title}
		{Realization of directional amplification in a microwave optomechanical
			device},\ }\href@noop {} {\bibfield  {journal} {\bibinfo  {journal} {Physical
				Review Applied}\ }\textbf {\bibinfo {volume} {11}},\ \bibinfo {pages}
		{034027} (\bibinfo {year} {2019})}\BibitemShut {NoStop}%
	\bibitem [{\citenamefont {Barzanjeh}\ \emph {et~al.}(2017)\citenamefont
		{Barzanjeh}, \citenamefont {Wulf}, \citenamefont {Peruzzo}, \citenamefont
		{Kalaee}, \citenamefont {Dieterle}, \citenamefont {Painter},\ and\
		\citenamefont {Fink}}]{barzanjeh2017mechanical}%
	\BibitemOpen
	\bibfield  {author} {\bibinfo {author} {\bibfnamefont {S.}~\bibnamefont
			{Barzanjeh}}, \bibinfo {author} {\bibfnamefont {M.}~\bibnamefont {Wulf}},
		\bibinfo {author} {\bibfnamefont {M.}~\bibnamefont {Peruzzo}}, \bibinfo
		{author} {\bibfnamefont {M.}~\bibnamefont {Kalaee}}, \bibinfo {author}
		{\bibfnamefont {P.}~\bibnamefont {Dieterle}}, \bibinfo {author}
		{\bibfnamefont {O.}~\bibnamefont {Painter}},\ and\ \bibinfo {author}
		{\bibfnamefont {J.~M.}\ \bibnamefont {Fink}},\ }\bibfield  {title} {\bibinfo
		{title} {Mechanical on-chip microwave circulator},\ }\href
	{https://www.nature.com/articles/s41467-017-01304-x} {\bibfield  {journal}
		{\bibinfo  {journal} {Nature communications}\ }\textbf {\bibinfo {volume}
			{8}},\ \bibinfo {pages} {1} (\bibinfo {year} {2017})}\BibitemShut {NoStop}%
	\bibitem [{\citenamefont {Teufel}\ \emph
		{et~al.}(2011{\natexlab{b}})\citenamefont {Teufel}, \citenamefont {Li},
		\citenamefont {Allman}, \citenamefont {Cicak}, \citenamefont {Sirois},
		\citenamefont {Whittaker},\ and\ \citenamefont
		{Simmonds}}]{teufel2011circuit}%
	\BibitemOpen
	\bibfield  {author} {\bibinfo {author} {\bibfnamefont {J.~D.}\ \bibnamefont
			{Teufel}}, \bibinfo {author} {\bibfnamefont {D.}~\bibnamefont {Li}}, \bibinfo
		{author} {\bibfnamefont {M.}~\bibnamefont {Allman}}, \bibinfo {author}
		{\bibfnamefont {K.}~\bibnamefont {Cicak}}, \bibinfo {author} {\bibfnamefont
			{A.}~\bibnamefont {Sirois}}, \bibinfo {author} {\bibfnamefont
			{J.}~\bibnamefont {Whittaker}},\ and\ \bibinfo {author} {\bibfnamefont
			{R.}~\bibnamefont {Simmonds}},\ }\bibfield  {title} {\bibinfo {title}
		{Circuit cavity electromechanics in the strong-coupling regime},\ }\href
	{https://www.nature.com/articles/nature09898} {\bibfield  {journal} {\bibinfo
			{journal} {Nature}\ }\textbf {\bibinfo {volume} {471}},\ \bibinfo {pages}
		{204} (\bibinfo {year} {2011}{\natexlab{b}})}\BibitemShut {NoStop}%
	\bibitem [{\citenamefont {Suh}\ \emph {et~al.}(2014)\citenamefont {Suh},
		\citenamefont {Weinstein}, \citenamefont {Lei}, \citenamefont {Wollman},
		\citenamefont {Steinke}, \citenamefont {Meystre}, \citenamefont {Clerk},\
		and\ \citenamefont {Schwab}}]{suh2014mechanically}%
	\BibitemOpen
	\bibfield  {author} {\bibinfo {author} {\bibfnamefont {J.}~\bibnamefont
			{Suh}}, \bibinfo {author} {\bibfnamefont {A.}~\bibnamefont {Weinstein}},
		\bibinfo {author} {\bibfnamefont {C.}~\bibnamefont {Lei}}, \bibinfo {author}
		{\bibfnamefont {E.}~\bibnamefont {Wollman}}, \bibinfo {author} {\bibfnamefont
			{S.}~\bibnamefont {Steinke}}, \bibinfo {author} {\bibfnamefont
			{P.}~\bibnamefont {Meystre}}, \bibinfo {author} {\bibfnamefont {A.~A.}\
			\bibnamefont {Clerk}},\ and\ \bibinfo {author} {\bibfnamefont
			{K.}~\bibnamefont {Schwab}},\ }\bibfield  {title} {\bibinfo {title}
		{Mechanically detecting and avoiding the quantum fluctuations of a microwave
			field},\ }\href {https://www.science.org/doi/full/10.1126/science.1253258}
	{\bibfield  {journal} {\bibinfo  {journal} {Science}\ }\textbf {\bibinfo
			{volume} {344}},\ \bibinfo {pages} {1262} (\bibinfo {year}
		{2014})}\BibitemShut {NoStop}%
	\bibitem [{\citenamefont {Toth}\ \emph {et~al.}(2017)\citenamefont {Toth},
		\citenamefont {Bernier}, \citenamefont {Nunnenkamp}, \citenamefont
		{Feofanov},\ and\ \citenamefont {Kippenberg}}]{toth2017dissipative}%
	\BibitemOpen
	\bibfield  {author} {\bibinfo {author} {\bibfnamefont {L.~D.}\ \bibnamefont
			{Toth}}, \bibinfo {author} {\bibfnamefont {N.~R.}\ \bibnamefont {Bernier}},
		\bibinfo {author} {\bibfnamefont {A.}~\bibnamefont {Nunnenkamp}}, \bibinfo
		{author} {\bibfnamefont {A.}~\bibnamefont {Feofanov}},\ and\ \bibinfo
		{author} {\bibfnamefont {T.}~\bibnamefont {Kippenberg}},\ }\bibfield  {title}
	{\bibinfo {title} {A dissipative quantum reservoir for microwave light using
			a mechanical oscillator},\ }\href {https://www.nature.com/articles/nphys4121}
	{\bibfield  {journal} {\bibinfo  {journal} {Nature Physics}\ }\textbf
		{\bibinfo {volume} {13}},\ \bibinfo {pages} {787} (\bibinfo {year}
		{2017})}\BibitemShut {NoStop}%
	\bibitem [{\citenamefont {Liu}\ \emph {et~al.}(2025)\citenamefont {Liu},
		\citenamefont {Sun}, \citenamefont {Liu}, \citenamefont {Wu}, \citenamefont
		{Sillanp{\"a}{\"a}},\ and\ \citenamefont {Li}}]{liu2025degeneracy}%
	\BibitemOpen
	\bibfield  {author} {\bibinfo {author} {\bibfnamefont {Y.}~\bibnamefont
			{Liu}}, \bibinfo {author} {\bibfnamefont {H.}~\bibnamefont {Sun}}, \bibinfo
		{author} {\bibfnamefont {Q.}~\bibnamefont {Liu}}, \bibinfo {author}
		{\bibfnamefont {H.}~\bibnamefont {Wu}}, \bibinfo {author} {\bibfnamefont
			{M.~A.}\ \bibnamefont {Sillanp{\"a}{\"a}}},\ and\ \bibinfo {author}
		{\bibfnamefont {T.}~\bibnamefont {Li}},\ }\bibfield  {title} {\bibinfo
		{title} {Degeneracy-breaking and long-lived multimode microwave
			electromechanical systems enabled by cubic silicon-carbide membrane
			crystals},\ }\href@noop {} {\bibfield  {journal} {\bibinfo  {journal} {Nature
				Communications}\ }\textbf {\bibinfo {volume} {16}},\ \bibinfo {pages} {1207}
		(\bibinfo {year} {2025})}\BibitemShut {NoStop}%
	\bibitem [{\citenamefont {Seis}\ \emph {et~al.}(2022)\citenamefont {Seis},
		\citenamefont {Capelle}, \citenamefont {Langman}, \citenamefont {Saarinen},
		\citenamefont {Planz},\ and\ \citenamefont {Schliesser}}]{seis2022ground}%
	\BibitemOpen
	\bibfield  {author} {\bibinfo {author} {\bibfnamefont {Y.}~\bibnamefont
			{Seis}}, \bibinfo {author} {\bibfnamefont {T.}~\bibnamefont {Capelle}},
		\bibinfo {author} {\bibfnamefont {E.}~\bibnamefont {Langman}}, \bibinfo
		{author} {\bibfnamefont {S.}~\bibnamefont {Saarinen}}, \bibinfo {author}
		{\bibfnamefont {E.}~\bibnamefont {Planz}},\ and\ \bibinfo {author}
		{\bibfnamefont {A.}~\bibnamefont {Schliesser}},\ }\bibfield  {title}
	{\bibinfo {title} {Ground state cooling of an ultracoherent electromechanical
			system},\ }\href {https://www.nature.com/articles/s41467-022-29115-9}
	{\bibfield  {journal} {\bibinfo  {journal} {Nature communications}\ }\textbf
		{\bibinfo {volume} {13}},\ \bibinfo {pages} {1} (\bibinfo {year}
		{2022})}\BibitemShut {NoStop}%
	\bibitem [{\citenamefont {T{\'o}th}(2018)}]{toth2018dissipation}%
	\BibitemOpen
	\bibfield  {author} {\bibinfo {author} {\bibfnamefont {L.~D.}\ \bibnamefont
			{T{\'o}th}},\ }\href
	{https://doi.org/http://dx.doi.org/10.5075/epfl-thesis-8709} {\emph {\bibinfo
			{title} {Dissipation as a resource in circuit quantum electromechanics}}},\
	\bibinfo {type} {Tech. Rep.}\ (\bibinfo  {institution} {EPFL},\ \bibinfo
	{year} {2018})\BibitemShut {NoStop}%
	\bibitem [{\citenamefont {Bereyhi}\ \emph {et~al.}(2019)\citenamefont
		{Bereyhi}, \citenamefont {Beccari}, \citenamefont {Fedorov}, \citenamefont
		{Ghadimi}, \citenamefont {Schilling}, \citenamefont {Wilson}, \citenamefont
		{Engelsen},\ and\ \citenamefont {Kippenberg}}]{bereyhi2019clamp}%
	\BibitemOpen
	\bibfield  {author} {\bibinfo {author} {\bibfnamefont {M.~J.}\ \bibnamefont
			{Bereyhi}}, \bibinfo {author} {\bibfnamefont {A.}~\bibnamefont {Beccari}},
		\bibinfo {author} {\bibfnamefont {S.~A.}\ \bibnamefont {Fedorov}}, \bibinfo
		{author} {\bibfnamefont {A.~H.}\ \bibnamefont {Ghadimi}}, \bibinfo {author}
		{\bibfnamefont {R.}~\bibnamefont {Schilling}}, \bibinfo {author}
		{\bibfnamefont {D.~J.}\ \bibnamefont {Wilson}}, \bibinfo {author}
		{\bibfnamefont {N.~J.}\ \bibnamefont {Engelsen}},\ and\ \bibinfo {author}
		{\bibfnamefont {T.~J.}\ \bibnamefont {Kippenberg}},\ }\bibfield  {title}
	{\bibinfo {title} {Clamp-tapering increases the quality factor of stressed
			nanobeams},\ }\href@noop {} {\bibfield  {journal} {\bibinfo  {journal} {Nano
				letters}\ }\textbf {\bibinfo {volume} {19}},\ \bibinfo {pages} {2329}
		(\bibinfo {year} {2019})}\BibitemShut {NoStop}%
	\bibitem [{\citenamefont {Burnett}\ \emph {et~al.}(2019)\citenamefont
		{Burnett}, \citenamefont {Bengtsson}, \citenamefont {Scigliuzzo},
		\citenamefont {Niepce}, \citenamefont {Kudra}, \citenamefont {Delsing},\ and\
		\citenamefont {Bylander}}]{burnett2019decoherence}%
	\BibitemOpen
	\bibfield  {author} {\bibinfo {author} {\bibfnamefont {J.~J.}\ \bibnamefont
			{Burnett}}, \bibinfo {author} {\bibfnamefont {A.}~\bibnamefont {Bengtsson}},
		\bibinfo {author} {\bibfnamefont {M.}~\bibnamefont {Scigliuzzo}}, \bibinfo
		{author} {\bibfnamefont {D.}~\bibnamefont {Niepce}}, \bibinfo {author}
		{\bibfnamefont {M.}~\bibnamefont {Kudra}}, \bibinfo {author} {\bibfnamefont
			{P.}~\bibnamefont {Delsing}},\ and\ \bibinfo {author} {\bibfnamefont
			{J.}~\bibnamefont {Bylander}},\ }\bibfield  {title} {\bibinfo {title}
		{Decoherence benchmarking of superconducting qubits},\ }\href@noop {}
	{\bibfield  {journal} {\bibinfo  {journal} {npj Quantum Information}\
		}\textbf {\bibinfo {volume} {5}},\ \bibinfo {pages} {54} (\bibinfo {year}
		{2019})}\BibitemShut {NoStop}%
	\bibitem [{\citenamefont {Gordon}\ \emph {et~al.}(2022)\citenamefont {Gordon},
		\citenamefont {Murray}, \citenamefont {Kurter}, \citenamefont {Sandberg},
		\citenamefont {Hall}, \citenamefont {Balakrishnan}, \citenamefont {Shelby},
		\citenamefont {Wacaser}, \citenamefont {Stabile}, \citenamefont {Sleight}
		\emph {et~al.}}]{gordon2022environmental}%
	\BibitemOpen
	\bibfield  {author} {\bibinfo {author} {\bibfnamefont {R.}~\bibnamefont
			{Gordon}}, \bibinfo {author} {\bibfnamefont {C.~E.}\ \bibnamefont {Murray}},
		\bibinfo {author} {\bibfnamefont {C.}~\bibnamefont {Kurter}}, \bibinfo
		{author} {\bibfnamefont {M.}~\bibnamefont {Sandberg}}, \bibinfo {author}
		{\bibfnamefont {S.}~\bibnamefont {Hall}}, \bibinfo {author} {\bibfnamefont
			{K.}~\bibnamefont {Balakrishnan}}, \bibinfo {author} {\bibfnamefont
			{R.}~\bibnamefont {Shelby}}, \bibinfo {author} {\bibfnamefont
			{B.}~\bibnamefont {Wacaser}}, \bibinfo {author} {\bibfnamefont
			{A.}~\bibnamefont {Stabile}}, \bibinfo {author} {\bibfnamefont
			{J.}~\bibnamefont {Sleight}}, \emph {et~al.},\ }\bibfield  {title} {\bibinfo
		{title} {Environmental radiation impact on lifetimes and quasiparticle
			tunneling rates of fixed-frequency transmon qubits},\ }\href@noop {}
	{\bibfield  {journal} {\bibinfo  {journal} {Applied Physics Letters}\
		}\textbf {\bibinfo {volume} {120}} (\bibinfo {year} {2022})}\BibitemShut
	{NoStop}%
	\bibitem [{\citenamefont {Kono}\ \emph {et~al.}(2024)\citenamefont {Kono},
		\citenamefont {Pan}, \citenamefont {Chegnizadeh}, \citenamefont {Wang},
		\citenamefont {Youssefi}, \citenamefont {Scigliuzzo},\ and\ \citenamefont
		{Kippenberg}}]{kono2024mechanically}%
	\BibitemOpen
	\bibfield  {author} {\bibinfo {author} {\bibfnamefont {S.}~\bibnamefont
			{Kono}}, \bibinfo {author} {\bibfnamefont {J.}~\bibnamefont {Pan}}, \bibinfo
		{author} {\bibfnamefont {M.}~\bibnamefont {Chegnizadeh}}, \bibinfo {author}
		{\bibfnamefont {X.}~\bibnamefont {Wang}}, \bibinfo {author} {\bibfnamefont
			{A.}~\bibnamefont {Youssefi}}, \bibinfo {author} {\bibfnamefont
			{M.}~\bibnamefont {Scigliuzzo}},\ and\ \bibinfo {author} {\bibfnamefont
			{T.~J.}\ \bibnamefont {Kippenberg}},\ }\bibfield  {title} {\bibinfo {title}
		{Mechanically induced correlated errors on superconducting qubits with
			relaxation times exceeding 0.4 ms},\ }\href@noop {} {\bibfield  {journal}
		{\bibinfo  {journal} {Nature Communications}\ }\textbf {\bibinfo {volume}
			{15}},\ \bibinfo {pages} {3950} (\bibinfo {year} {2024})}\BibitemShut
	{NoStop}%
	\bibitem [{\citenamefont {Place}\ \emph {et~al.}(2021)\citenamefont {Place},
		\citenamefont {Rodgers}, \citenamefont {Mundada}, \citenamefont {Smitham},
		\citenamefont {Fitzpatrick}, \citenamefont {Leng}, \citenamefont {Premkumar},
		\citenamefont {Bryon}, \citenamefont {Vrajitoarea}, \citenamefont {Sussman}
		\emph {et~al.}}]{place2021new}%
	\BibitemOpen
	\bibfield  {author} {\bibinfo {author} {\bibfnamefont {A.~P.}\ \bibnamefont
			{Place}}, \bibinfo {author} {\bibfnamefont {L.~V.}\ \bibnamefont {Rodgers}},
		\bibinfo {author} {\bibfnamefont {P.}~\bibnamefont {Mundada}}, \bibinfo
		{author} {\bibfnamefont {B.~M.}\ \bibnamefont {Smitham}}, \bibinfo {author}
		{\bibfnamefont {M.}~\bibnamefont {Fitzpatrick}}, \bibinfo {author}
		{\bibfnamefont {Z.}~\bibnamefont {Leng}}, \bibinfo {author} {\bibfnamefont
			{A.}~\bibnamefont {Premkumar}}, \bibinfo {author} {\bibfnamefont
			{J.}~\bibnamefont {Bryon}}, \bibinfo {author} {\bibfnamefont
			{A.}~\bibnamefont {Vrajitoarea}}, \bibinfo {author} {\bibfnamefont
			{S.}~\bibnamefont {Sussman}}, \emph {et~al.},\ }\bibfield  {title} {\bibinfo
		{title} {New material platform for superconducting transmon qubits with
			coherence times exceeding 0.3 milliseconds},\ }\href@noop {} {\bibfield
		{journal} {\bibinfo  {journal} {Nature communications}\ }\textbf {\bibinfo
			{volume} {12}},\ \bibinfo {pages} {1779} (\bibinfo {year}
		{2021})}\BibitemShut {NoStop}%
	\bibitem [{\citenamefont {Crowley}\ \emph {et~al.}(2023)\citenamefont
		{Crowley}, \citenamefont {McLellan}, \citenamefont {Dutta}, \citenamefont
		{Shumiya}, \citenamefont {Place}, \citenamefont {Le}, \citenamefont {Gang},
		\citenamefont {Madhavan}, \citenamefont {Bland}, \citenamefont {Chang} \emph
		{et~al.}}]{crowley2023disentangling}%
	\BibitemOpen
	\bibfield  {author} {\bibinfo {author} {\bibfnamefont {K.~D.}\ \bibnamefont
			{Crowley}}, \bibinfo {author} {\bibfnamefont {R.~A.}\ \bibnamefont
			{McLellan}}, \bibinfo {author} {\bibfnamefont {A.}~\bibnamefont {Dutta}},
		\bibinfo {author} {\bibfnamefont {N.}~\bibnamefont {Shumiya}}, \bibinfo
		{author} {\bibfnamefont {A.~P.}\ \bibnamefont {Place}}, \bibinfo {author}
		{\bibfnamefont {X.~H.}\ \bibnamefont {Le}}, \bibinfo {author} {\bibfnamefont
			{Y.}~\bibnamefont {Gang}}, \bibinfo {author} {\bibfnamefont {T.}~\bibnamefont
			{Madhavan}}, \bibinfo {author} {\bibfnamefont {M.~P.}\ \bibnamefont {Bland}},
		\bibinfo {author} {\bibfnamefont {R.}~\bibnamefont {Chang}}, \emph {et~al.},\
	}\bibfield  {title} {\bibinfo {title} {Disentangling losses in tantalum
			superconducting circuits},\ }\href@noop {} {\bibfield  {journal} {\bibinfo
			{journal} {Physical Review X}\ }\textbf {\bibinfo {volume} {13}},\ \bibinfo
		{pages} {041005} (\bibinfo {year} {2023})}\BibitemShut {NoStop}%
	\bibitem [{\citenamefont {Murray}(2021)}]{murray2021material}%
	\BibitemOpen
	\bibfield  {author} {\bibinfo {author} {\bibfnamefont {C.~E.}\ \bibnamefont
			{Murray}},\ }\bibfield  {title} {\bibinfo {title} {Material matters in
			superconducting qubits},\ }\href@noop {} {\bibfield  {journal} {\bibinfo
			{journal} {Materials Science and Engineering: R: Reports}\ }\textbf {\bibinfo
			{volume} {146}},\ \bibinfo {pages} {100646} (\bibinfo {year}
		{2021})}\BibitemShut {NoStop}%
	\bibitem [{\citenamefont {Weis}\ \emph {et~al.}(2010)\citenamefont {Weis},
		\citenamefont {Rivi{\`e}re}, \citenamefont {Del{\'e}glise}, \citenamefont
		{Gavartin}, \citenamefont {Arcizet}, \citenamefont {Schliesser},\ and\
		\citenamefont {Kippenberg}}]{weis2010optomechanically}%
	\BibitemOpen
	\bibfield  {author} {\bibinfo {author} {\bibfnamefont {S.}~\bibnamefont
			{Weis}}, \bibinfo {author} {\bibfnamefont {R.}~\bibnamefont {Rivi{\`e}re}},
		\bibinfo {author} {\bibfnamefont {S.}~\bibnamefont {Del{\'e}glise}}, \bibinfo
		{author} {\bibfnamefont {E.}~\bibnamefont {Gavartin}}, \bibinfo {author}
		{\bibfnamefont {O.}~\bibnamefont {Arcizet}}, \bibinfo {author} {\bibfnamefont
			{A.}~\bibnamefont {Schliesser}},\ and\ \bibinfo {author} {\bibfnamefont
			{T.~J.}\ \bibnamefont {Kippenberg}},\ }\bibfield  {title} {\bibinfo {title}
		{Optomechanically induced transparency},\ }\href
	{https://www.science.org/doi/full/10.1126/science.1195596} {\bibfield
		{journal} {\bibinfo  {journal} {Science}\ }\textbf {\bibinfo {volume}
			{330}},\ \bibinfo {pages} {1520} (\bibinfo {year} {2010})}\BibitemShut
	{NoStop}%
	\bibitem [{\citenamefont {Macklin}\ \emph {et~al.}(2015)\citenamefont
		{Macklin}, \citenamefont {O’Brien}, \citenamefont {Hover}, \citenamefont
		{Schwartz}, \citenamefont {Bolkhovsky} \emph {et~al.}}]{macklin2015near}%
	\BibitemOpen
	\bibfield  {author} {\bibinfo {author} {\bibfnamefont {C.}~\bibnamefont
			{Macklin}}, \bibinfo {author} {\bibfnamefont {K.}~\bibnamefont {O’Brien}},
		\bibinfo {author} {\bibfnamefont {D.}~\bibnamefont {Hover}}, \bibinfo
		{author} {\bibfnamefont {M.}~\bibnamefont {Schwartz}}, \bibinfo {author}
		{\bibfnamefont {V.}~\bibnamefont {Bolkhovsky}}, \emph {et~al.},\ }\bibfield
	{title} {\bibinfo {title} {A near--quantum-limited josephson traveling-wave
			parametric amplifier},\ }\href
	{https://science.sciencemag.org/content/350/6258/307} {\bibfield  {journal}
		{\bibinfo  {journal} {Science}\ }\textbf {\bibinfo {volume} {350}},\ \bibinfo
		{pages} {307} (\bibinfo {year} {2015})}\BibitemShut {NoStop}%
	\bibitem [{\citenamefont {Boussaha}\ \emph {et~al.}(2020)\citenamefont
		{Boussaha}, \citenamefont {Beldi}, \citenamefont {Monfardini}, \citenamefont
		{Hu}, \citenamefont {Calvo}, \citenamefont {Chaumont}, \citenamefont
		{L{\'e}vy-Bertrand}, \citenamefont {Vacelet}, \citenamefont {Traini},
		\citenamefont {Firminy} \emph {et~al.}}]{boussaha2020development}%
	\BibitemOpen
	\bibfield  {author} {\bibinfo {author} {\bibfnamefont {F.}~\bibnamefont
			{Boussaha}}, \bibinfo {author} {\bibfnamefont {S.}~\bibnamefont {Beldi}},
		\bibinfo {author} {\bibfnamefont {A.}~\bibnamefont {Monfardini}}, \bibinfo
		{author} {\bibfnamefont {J.}~\bibnamefont {Hu}}, \bibinfo {author}
		{\bibfnamefont {M.}~\bibnamefont {Calvo}}, \bibinfo {author} {\bibfnamefont
			{C.}~\bibnamefont {Chaumont}}, \bibinfo {author} {\bibfnamefont
			{F.}~\bibnamefont {L{\'e}vy-Bertrand}}, \bibinfo {author} {\bibfnamefont
			{T.}~\bibnamefont {Vacelet}}, \bibinfo {author} {\bibfnamefont
			{A.}~\bibnamefont {Traini}}, \bibinfo {author} {\bibfnamefont
			{J.}~\bibnamefont {Firminy}}, \emph {et~al.},\ }\bibfield  {title} {\bibinfo
		{title} {Development of tin vacuum-gap capacitor lumped-element kinetic
			inductance detectors},\ }\href@noop {} {\bibfield  {journal} {\bibinfo
			{journal} {Journal of Low Temperature Physics}\ }\textbf {\bibinfo {volume}
			{199}},\ \bibinfo {pages} {994} (\bibinfo {year} {2020})}\BibitemShut
	{NoStop}%
	\bibitem [{\citenamefont {Zemlicka}\ \emph {et~al.}(2023)\citenamefont
		{Zemlicka}, \citenamefont {Redchenko}, \citenamefont {Peruzzo}, \citenamefont
		{Hassani}, \citenamefont {Trioni}, \citenamefont {Barzanjeh},\ and\
		\citenamefont {Fink}}]{zemlicka2023compact}%
	\BibitemOpen
	\bibfield  {author} {\bibinfo {author} {\bibfnamefont {M.}~\bibnamefont
			{Zemlicka}}, \bibinfo {author} {\bibfnamefont {E.}~\bibnamefont {Redchenko}},
		\bibinfo {author} {\bibfnamefont {M.}~\bibnamefont {Peruzzo}}, \bibinfo
		{author} {\bibfnamefont {F.}~\bibnamefont {Hassani}}, \bibinfo {author}
		{\bibfnamefont {A.}~\bibnamefont {Trioni}}, \bibinfo {author} {\bibfnamefont
			{S.}~\bibnamefont {Barzanjeh}},\ and\ \bibinfo {author} {\bibfnamefont
			{J.~M.}\ \bibnamefont {Fink}},\ }\bibfield  {title} {\bibinfo {title}
		{Compact vacuum-gap transmon qubits: Selective and sensitive probes for
			superconductor surface losses},\ }\href@noop {} {\bibfield  {journal}
		{\bibinfo  {journal} {Physical Review Applied}\ }\textbf {\bibinfo {volume}
			{20}},\ \bibinfo {pages} {044054} (\bibinfo {year} {2023})}\BibitemShut
	{NoStop}%
	\bibitem [{\citenamefont {Fedorov}(2020)}]{fedorov2020mechanical}%
	\BibitemOpen
	\bibfield  {author} {\bibinfo {author} {\bibfnamefont {S.}~\bibnamefont
			{Fedorov}},\ }\href {http://dx.doi.org/10.5075/epfl-thesis-10421} {\emph
		{\bibinfo {title} {Mechanical resonators with high dissipation dilution in
				precision and quantum measurements}}},\ \bibinfo {type} {Tech. Rep.}\
	(\bibinfo  {institution} {EPFL},\ \bibinfo {year} {2020})\BibitemShut
	{NoStop}%
	\bibitem [{\citenamefont {Fedorov}\ \emph {et~al.}(2019)\citenamefont
		{Fedorov}, \citenamefont {Engelsen}, \citenamefont {Ghadimi}, \citenamefont
		{Bereyhi}, \citenamefont {Schilling}, \citenamefont {Wilson},\ and\
		\citenamefont {Kippenberg}}]{fedorov2019generalized}%
	\BibitemOpen
	\bibfield  {author} {\bibinfo {author} {\bibfnamefont {S.~A.}\ \bibnamefont
			{Fedorov}}, \bibinfo {author} {\bibfnamefont {N.~J.}\ \bibnamefont
			{Engelsen}}, \bibinfo {author} {\bibfnamefont {A.~H.}\ \bibnamefont
			{Ghadimi}}, \bibinfo {author} {\bibfnamefont {M.~J.}\ \bibnamefont
			{Bereyhi}}, \bibinfo {author} {\bibfnamefont {R.}~\bibnamefont {Schilling}},
		\bibinfo {author} {\bibfnamefont {D.~J.}\ \bibnamefont {Wilson}},\ and\
		\bibinfo {author} {\bibfnamefont {T.~J.}\ \bibnamefont {Kippenberg}},\
	}\bibfield  {title} {\bibinfo {title} {Generalized dissipation dilution in
			strained mechanical resonators},\ }\href
	{https://journals.aps.org/prb/abstract/10.1103/PhysRevB.99.054107} {\bibfield
		{journal} {\bibinfo  {journal} {Physical Review B}\ }\textbf {\bibinfo
			{volume} {99}},\ \bibinfo {pages} {054107} (\bibinfo {year}
		{2019})}\BibitemShut {NoStop}%
	\bibitem [{\citenamefont {Heinrich}\ \emph {et~al.}(2011)\citenamefont
		{Heinrich}, \citenamefont {Ludwig}, \citenamefont {Qian}, \citenamefont
		{Kubala},\ and\ \citenamefont {Marquardt}}]{heinrich2011collective}%
	\BibitemOpen
	\bibfield  {author} {\bibinfo {author} {\bibfnamefont {G.}~\bibnamefont
			{Heinrich}}, \bibinfo {author} {\bibfnamefont {M.}~\bibnamefont {Ludwig}},
		\bibinfo {author} {\bibfnamefont {J.}~\bibnamefont {Qian}}, \bibinfo {author}
		{\bibfnamefont {B.}~\bibnamefont {Kubala}},\ and\ \bibinfo {author}
		{\bibfnamefont {F.}~\bibnamefont {Marquardt}},\ }\bibfield  {title} {\bibinfo
		{title} {Collective dynamics in optomechanical arrays},\ }\href
	{https://journals.aps.org/prl/abstract/10.1103/PhysRevLett.107.043603}
	{\bibfield  {journal} {\bibinfo  {journal} {Physical review letters}\
		}\textbf {\bibinfo {volume} {107}},\ \bibinfo {pages} {043603} (\bibinfo
		{year} {2011})}\BibitemShut {NoStop}%
	\bibitem [{\citenamefont {Xuereb}\ \emph {et~al.}(2012)\citenamefont {Xuereb},
		\citenamefont {Genes},\ and\ \citenamefont {Dantan}}]{xuereb2012strong}%
	\BibitemOpen
	\bibfield  {author} {\bibinfo {author} {\bibfnamefont {A.}~\bibnamefont
			{Xuereb}}, \bibinfo {author} {\bibfnamefont {C.}~\bibnamefont {Genes}},\ and\
		\bibinfo {author} {\bibfnamefont {A.}~\bibnamefont {Dantan}},\ }\bibfield
	{title} {\bibinfo {title} {Strong coupling and long-range collective
			interactions in optomechanical arrays},\ }\href
	{https://journals.aps.org/prl/abstract/10.1103/PhysRevLett.109.223601}
	{\bibfield  {journal} {\bibinfo  {journal} {Physical review letters}\
		}\textbf {\bibinfo {volume} {109}},\ \bibinfo {pages} {223601} (\bibinfo
		{year} {2012})}\BibitemShut {NoStop}%
	\bibitem [{\citenamefont {Ludwig}\ and\ \citenamefont
		{Marquardt}(2013)}]{ludwig2013quantum}%
	\BibitemOpen
	\bibfield  {author} {\bibinfo {author} {\bibfnamefont {M.}~\bibnamefont
			{Ludwig}}\ and\ \bibinfo {author} {\bibfnamefont {F.}~\bibnamefont
			{Marquardt}},\ }\bibfield  {title} {\bibinfo {title} {Quantum many-body
			dynamics in optomechanical arrays},\ }\href
	{https://journals.aps.org/prl/abstract/10.1103/PhysRevLett.111.073603}
	{\bibfield  {journal} {\bibinfo  {journal} {Physical review letters}\
		}\textbf {\bibinfo {volume} {111}},\ \bibinfo {pages} {073603} (\bibinfo
		{year} {2013})}\BibitemShut {NoStop}%
	\bibitem [{\citenamefont {Raeisi}\ and\ \citenamefont
		{Marquardt}(2020)}]{raeisi2020quench}%
	\BibitemOpen
	\bibfield  {author} {\bibinfo {author} {\bibfnamefont {S.}~\bibnamefont
			{Raeisi}}\ and\ \bibinfo {author} {\bibfnamefont {F.}~\bibnamefont
			{Marquardt}},\ }\bibfield  {title} {\bibinfo {title} {Quench dynamics in
			one-dimensional optomechanical arrays},\ }\href
	{https://journals.aps.org/pra/abstract/10.1103/PhysRevA.101.023814}
	{\bibfield  {journal} {\bibinfo  {journal} {Physical Review A}\ }\textbf
		{\bibinfo {volume} {101}},\ \bibinfo {pages} {023814} (\bibinfo {year}
		{2020})}\BibitemShut {NoStop}%
	\bibitem [{\citenamefont {Peano}\ \emph {et~al.}(2015)\citenamefont {Peano},
		\citenamefont {Brendel}, \citenamefont {Schmidt},\ and\ \citenamefont
		{Marquardt}}]{peano2015topological}%
	\BibitemOpen
	\bibfield  {author} {\bibinfo {author} {\bibfnamefont {V.}~\bibnamefont
			{Peano}}, \bibinfo {author} {\bibfnamefont {C.}~\bibnamefont {Brendel}},
		\bibinfo {author} {\bibfnamefont {M.}~\bibnamefont {Schmidt}},\ and\ \bibinfo
		{author} {\bibfnamefont {F.}~\bibnamefont {Marquardt}},\ }\bibfield  {title}
	{\bibinfo {title} {Topological phases of sound and light},\ }\href
	{https://journals.aps.org/prx/abstract/10.1103/PhysRevX.5.031011} {\bibfield
		{journal} {\bibinfo  {journal} {Physical Review X}\ }\textbf {\bibinfo
			{volume} {5}},\ \bibinfo {pages} {031011} (\bibinfo {year}
		{2015})}\BibitemShut {NoStop}%
	\bibitem [{\citenamefont {Zangeneh-Nejad}\ and\ \citenamefont
		{Fleury}(2020)}]{zangeneh2020topological}%
	\BibitemOpen
	\bibfield  {author} {\bibinfo {author} {\bibfnamefont {F.}~\bibnamefont
			{Zangeneh-Nejad}}\ and\ \bibinfo {author} {\bibfnamefont {R.}~\bibnamefont
			{Fleury}},\ }\bibfield  {title} {\bibinfo {title} {Topological
			optomechanically induced transparency},\ }\href
	{https://www.osapublishing.org/ol/fulltext.cfm?uri=ol-45-21-5966&id=441914}
	{\bibfield  {journal} {\bibinfo  {journal} {Optics Letters}\ }\textbf
		{\bibinfo {volume} {45}},\ \bibinfo {pages} {5966} (\bibinfo {year}
		{2020})}\BibitemShut {NoStop}%
	\bibitem [{\citenamefont {Asb{\'o}th}\ \emph {et~al.}(2016)\citenamefont
		{Asb{\'o}th}, \citenamefont {Oroszl{\'a}ny},\ and\ \citenamefont
		{P{\'a}lyi}}]{asboth2016short}%
	\BibitemOpen
	\bibfield  {author} {\bibinfo {author} {\bibfnamefont {J.~K.}\ \bibnamefont
			{Asb{\'o}th}}, \bibinfo {author} {\bibfnamefont {L.}~\bibnamefont
			{Oroszl{\'a}ny}},\ and\ \bibinfo {author} {\bibfnamefont {A.}~\bibnamefont
			{P{\'a}lyi}},\ }\bibfield  {title} {\bibinfo {title} {A short course on
			topological insulators},\ }\href {https://arxiv.org/abs/1509.02295}
	{\bibfield  {journal} {\bibinfo  {journal} {Lecture notes in physics}\
		}\textbf {\bibinfo {volume} {919}},\ \bibinfo {pages} {997} (\bibinfo {year}
		{2016})}\BibitemShut {NoStop}%
	\bibitem [{\citenamefont {Ozawa}\ \emph {et~al.}(2019)\citenamefont {Ozawa},
		\citenamefont {Price}, \citenamefont {Amo}, \citenamefont {Goldman},
		\citenamefont {Hafezi}, \citenamefont {Lu}, \citenamefont {Rechtsman},
		\citenamefont {Schuster}, \citenamefont {Simon}, \citenamefont {Zilberberg}
		\emph {et~al.}}]{ozawa2019topological}%
	\BibitemOpen
	\bibfield  {author} {\bibinfo {author} {\bibfnamefont {T.}~\bibnamefont
			{Ozawa}}, \bibinfo {author} {\bibfnamefont {H.~M.}\ \bibnamefont {Price}},
		\bibinfo {author} {\bibfnamefont {A.}~\bibnamefont {Amo}}, \bibinfo {author}
		{\bibfnamefont {N.}~\bibnamefont {Goldman}}, \bibinfo {author} {\bibfnamefont
			{M.}~\bibnamefont {Hafezi}}, \bibinfo {author} {\bibfnamefont
			{L.}~\bibnamefont {Lu}}, \bibinfo {author} {\bibfnamefont {M.~C.}\
			\bibnamefont {Rechtsman}}, \bibinfo {author} {\bibfnamefont {D.}~\bibnamefont
			{Schuster}}, \bibinfo {author} {\bibfnamefont {J.}~\bibnamefont {Simon}},
		\bibinfo {author} {\bibfnamefont {O.}~\bibnamefont {Zilberberg}}, \emph
		{et~al.},\ }\bibfield  {title} {\bibinfo {title} {Topological photonics},\
	}\href {https://journals.aps.org/rmp/abstract/10.1103/RevModPhys.91.015006}
	{\bibfield  {journal} {\bibinfo  {journal} {Reviews of Modern Physics}\
		}\textbf {\bibinfo {volume} {91}},\ \bibinfo {pages} {015006} (\bibinfo
		{year} {2019})}\BibitemShut {NoStop}%
	\bibitem [{\citenamefont {Jouanny}\ \emph {et~al.}(2024)\citenamefont
		{Jouanny}, \citenamefont {Frasca}, \citenamefont {Weibel}, \citenamefont
		{Peyruchat}, \citenamefont {Scigliuzzo}, \citenamefont {Oppliger},
		\citenamefont {De~Palma}, \citenamefont {Sbroggio}, \citenamefont {Beaulieu},
		\citenamefont {Zilberberg} \emph {et~al.}}]{jouanny2024band}%
	\BibitemOpen
	\bibfield  {author} {\bibinfo {author} {\bibfnamefont {V.}~\bibnamefont
			{Jouanny}}, \bibinfo {author} {\bibfnamefont {S.}~\bibnamefont {Frasca}},
		\bibinfo {author} {\bibfnamefont {V.~J.}\ \bibnamefont {Weibel}}, \bibinfo
		{author} {\bibfnamefont {L.}~\bibnamefont {Peyruchat}}, \bibinfo {author}
		{\bibfnamefont {M.}~\bibnamefont {Scigliuzzo}}, \bibinfo {author}
		{\bibfnamefont {F.}~\bibnamefont {Oppliger}}, \bibinfo {author}
		{\bibfnamefont {F.}~\bibnamefont {De~Palma}}, \bibinfo {author}
		{\bibfnamefont {D.}~\bibnamefont {Sbroggio}}, \bibinfo {author}
		{\bibfnamefont {G.}~\bibnamefont {Beaulieu}}, \bibinfo {author}
		{\bibfnamefont {O.}~\bibnamefont {Zilberberg}}, \emph {et~al.},\ }\bibfield
	{title} {\bibinfo {title} {Band engineering and study of disorder using
			topology in compact high kinetic inductance cavity arrays},\ }\href@noop {}
	{\bibfield  {journal} {\bibinfo  {journal} {arXiv preprint arXiv:2403.18150}\
		} (\bibinfo {year} {2024})}\BibitemShut {NoStop}%
	\bibitem [{\citenamefont {Osman}\ \emph {et~al.}(2023)\citenamefont {Osman},
		\citenamefont {Fern{\'a}ndez-Pend{\'a}s}, \citenamefont {Warren},
		\citenamefont {Kosen}, \citenamefont {Scigliuzzo}, \citenamefont
		{Frisk~Kockum}, \citenamefont {Tancredi}, \citenamefont {Fadavi~Roudsari},\
		and\ \citenamefont {Bylander}}]{osman2023mitigation}%
	\BibitemOpen
	\bibfield  {author} {\bibinfo {author} {\bibfnamefont {A.}~\bibnamefont
			{Osman}}, \bibinfo {author} {\bibfnamefont {J.}~\bibnamefont
			{Fern{\'a}ndez-Pend{\'a}s}}, \bibinfo {author} {\bibfnamefont
			{C.}~\bibnamefont {Warren}}, \bibinfo {author} {\bibfnamefont
			{S.}~\bibnamefont {Kosen}}, \bibinfo {author} {\bibfnamefont
			{M.}~\bibnamefont {Scigliuzzo}}, \bibinfo {author} {\bibfnamefont
			{A.}~\bibnamefont {Frisk~Kockum}}, \bibinfo {author} {\bibfnamefont
			{G.}~\bibnamefont {Tancredi}}, \bibinfo {author} {\bibfnamefont
			{A.}~\bibnamefont {Fadavi~Roudsari}},\ and\ \bibinfo {author} {\bibfnamefont
			{J.}~\bibnamefont {Bylander}},\ }\bibfield  {title} {\bibinfo {title}
		{Mitigation of frequency collisions in superconducting quantum processors},\
	}\href@noop {} {\bibfield  {journal} {\bibinfo  {journal} {Physical Review
				Research}\ }\textbf {\bibinfo {volume} {5}},\ \bibinfo {pages} {043001}
		(\bibinfo {year} {2023})}\BibitemShut {NoStop}%
	\bibitem [{\citenamefont {Li}\ \emph {et~al.}(2021)\citenamefont {Li},
		\citenamefont {Ou}, \citenamefont {Lei},\ and\ \citenamefont
		{Liu}}]{li2021cavity}%
	\BibitemOpen
	\bibfield  {author} {\bibinfo {author} {\bibfnamefont {B.-B.}\ \bibnamefont
			{Li}}, \bibinfo {author} {\bibfnamefont {L.}~\bibnamefont {Ou}}, \bibinfo
		{author} {\bibfnamefont {Y.}~\bibnamefont {Lei}},\ and\ \bibinfo {author}
		{\bibfnamefont {Y.-C.}\ \bibnamefont {Liu}},\ }\bibfield  {title} {\bibinfo
		{title} {Cavity optomechanical sensing},\ }\href
	{https://doi.org/10.1515/nanoph-2021-0256} {\bibfield  {journal} {\bibinfo
			{journal} {Nanophotonics}\ }\textbf {\bibinfo {volume} {10}},\ \bibinfo
		{pages} {2799} (\bibinfo {year} {2021})}\BibitemShut {NoStop}%
	\bibitem [{\citenamefont {Gely}\ and\ \citenamefont
		{Steele}(2021{\natexlab{a}})}]{gely2021phonon}%
	\BibitemOpen
	\bibfield  {author} {\bibinfo {author} {\bibfnamefont {M.~F.}\ \bibnamefont
			{Gely}}\ and\ \bibinfo {author} {\bibfnamefont {G.~A.}\ \bibnamefont
			{Steele}},\ }\bibfield  {title} {\bibinfo {title} {Phonon-number resolution
			of voltage-biased mechanical oscillators with weakly anharmonic
			superconducting circuits},\ }\href
	{https://journals.aps.org/pra/abstract/10.1103/PhysRevA.104.053509}
	{\bibfield  {journal} {\bibinfo  {journal} {Physical Review A}\ }\textbf
		{\bibinfo {volume} {104}},\ \bibinfo {pages} {053509} (\bibinfo {year}
		{2021}{\natexlab{a}})}\BibitemShut {NoStop}%
	\bibitem [{\citenamefont {Wallucks}\ \emph {et~al.}(2020)\citenamefont
		{Wallucks}, \citenamefont {Marinkovi{\'c}}, \citenamefont {Hensen},
		\citenamefont {Stockill},\ and\ \citenamefont
		{Gr{\"o}blacher}}]{wallucks2020quantum}%
	\BibitemOpen
	\bibfield  {author} {\bibinfo {author} {\bibfnamefont {A.}~\bibnamefont
			{Wallucks}}, \bibinfo {author} {\bibfnamefont {I.}~\bibnamefont
			{Marinkovi{\'c}}}, \bibinfo {author} {\bibfnamefont {B.}~\bibnamefont
			{Hensen}}, \bibinfo {author} {\bibfnamefont {R.}~\bibnamefont {Stockill}},\
		and\ \bibinfo {author} {\bibfnamefont {S.}~\bibnamefont {Gr{\"o}blacher}},\
	}\bibfield  {title} {\bibinfo {title} {A quantum memory at telecom
			wavelengths},\ }\href {https://www.nature.com/articles/s41567-020-0891-z}
	{\bibfield  {journal} {\bibinfo  {journal} {Nature Physics}\ }\textbf
		{\bibinfo {volume} {16}},\ \bibinfo {pages} {772} (\bibinfo {year}
		{2020})}\BibitemShut {NoStop}%
	\bibitem [{\citenamefont {Pechal}\ \emph {et~al.}(2018)\citenamefont {Pechal},
		\citenamefont {Arrangoiz-Arriola},\ and\ \citenamefont
		{Safavi-Naeini}}]{pechal2018superconducting}%
	\BibitemOpen
	\bibfield  {author} {\bibinfo {author} {\bibfnamefont {M.}~\bibnamefont
			{Pechal}}, \bibinfo {author} {\bibfnamefont {P.}~\bibnamefont
			{Arrangoiz-Arriola}},\ and\ \bibinfo {author} {\bibfnamefont {A.~H.}\
			\bibnamefont {Safavi-Naeini}},\ }\bibfield  {title} {\bibinfo {title}
		{Superconducting circuit quantum computing with nanomechanical resonators as
			storage},\ }\href
	{https://iopscience.iop.org/article/10.1088/2058-9565/aadc6c/meta} {\bibfield
		{journal} {\bibinfo  {journal} {Quantum Science and Technology}\ }\textbf
		{\bibinfo {volume} {4}},\ \bibinfo {pages} {015006} (\bibinfo {year}
		{2018})}\BibitemShut {NoStop}%
	\bibitem [{\citenamefont {Liu}\ \emph {et~al.}(2021)\citenamefont {Liu},
		\citenamefont {Mummery}, \citenamefont {Zhou},\ and\ \citenamefont
		{Sillanp{\"a}{\"a}}}]{liu2021gravitational}%
	\BibitemOpen
	\bibfield  {author} {\bibinfo {author} {\bibfnamefont {Y.}~\bibnamefont
			{Liu}}, \bibinfo {author} {\bibfnamefont {J.}~\bibnamefont {Mummery}},
		\bibinfo {author} {\bibfnamefont {J.}~\bibnamefont {Zhou}},\ and\ \bibinfo
		{author} {\bibfnamefont {M.~A.}\ \bibnamefont {Sillanp{\"a}{\"a}}},\
	}\bibfield  {title} {\bibinfo {title} {Gravitational forces between
			nonclassical mechanical oscillators},\ }\href
	{https://journals.aps.org/prapplied/abstract/10.1103/PhysRevApplied.15.034004}
	{\bibfield  {journal} {\bibinfo  {journal} {Physical Review Applied}\
		}\textbf {\bibinfo {volume} {15}},\ \bibinfo {pages} {034004} (\bibinfo
		{year} {2021})}\BibitemShut {NoStop}%
	\bibitem [{\citenamefont {Gely}\ and\ \citenamefont
		{Steele}(2021{\natexlab{b}})}]{gely2021superconducting}%
	\BibitemOpen
	\bibfield  {author} {\bibinfo {author} {\bibfnamefont {M.~F.}\ \bibnamefont
			{Gely}}\ and\ \bibinfo {author} {\bibfnamefont {G.~A.}\ \bibnamefont
			{Steele}},\ }\bibfield  {title} {\bibinfo {title} {Superconducting
			electro-mechanics to test di{\'o}si--penrose effects of general relativity in
			massive superpositions},\ }\href
	{https://avs.scitation.org/doi/full/10.1116/5.0050988} {\bibfield  {journal}
		{\bibinfo  {journal} {AVS Quantum Science}\ }\textbf {\bibinfo {volume}
			{3}},\ \bibinfo {pages} {035601} (\bibinfo {year}
		{2021}{\natexlab{b}})}\BibitemShut {NoStop}%
	\bibitem [{\citenamefont {Marinkovi{\'c}}\ \emph {et~al.}(2018)\citenamefont
		{Marinkovi{\'c}}, \citenamefont {Wallucks}, \citenamefont {Riedinger},
		\citenamefont {Hong}, \citenamefont {Aspelmeyer},\ and\ \citenamefont
		{Gr{\"o}blacher}}]{marinkovic2018optomechanical}%
	\BibitemOpen
	\bibfield  {author} {\bibinfo {author} {\bibfnamefont {I.}~\bibnamefont
			{Marinkovi{\'c}}}, \bibinfo {author} {\bibfnamefont {A.}~\bibnamefont
			{Wallucks}}, \bibinfo {author} {\bibfnamefont {R.}~\bibnamefont {Riedinger}},
		\bibinfo {author} {\bibfnamefont {S.}~\bibnamefont {Hong}}, \bibinfo {author}
		{\bibfnamefont {M.}~\bibnamefont {Aspelmeyer}},\ and\ \bibinfo {author}
		{\bibfnamefont {S.}~\bibnamefont {Gr{\"o}blacher}},\ }\bibfield  {title}
	{\bibinfo {title} {Optomechanical bell test},\ }\href
	{https://journals.aps.org/prl/abstract/10.1103/PhysRevLett.121.220404}
	{\bibfield  {journal} {\bibinfo  {journal} {Physical review letters}\
		}\textbf {\bibinfo {volume} {121}},\ \bibinfo {pages} {220404} (\bibinfo
		{year} {2018})}\BibitemShut {NoStop}%
	\bibitem [{\citenamefont {Hong}\ \emph {et~al.}(2017)\citenamefont {Hong},
		\citenamefont {Riedinger}, \citenamefont {Marinkovi{\'c}}, \citenamefont
		{Wallucks}, \citenamefont {Hofer}, \citenamefont {Norte}, \citenamefont
		{Aspelmeyer},\ and\ \citenamefont {Gr{\"o}blacher}}]{hong2017hanbury}%
	\BibitemOpen
	\bibfield  {author} {\bibinfo {author} {\bibfnamefont {S.}~\bibnamefont
			{Hong}}, \bibinfo {author} {\bibfnamefont {R.}~\bibnamefont {Riedinger}},
		\bibinfo {author} {\bibfnamefont {I.}~\bibnamefont {Marinkovi{\'c}}},
		\bibinfo {author} {\bibfnamefont {A.}~\bibnamefont {Wallucks}}, \bibinfo
		{author} {\bibfnamefont {S.~G.}\ \bibnamefont {Hofer}}, \bibinfo {author}
		{\bibfnamefont {R.~A.}\ \bibnamefont {Norte}}, \bibinfo {author}
		{\bibfnamefont {M.}~\bibnamefont {Aspelmeyer}},\ and\ \bibinfo {author}
		{\bibfnamefont {S.}~\bibnamefont {Gr{\"o}blacher}},\ }\bibfield  {title}
	{\bibinfo {title} {Hanbury brown and twiss interferometry of single phonons
			from an optomechanical resonator},\ }\href
	{https://doi.org/10.1126/science.aan7939} {\bibfield  {journal} {\bibinfo
			{journal} {Science}\ }\textbf {\bibinfo {volume} {358}},\ \bibinfo {pages}
		{203} (\bibinfo {year} {2017})}\BibitemShut {NoStop}%
	\bibitem [{\citenamefont {Fiaschi}\ \emph {et~al.}(2021)\citenamefont
		{Fiaschi}, \citenamefont {Hensen}, \citenamefont {Wallucks}, \citenamefont
		{Benevides}, \citenamefont {Li}, \citenamefont {Alegre},\ and\ \citenamefont
		{Gr{\"o}blacher}}]{fiaschi2021optomechanical}%
	\BibitemOpen
	\bibfield  {author} {\bibinfo {author} {\bibfnamefont {N.}~\bibnamefont
			{Fiaschi}}, \bibinfo {author} {\bibfnamefont {B.}~\bibnamefont {Hensen}},
		\bibinfo {author} {\bibfnamefont {A.}~\bibnamefont {Wallucks}}, \bibinfo
		{author} {\bibfnamefont {R.}~\bibnamefont {Benevides}}, \bibinfo {author}
		{\bibfnamefont {J.}~\bibnamefont {Li}}, \bibinfo {author} {\bibfnamefont
			{T.~P.~M.}\ \bibnamefont {Alegre}},\ and\ \bibinfo {author} {\bibfnamefont
			{S.}~\bibnamefont {Gr{\"o}blacher}},\ }\bibfield  {title} {\bibinfo {title}
		{Optomechanical quantum teleportation},\ }\href
	{https://www.nature.com/articles/s41566-021-00866-z} {\bibfield  {journal}
		{\bibinfo  {journal} {Nature Photonics}\ }\textbf {\bibinfo {volume} {15}},\
		\bibinfo {pages} {817} (\bibinfo {year} {2021})}\BibitemShut {NoStop}%
	\bibitem [{\citenamefont {Carney}\ \emph {et~al.}(2021)\citenamefont {Carney},
		\citenamefont {Krnjaic}, \citenamefont {Moore}, \citenamefont {Regal},
		\citenamefont {Afek}, \citenamefont {Bhave}, \citenamefont {Brubaker},
		\citenamefont {Corbitt}, \citenamefont {Cripe}, \citenamefont {Crisosto}
		\emph {et~al.}}]{carney2021mechanical}%
	\BibitemOpen
	\bibfield  {author} {\bibinfo {author} {\bibfnamefont {D.}~\bibnamefont
			{Carney}}, \bibinfo {author} {\bibfnamefont {G.}~\bibnamefont {Krnjaic}},
		\bibinfo {author} {\bibfnamefont {D.~C.}\ \bibnamefont {Moore}}, \bibinfo
		{author} {\bibfnamefont {C.~A.}\ \bibnamefont {Regal}}, \bibinfo {author}
		{\bibfnamefont {G.}~\bibnamefont {Afek}}, \bibinfo {author} {\bibfnamefont
			{S.}~\bibnamefont {Bhave}}, \bibinfo {author} {\bibfnamefont
			{B.}~\bibnamefont {Brubaker}}, \bibinfo {author} {\bibfnamefont
			{T.}~\bibnamefont {Corbitt}}, \bibinfo {author} {\bibfnamefont
			{J.}~\bibnamefont {Cripe}}, \bibinfo {author} {\bibfnamefont
			{N.}~\bibnamefont {Crisosto}}, \emph {et~al.},\ }\bibfield  {title} {\bibinfo
		{title} {Mechanical quantum sensing in the search for dark matter},\ }\href
	{https://iopscience.iop.org/article/10.1088/2058-9565/abcfcd/meta} {\bibfield
		{journal} {\bibinfo  {journal} {Quantum Science and Technology}\ }\textbf
		{\bibinfo {volume} {6}},\ \bibinfo {pages} {024002} (\bibinfo {year}
		{2021})}\BibitemShut {NoStop}%
	\bibitem [{\citenamefont {Manley}\ \emph {et~al.}(2021)\citenamefont {Manley},
		\citenamefont {Chowdhury}, \citenamefont {Grin}, \citenamefont {Singh},\ and\
		\citenamefont {Wilson}}]{manley2021searching}%
	\BibitemOpen
	\bibfield  {author} {\bibinfo {author} {\bibfnamefont {J.}~\bibnamefont
			{Manley}}, \bibinfo {author} {\bibfnamefont {M.~D.}\ \bibnamefont
			{Chowdhury}}, \bibinfo {author} {\bibfnamefont {D.}~\bibnamefont {Grin}},
		\bibinfo {author} {\bibfnamefont {S.}~\bibnamefont {Singh}},\ and\ \bibinfo
		{author} {\bibfnamefont {D.~J.}\ \bibnamefont {Wilson}},\ }\bibfield  {title}
	{\bibinfo {title} {Searching for vector dark matter with an optomechanical
			accelerometer},\ }\href
	{https://journals.aps.org/prl/abstract/10.1103/PhysRevLett.126.061301}
	{\bibfield  {journal} {\bibinfo  {journal} {Physical review letters}\
		}\textbf {\bibinfo {volume} {126}},\ \bibinfo {pages} {061301} (\bibinfo
		{year} {2021})}\BibitemShut {NoStop}%
	\bibitem [{\citenamefont {Youssefi}\ \emph {et~al.}(2025)\citenamefont
		{Youssefi}, \citenamefont {Chegnizadeh}, \citenamefont {Scigliuzzo},\ and\
		\citenamefont {Kippenberg}}]{chegnizadeh_2025_15413211}%
	\BibitemOpen
	\bibfield  {author} {\bibinfo {author} {\bibfnamefont {A.}~\bibnamefont
			{Youssefi}}, \bibinfo {author} {\bibfnamefont {M.}~\bibnamefont
			{Chegnizadeh}}, \bibinfo {author} {\bibfnamefont {M.}~\bibnamefont
			{Scigliuzzo}},\ and\ \bibinfo {author} {\bibfnamefont {T.~J.}\ \bibnamefont
			{Kippenberg}},\ }\bibfield  {title} {\bibinfo {title} {Data and codes for the
			article "compact superconducting vacuum-gap capacitors with low microwave
			loss and high mechanical coherence for scalable quantum circuits"},\ }\href
	{https://doi.org/10.5281/zenodo.15413211} {10.5281/zenodo.15413211} (\bibinfo
	{year} {2025})\BibitemShut {NoStop}%
	\bibitem [{\citenamefont {Bernier}(2019)}]{bernier2019multimode}%
	\BibitemOpen
	\bibfield  {author} {\bibinfo {author} {\bibfnamefont {N.~R.}\ \bibnamefont
			{Bernier}},\ }\href
	{https://doi.org/http://dx.doi.org/10.5075/epfl-thesis-9217} {\emph {\bibinfo
			{title} {Multimode microwave circuit optomechanics as a platform to study
				coupled quantum harmonic oscillators}}},\ \bibinfo {type} {Tech. Rep.}\
	(\bibinfo  {institution} {EPFL},\ \bibinfo {year} {2019})\BibitemShut
	{NoStop}%
	\bibitem [{\citenamefont {Gardner}\ and\ \citenamefont
		{Flinn}(1990)}]{gardner1990mechanical}%
	\BibitemOpen
	\bibfield  {author} {\bibinfo {author} {\bibfnamefont {D.~S.}\ \bibnamefont
			{Gardner}}\ and\ \bibinfo {author} {\bibfnamefont {P.~A.}\ \bibnamefont
			{Flinn}},\ }\bibfield  {title} {\bibinfo {title} {Mechanical stress as a
			function of temperature for aluminum alloy films},\ }\href
	{https://doi.org/https://doi.org/10.1063/1.345611} {\bibfield  {journal}
		{\bibinfo  {journal} {Journal of Applied Physics}\ }\textbf {\bibinfo
			{volume} {67}},\ \bibinfo {pages} {1831} (\bibinfo {year}
		{1990})}\BibitemShut {NoStop}%
	\bibitem [{\citenamefont {Courtney}(2005)}]{courtney2005mechanical}%
	\BibitemOpen
	\bibfield  {author} {\bibinfo {author} {\bibfnamefont {T.~H.}\ \bibnamefont
			{Courtney}},\ }\href
	{https://doi.org/https://doi.org/10.1002/crat.2170270407} {\emph {\bibinfo
			{title} {Mechanical behavior of materials}}}\ (\bibinfo  {publisher}
	{Waveland Press},\ \bibinfo {year} {2005})\BibitemShut {NoStop}%
	\bibitem [{\citenamefont {Ekin}(2006)}]{ekin2006experimental}%
	\BibitemOpen
	\bibfield  {author} {\bibinfo {author} {\bibfnamefont {J.}~\bibnamefont
			{Ekin}},\ }\href
	{https://doi.org/https://doi.org/10.1093/acprof:oso/9780198570547.001.0001}
	{\emph {\bibinfo {title} {Experimental techniques for low-temperature
				measurements: cryostat design, material properties and superconductor
				critical-current testing}}}\ (\bibinfo  {publisher} {Oxford university
		press},\ \bibinfo {year} {2006})\BibitemShut {NoStop}%
	\bibitem [{\citenamefont {Richter}\ \emph {et~al.}(2009)\citenamefont
		{Richter}, \citenamefont {Hillerich}, \citenamefont {Gianola}, \citenamefont
		{Monig}, \citenamefont {Kraft},\ and\ \citenamefont
		{Volkert}}]{richter2009ultrahigh}%
	\BibitemOpen
	\bibfield  {author} {\bibinfo {author} {\bibfnamefont {G.}~\bibnamefont
			{Richter}}, \bibinfo {author} {\bibfnamefont {K.}~\bibnamefont {Hillerich}},
		\bibinfo {author} {\bibfnamefont {D.~S.}\ \bibnamefont {Gianola}}, \bibinfo
		{author} {\bibfnamefont {R.}~\bibnamefont {Monig}}, \bibinfo {author}
		{\bibfnamefont {O.}~\bibnamefont {Kraft}},\ and\ \bibinfo {author}
		{\bibfnamefont {C.~A.}\ \bibnamefont {Volkert}},\ }\bibfield  {title}
	{\bibinfo {title} {Ultrahigh strength single crystalline nanowhiskers grown
			by physical vapor deposition},\ }\href
	{https://doi.org/https://doi.org/10.1021/nl9015107} {\bibfield  {journal}
		{\bibinfo  {journal} {Nano Letters}\ }\textbf {\bibinfo {volume} {9}},\
		\bibinfo {pages} {3048} (\bibinfo {year} {2009})}\BibitemShut {NoStop}%
	\bibitem [{\citenamefont {Steinwall}\ and\ \citenamefont
		{Johnson}(1990)}]{steinwall1990mechanical}%
	\BibitemOpen
	\bibfield  {author} {\bibinfo {author} {\bibfnamefont {J.~E.}\ \bibnamefont
			{Steinwall}}\ and\ \bibinfo {author} {\bibfnamefont {H.}~\bibnamefont
			{Johnson}},\ }\bibfield  {title} {\bibinfo {title} {Mechanical properties of
			thin film aluminum fibers: Grain size effects},\ }\bibfield  {journal}
	{\bibinfo  {journal} {MRS Online Proceedings Library (OPL)}\ }\textbf
	{\bibinfo {volume} {188}},\ \href
	{https://doi.org/https://doi.org/10.1557/PROC-188-177}
	{https://doi.org/10.1557/PROC-188-177} (\bibinfo {year} {1990})\BibitemShut
	{NoStop}%
	\bibitem [{\citenamefont {Read}\ \emph {et~al.}(2022)\citenamefont {Read},
		\citenamefont {Chapman}, \citenamefont {Lei}, \citenamefont {Curtis},
		\citenamefont {Ganjam}, \citenamefont {Krayzman}, \citenamefont {Frunzio},\
		and\ \citenamefont {Schoelkopf}}]{read2022precision}%
	\BibitemOpen
	\bibfield  {author} {\bibinfo {author} {\bibfnamefont {A.~P.}\ \bibnamefont
			{Read}}, \bibinfo {author} {\bibfnamefont {B.~J.}\ \bibnamefont {Chapman}},
		\bibinfo {author} {\bibfnamefont {C.~U.}\ \bibnamefont {Lei}}, \bibinfo
		{author} {\bibfnamefont {J.~C.}\ \bibnamefont {Curtis}}, \bibinfo {author}
		{\bibfnamefont {S.}~\bibnamefont {Ganjam}}, \bibinfo {author} {\bibfnamefont
			{L.}~\bibnamefont {Krayzman}}, \bibinfo {author} {\bibfnamefont
			{L.}~\bibnamefont {Frunzio}},\ and\ \bibinfo {author} {\bibfnamefont {R.~J.}\
			\bibnamefont {Schoelkopf}},\ }\bibfield  {title} {\bibinfo {title} {Precision
			measurement of the microwave dielectric loss of sapphire in the quantum
			regime with parts-per-billion sensitivity},\ }\href
	{https://arxiv.org/abs/2206.14334} {\bibfield  {journal} {\bibinfo  {journal}
			{arXiv preprint arXiv:2206.14334}\ } (\bibinfo {year} {2022})}\BibitemShut
	{NoStop}%
	\bibitem [{\citenamefont {Jeong}\ \emph {et~al.}(2002)\citenamefont {Jeong},
		\citenamefont {Kim}, \citenamefont {Kim},\ and\ \citenamefont
		{Yeom}}]{jeong2002study}%
	\BibitemOpen
	\bibfield  {author} {\bibinfo {author} {\bibfnamefont {C.~H.}\ \bibnamefont
			{Jeong}}, \bibinfo {author} {\bibfnamefont {D.~W.}\ \bibnamefont {Kim}},
		\bibinfo {author} {\bibfnamefont {K.~N.}\ \bibnamefont {Kim}},\ and\ \bibinfo
		{author} {\bibfnamefont {G.~Y.}\ \bibnamefont {Yeom}},\ }\bibfield  {title}
	{\bibinfo {title} {A study of sapphire etching characteristics using
			bcl3-based inductively coupled plasmas},\ }\href
	{https://doi.org/10.1143/JJAP.41.6206} {\bibfield  {journal} {\bibinfo
			{journal} {Japanese journal of applied physics}\ }\textbf {\bibinfo {volume}
			{41}},\ \bibinfo {pages} {6206} (\bibinfo {year} {2002})}\BibitemShut
	{NoStop}%
\end{thebibliography}
%apsrev4-2.bst 2019-01-14 (MD) hand-edited version of apsrev4-1.bst
%Control: key (0)
%Control: author (8) initials jnrlst
%Control: editor formatted (1) identically to author
%Control: production of article title (0) allowed
%Control: page (0) single
%Control: year (1) truncated
%Control: production of eprint (0) enabled
%

%apsrev4-2.bst 2019-01-14 (MD) hand-edited version of apsrev4-1.bst
%Control: key (0)
%Control: author (8) initials jnrlst
%Control: editor formatted (1) identically to author
%Control: production of article title (0) allowed
%Control: page (0) single
%Control: year (1) truncated
%Control: production of eprint (0) enabled

%%=========================================== SI =========================
 %%=========================================== SI =========================
  %%=========================================== SI =========================
%\let\addcontentsline\oldaddcontentsline% Restore \addcontentsline

\clearpage
\onecolumngrid

%%%%%%% TITLE BLOCK OF SUPPLEMENTARY %%%%%%%
\begin{center}
	\large{\textbf{Supplementary Information for: Compact superconducting vacuum-gap capacitors with low microwave loss and high mechanical coherence for scalable quantum circuits}}
\end{center}
\begin{center}
	Amir Youssefi$^{1,2,3*}$, Mahdi Chegnizadeh$^{2,3,*}$, Marco Scigliuzzo$^{2,3}$,  and Tobias J. Kippenberg$^{2,3,\dagger}$\\[.5cm]
	{\itshape
	$^{1}$EDWATEC SA, EPFL Innovation Park, Lausanne, Switzerland.\\
	$^{2}$Institute of Physics, Swiss Federal Institute of Technology Lausanne (EPFL), Lausanne, Switzerland.\\
	$^{3}$Institute of Electrical and Micro Engineering, Swiss Federal Institute of Technology Lausanne (EPFL), Lausanne, Switzerland.\\
	}\vspace{0.5cm}
	$^*${\small These authors contributed equally to this work.}\\
	$^\dagger${\small Electronic address: tobias.kippenberg@epfl.ch}\\
\end{center}

%%%%%% RESET EQUATION NUMBERS ETC. %%%%%%%%
\setcounter{equation}{0}
\renewcommand{\thefigure}{S\arabic{figure}}
\renewcommand{\theHfigure}{S\arabic{figure}}
\setcounter{figure}{0}
\setcounter{table}{0}

\setcounter{subsection}{0}
\setcounter{section}{0}

%\tableofcontents
%\input Arxiv.toc

%\pagebreak

	\section{Batch-to-batch reproducibility}

\begin{figure*}[h] 
	\centering 
	\includegraphics{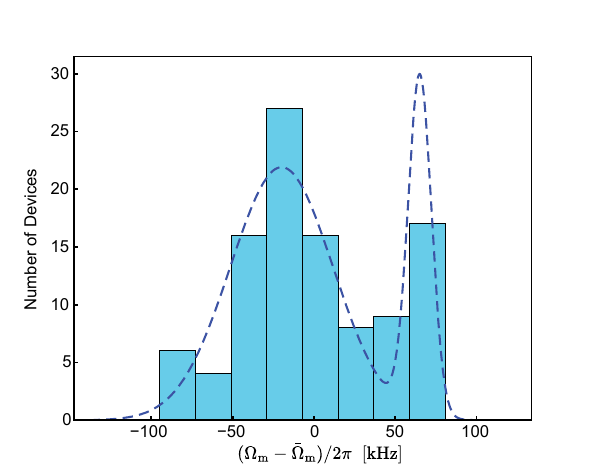}
	\caption{\textbf{Mechanical frequency distribution in multiple batches}. 
		The mechanical frequency of identical samples with 70~$\mu$m radius across the wafers realized in three different batches CCv3, CCv4, and CSv1, which are fabricated on September 2022, May 2023, and September 2024, respectively. The dashed line represents the best fit to two Guassian distributions. 
	}
	\label{fig:freq_col} 
\end{figure*}

We have measured the mechanical frequencies of devices with 103 identical drum geometry fabricated on three different batches. The average frequency is around 2~MHz, however the distribution is bimodal. The total standard deviation yields at 45~kHz. The distribution of frequencies is shown in \figref{fig:freq_col}.

\section{Suspended superconductor compared to metallized dielectric membranes}
We report quality factors as high as 40 million for suspended drum completely realized in superconductor materials that forms a capacitor with a second electrode. To achieve similar electric circuit, a dielectric suspended membrane can be metalized with a thin film of superconductor.

\begin{figure*}[h] 
	\centering 
	\includegraphics{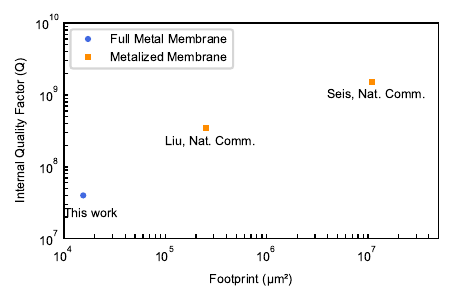}
	\caption{\textbf{Comparison of the mechanical quality factors in different platforms}. The value of quality factor for our drum with the blue marker is compareed to the recent one for metalized membrane in \cite{liu2025degeneracy} and \cite{seis2022ground}. 
	}
	\label{fig:comparison_q} 
\end{figure*}

It is important to notice that the larger quality factor achieved with metalized membrane usually comes at the cost of much larger footprint on chip. 

\section{Lithography and dose test}
\begin{figure*}[h] 
	\centering 
	\includegraphics{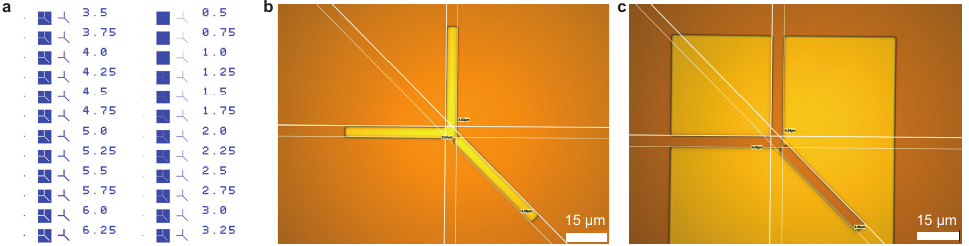}
	\caption{\textbf{Dose test.} \textbf{a}, A dose test pattern containing lines with various angles and widths, both for positive and inverted lithography jobs. The circuit wires are written in the inverted mode using positive resist for aluminum etching, while the trench itself is patterned without inversion. The pattern is written with different doses and depths of focus in direct (mask-less) lithography. We extract The critical dimensions of the wire and trenches using the optimized dose. \textbf{b, c}, Examples of positive and inverted patterns after dose test resist development.}
	\label{fig:fab_dose} 
\end{figure*}

To accurately calibrate the exposure parameters in the optical lithography we conduct a dose test. An example of a dose test pattern is shown in Fig.~\ref{fig:fab_dose}.

It is worth noting that the perfect exposure dose is different when etching silicon trench or etching aluminum film because of the different reflectivity of the surface. This needs to be tuned by separate dose tests with and without the aluminum layer.

The critical dimensions can be achieved by direct laser writing is CD$\approx 1 \mu$m minimum thickness of a pattern and pattern size fluctuation of $\Delta_\mathrm{d}\approx 500$~nm. Compared to the sizes of drumhead capacitors and spiral inductors, these CD and SF are sufficiently low resulting in mechanical frequency disorder of $\frac{\Delta\Omega_\mathrm{m}}{\Omega_\mathrm{m}} = \frac{\Delta_\mathrm{d}}{R} \approx 1\%$ for a trench radius of $R=50 \mu$m. However, when size fluctuation of drums and trenches matter -- e.g., to observe collective mechanical phenomena when degenerate mechanical modes are desired -- we may consider electron beam lithography ($\Delta_\mathrm{d}\approx 5$~nm) or deep UV lithography ($\Delta_\mathrm{d}\approx 50$~nm). This may result in smaller disorder of mechanical frequencies down to $\sim0.01\%$ and $\sim0.1\%$ for e-beam and DUV, respectively~\cite{chegnizadeh2024quantum}. Nevertheless, the microwave frequencies are more robust to lateral size fluctuations since the spiral inductor is a relatively large structure with a less concentrated electromagnetic field (compared with meander inductors or interrogated capacitors), and the value of inductance is less sensitive to the thickness disorder of the wire.

\section{Chemical mechanical polishing}
\begin{figure*}[h] 
	\centering 
	\includegraphics[width=\textwidth]{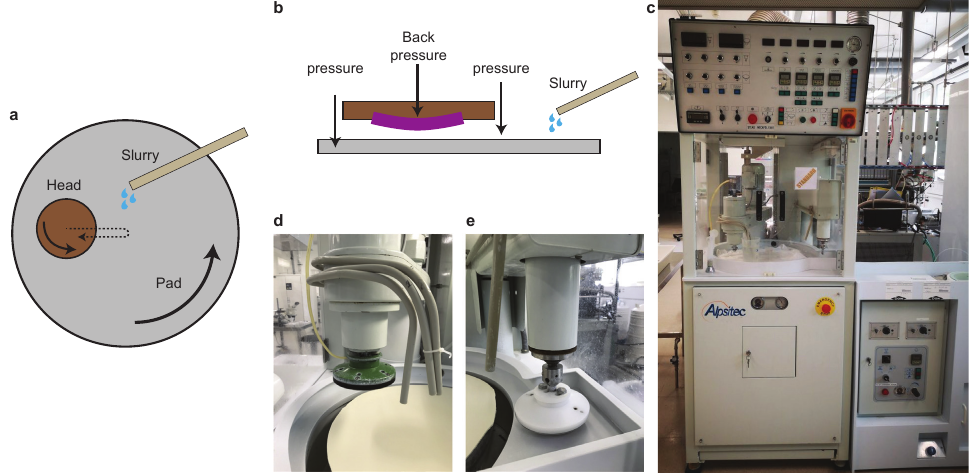} 
	\caption[CMP tool]{\textbf{Chemical mechanical polishing tool.} \textbf{a,} Top view schematics of CMP tool. A rotating head holding the wafer is moving back and forth to polish the wafer on a big rotating pad with rough surface. The pad is moisturized with a liquid slurry containing abrasive nano-particles. \textbf{b,} The side view schematics showing the head pressure and the back pressure used to compensate the wafer's bow. \textbf{c,} A photo of the CMP tool used in EPFL CMI cleanroom. \textbf{d,} Photo of the polishing head and slurry nozzle. \textbf{e,} Photo of the conditioning head used to clean and prepare the polishing pad after each run.}
	\label{fig:fab_CMP_tool} 
\end{figure*}

\begin{figure*}[h!] 
	\centering 
	\includegraphics[width=\textwidth]{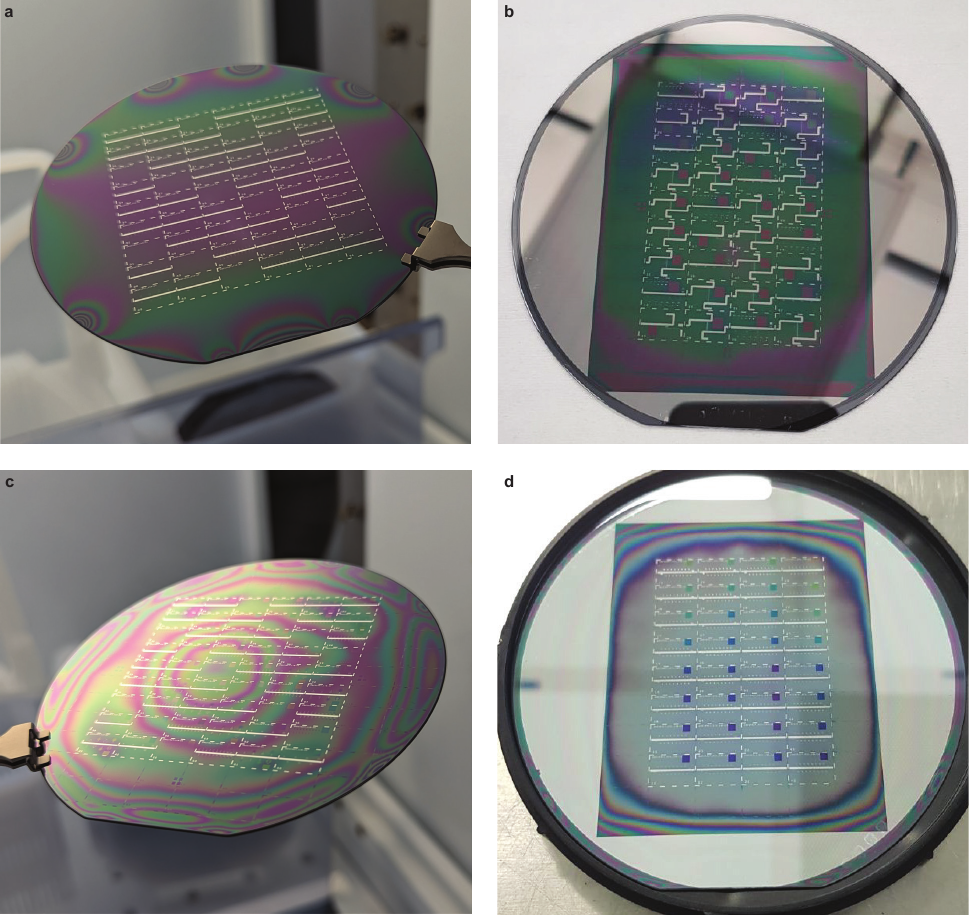} 
	\caption[CMP uniformity]{\textbf{CMP uniformity.} Photos show different wafers after CMP planarization. \textbf{a, b,} Showing successful CMP planarization with low non-uniformity around the center of 10~cm wafers. The non-uniformity can be quantitatively measured using optical reflectometer using the big square trenches on every chip. In addition, qualitatively the uniformity can be inspected by color change in fringes forming on the wafer due to the thickness variation of the remaining oxide film. \textbf{c,} CMP result when the back-pressure was not properly tuned to compensate the etch rate variation. \textbf{d,} It is recommended to remove the excess metal throughout the wafer to increase uniformity. In this wafer, although the uniformity is acceptable at the center, the excess metal imposes thickness variations on the edge chips. the photo is taken after IBE etch-back.}
	\label{fig:fab_CMP_wafers} 
\end{figure*}

As shown in Fig.~\ref{fig:fab_CMP_tool}, the tool consists of a big rotating polishing pad, a rotating head that holds the wafer, a slurry nozzle, and a conditioning head. The wafer will be fixed upside-down to the head. Then the head will bring the wafer close to the polishing pad and press it on the pad. Both pad and head rotate while the head also laterally moves around the pad. The slurry nozzle pours a little slurry on the pad to facilitate polishing. The conditioning head is separately used to clean and polish the big rotating pad itself after each planarization run. 
Although the CMP tool often has many knobs to tune different parameters, the most important ones to manipulate for a successful CMP planarization are the pad rotation speed, the head rotation speed, the head pressure on the pad, the back-pressure (for holding the wafer and tune its bow), the slurry rate, and the polishing time. Normally the tool does the polishing in three steps. The first step is surface preparation. The second is the longest step for polishing and the last for cleaning by replacing the slurry with water. After each run, conditioning is required to clean the pad.
The effect of each parameter on the etch rate and polishing is explained in the following. Note that all numbers provided are tuned for the tool we use (ALPSITEC$^\text{®}$ MECAPOL E 460) and may vary for other machines.
\begin{itemize}
	\item \textbf{Pad and head rotation speeds:} higher speed increases both etch rate and polishing rate, normally set around 80~rpm in our case.
	\item \textbf{Slurry rate:} higher rate increases the chemical etching faster than mechanical polishing rate. It is normally set to a low rate to just wet the pad. However, very low rate significantly increases the non-uniformity of the polishing due to the high friction.
	\item \textbf{Head pressure:} higher rate increases the polishing rate more than the etch rate. normally set around 0.5 Bars. Using very high pressures deforms the wafer and increases etching non-uniformity.
	\item \textbf{Back-pressure:} The wafer has a slightly bowed structure causing higher pressure on the edges than the center, which results in non-uniform polishing. In order to compensate for the wafer bow, back-pressure is applied to the wafer's backside which will equalize the center to edge etch rate difference. This is a crucial parameter to manipulate to reduce the non-uniformity.
\end{itemize}
Uniformity of CMP for different wafers and recipes is shown in Fig.~\ref{fig:fab_CMP_wafers}. 

Here we provide a short technical note on the CMP operation procedure we use to polish the wafers:
\begin{itemize}
	\item We first set up the tool, prepare the slurry (often 30N50, a basic slurry made of colloidal SiO$_2$ particles for dielectric polishing), and run a 1-minute pad conditioning. We dilute the slurry with water (1:1) to decrease the etch rate and increase the mechanical polishing effect.
	\item We start with plain dummy silicon wafers covered with the same oxide layer grown together with the main wafers in LTO. These plain dummy wafers have no pattern and are just used to optimize the etching rate and uniformity of the CMP. We measure their uniformity and average oxide thickness optically and run a single CMP cycle with optimized parameters from our previous experiences. After the run, we simply rinse and dry the plain dummy wafer and measure it again optically. With this, we can extract the average etch rate as well as the non-uniformity of the etch. Depending on these two values, we modify the back pressure (to increase uniformity), head pressure, and slurry rate (to change the etch rate). We then use another clean and plain dummy wafer with the modified parameters and continue this iteration until we reach the minimum possible non-uniformity (usually around 1\% non-uniformity after a single CMP run).
	\item After the uniformity optimization, we use another set of patterned dummy wafers (with exactly the same trench depth and pattern of the trenches we have on the design) to check the planarization rate. We measure the initial topography with a mechanical profilometer. After running an optimized CMP cycle, we measure the topography again to see how much the trench depth is reduced on the sacrificial layer. We continue running CMP cycles to ensure the topography can be reduced to less than 10~nm in a reasonable cycle number (not etching more than 5 times the trench thickness and not reaching closer than 500~nm to the substrate surface to avoid delamination). If the process does not work properly or results in a low planarization rate, we go back to the first step and increase the head pressure or modify the slurry rate and repeat the procedure.
	\item When all the suitable parameters are achieved, we switch to the main wafers. We run the CMP for several cycles on the wafer, including a 30~s pad conditioning between each cycle until we reach to the residual topography tolerance. After that we immediately run the post-CMP cleaning.
	\item Finally, we measure the residual thickness of SiO$_2$ on the big square trenches we discussed in the main text and extract a chip-wise map of thickness over the wafer area.
\end{itemize}

Because of the CMP planarization, all the topography information of the wafer is removed, and after covering the wafer with 200~nm reflective aluminum, it will be challenging to find markers for lithography, which are buried below the sacrificial layer and Al. To avoid this problem, we etch big openings on the markers (defined in the trench layer as the first pattern) during SiO$_2$ opening step to make them visible in the next lithography step.

We finally note that the CMP etch rate versus time is nonlinear due to the pad softening and heating up (this means that running the CMP two times with a conditioning in between gives a different etch rate of running it once with double of the time). The polishing rate also depends on parameters and very nonlinear, often higher when topography is deep and lower when it gets shallower. 

%\begin{figure*}[h!] 
%	\centering 
%	\includegraphics[width=\textwidth]{Figures/Fig_Fab_Plassys} 
%	\caption[Removing native aluminum oxide in Plassys]{\textbf{Removing native aluminum oxide in Plassys.} \textbf{a,} The schematic fabrication process shows the native Al$_2$O$_3$ layer in galvanic connection can be removed using Plassys$^\text{®}$ evaporator using argon ion milling. The wafer is then transferred to the evaporation chamber under ultra-high vacuum for the top Al layer deposition. \textbf{b,} Shows the Plassys machine we use in EPFL-CMi. The load-lock chamber is equipped with an ion beam source. The transfer arm moves the wafer between different chambers isolated with gates. The oxidation chamber is dedicated for Josephson junction fabrication and not used in this work.}
%	\label{fig:fab_Plassys} 
%\end{figure*}

\section{Dicing}
Now the wafer is ready to be diced into chips for the last step, the HF release. The size of chips in our design is 9.5~mm$\times$6.5~mm. To dice the wafer, we first spin coat it with a thick resist ($15 \mu$m AZ$^\text{®}$ 10XT-60) to protect the circuits from Si debris and other contaminations during dicing. We use 100~$\mu$m Nickel blade with 35000~rpm rotation speed and 5~mm/s cutting speed (Disco$^\text{®}$ DAD321). On the wafer design, we defined dashed lines in the trench layer and bottom Al layer to as chip border boxes to be used in dicing alignment. After dicing, the chips will be gently detached from the UV tape used for dicing and will be sorted in a Teflon chip holder for UFT resist stripping. We keep chips for more than 20 minutes in a UFT clean bath, then rinse and dry them manually (with a pressurized air nozzle). Afterward, we use 200~Watt and  200~sccm oxygen plasma (Tepla$^\text{®}$ GiGAbatch) for a few minutes to clean any remaining resist residue from the chips. 
We fabricated a dedicated silicon wafer chip holder by deep etching (Bosch DRIE) a Si wafer pattern 200~$\mu$m rectangular pads corresponding to our chip size. We locate chips inside these pads during oxygen plasma and HF release to minimize the risk of chip flipping during chamber pump down.

\section{Packaging}
\begin{figure*}[h!] 
	\centering 
	\includegraphics[width=\textwidth]{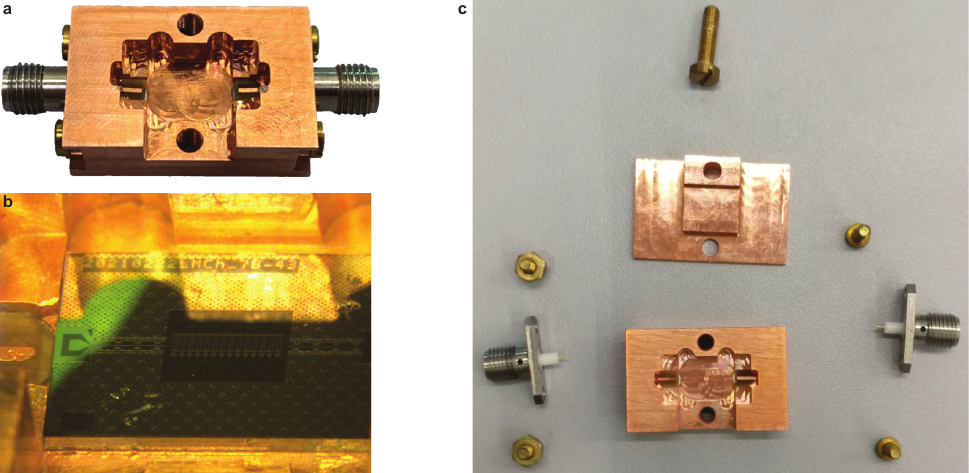} 
	\caption[Chip box]{\textbf{Chip box.} \textbf{a}, Photo of the chip box made out of oxygen free high conductivity copper. The box has two SMA coaxial outputs soldered to short micro-strip PCBs glued to the box. \textbf{b}, A chip will be glued by conductive silver paint inside the box. \textbf{c}, Photo of elements of a sample box before assembling. the lid will be closed to ensure the superconducting sample is light-tight during the experiment.}
	\label{fig:fab_box} 
\end{figure*}
After the release, the chips are ready for packaging. We need to handle chips carefully to avoid the risk of collapse due to mechanical shocks or electrostatic discharge. The electro-mechanical devices can be inspected using an optical microscope -if we have compressive stress in Al top layer, the dome shape of buckled drums is clearly visible under a 10X microscope aperture. More systematically, we use an optical profilometer (Sensofar$^\text{®}$ S-Neox or Bruker$^\text{®}$ Contour X) to measure the surface topography of the drum (especially when it does not buckle due to stress) and make sure it is successfully released.

We use a machined copper box as the chip holder made out of oxygen-free high-conductivity copper to ensure good thermal contact at mK temperatures developed by L.D. Toth~\cite{toth2018dissipation} and N.R. Bernier~\cite{bernier2019multimode}. The sample holder has two SMA connectors soldered to a micro-strip line defined on a short piece of printed circuit board which is glued inside the box (Fig.~{\ref{fig:fab_box}}). The PCBs will be electrically connected to the micro-strip feed line on the chip by wire bonds. The PCBs are permanently glued to the copper box using a conductive epoxy (Epo-Tek$^\text{®}$ H20E). The chip holder can be cleaned using a fiber brush, followed by Isopropanol cleaning. The copper box gets oxidized in the span of time, and can be cleaned by sonication in diluted acetic acid.

\begin{figure*}[hb!] 
	\centering 	\includegraphics[width=\textwidth]{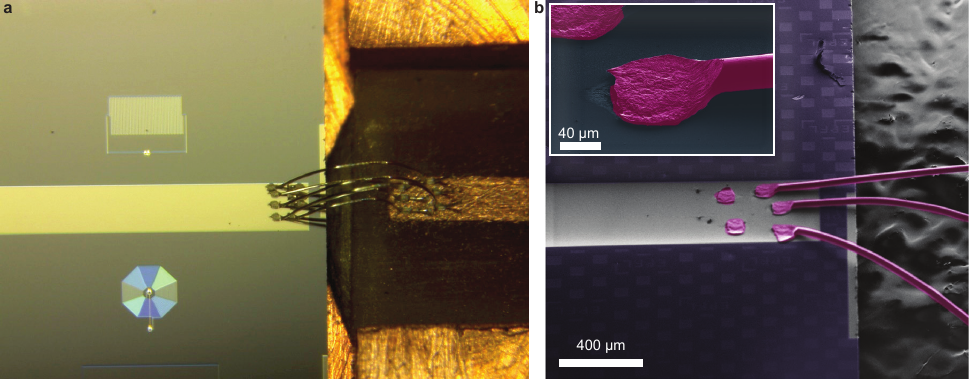} 
	\caption[Wire bonds]{\textbf{Wire bonds.} \textbf{a}, Microscope image of a chip with micro-strip waveguide connected to the copper center of PCB micro-strip waveguide through $35 \mu$m aluminum wire bonds. We normally use more than 5 bonds to ensure the impedance matching between the sample and PCB. \textbf{b}, SEM of $35 \mu$m aluminum wire bonds connected to the 100~nm aluminum bottom plate. If the parameters of bonding such as ultra-sonic power, press time, and force are not optimized, the bond will be detached and leaves a $~100\times100 \mu\mathrm{m}^2$ foot print on the chip which cannot be used for a second bonding anymore.}
	\label{fig:fab_wire_bond} 
\end{figure*}

We use silver conductive paint (RS$^\text{®}$ Pro) to mount the chip inside the box to make a good thermal connection between the chip and the box and maintain the electrical boundary condition.
Then we use wire bonder (F\&S$^\text{®}$ Bondtec 56i or TPT$^\text{®}$ HB10) to connect the feed line on the chip to the PCB using aluminum 25~$\mu$m diameter wires. We typically use more than five bonds on each side to ensure the 50-ohm impedance matching connection. Using a smaller number of wires showed impedance mismatch resulting in standing waves for micro-strip feed lines. After wire bonding, the electric connection will be tested by an Ohm-meter (typically shows $\sim$4 Ohms between two cores of the SMA connector using micro-strip feed line), and the lid of the box will be closed and tightened by a brass screw, and the device is ready for low-temperature measurements (Fig.~{\ref{fig:fab_wire_bond}}). 

Exposing samples to air during the packaging will not degrade them (except for dust contamination on the drums, which can be removed by gently blowing them with pressurized air). However, we recommend avoiding abrupt temperature and humidity changes during the transfer and keeping chips in Nitrogen boxes for long-time storage.

\section{Aluminum thin films}
\label{sec:fab:Al_film}
As expected, the quality, stress, and roughness of the aluminum thin film used for the top layer influences both the release step and the low-temperature mechanical quality factor of drumheads. Here discuss the effect of deposition and post-deposition techniques to manipulate such parameters.

\subsection{Deposition method}
Aluminum can be deposited by either electron beam evaporation or sputtering techniques as physical depositions. The evaporation is done under high vacuum ($10^{-6} - 10^{-8}$~mBar), where an electron beam is hitting a crucible to emit Al atoms. The deposition rate is controlled by the electron-beam power and the distance of the wafer to the crucible. In sputtering, a crucible of aluminum will be bombarded by plasma Argon ions. The detached Al atoms will be sputtered on the wafer. Sputtering normally gives better step coverage compared to evaporation. The deposition rate and film properties can be controlled by argon flow and the source power. The pressure of the chamber in sputtering is typically around $10^{-3}$~mBar, depending on the argon flow.

\subsection{Low-temperature stress of aluminum films}
Although the aluminum thin film can have compressive stress at room temperature, the significant difference between the thermal expansion rate of Al and Si results in the shrinking of Al films faster than Si and induces tensile stress at cryogenic temperatures. The following relation can estimate the final low-temperature stress of the film:
\begin{equation}
	\sigma_\mathrm{Al} = \sigma_\mathrm{Al}^\mathrm{RT} + \int^{300 \text{ K}}_{10 \text{ mK}}{Y^\mathrm{Al}_{(T)} (\alpha^\mathrm{Al}_{(T)} - \alpha^\mathrm{Si}_{(T)}) dT} \ \approx 300 \text{ MPa} + \sigma_\mathrm{Al}^\mathrm{RT},
\end{equation}
Where $Y^\mathrm{Al}$ is Young's modulus of aluminum and $\alpha$ is the thermal expansion rate of Al or Si, respectively. The relatively big difference between two expansion rates ($\alpha^\mathrm{Al}_{(20 \mathrm{K})} = 23.1 \times 10^{-6}/$°C and $\alpha^\mathrm{Si}_{(20 \mathrm{K})} = 2.6 \times 10^{-6}/$°C) results in considerable stress change at lower temperatures as well as high sensitivity of deposition induced initial stress to the deposition temperature and thermalization. The initial room temperature stress, $\sigma_\mathrm{Al}^\mathrm{RT}$, varies depending on the deposition conditions. The stress of a thin film can be calculated by measuring the change of the wafer's bow - i.e., the curvature of the wafer - before and after deposition using the following relation:
\begin{equation}
	\sigma_\mathrm{film} = - \frac{Y_\mathrm{sub}}{6(1-\nu_\mathrm{sub})} \frac{t_\mathrm{sub}^2}{t_\mathrm{film}} \left(\frac{1}{R_\mathrm{sub+film}} - \frac{1}{R_\mathrm{sub}}\right),
\end{equation}
Where $\nu$ shows the Poisson's ratio, $t$ is the thickness, and $R$ is the wafer's bow. The wafer's bow can be measured optically by sweeping a laser on the wafer in stress measurement tool (Toho Technology$^\text{®}$ FLX 2320-S). 

\textbf{High temperature aluminum deposition}

\begin{figure*}[h] 
	\centering 
	\includegraphics[width=\textwidth]{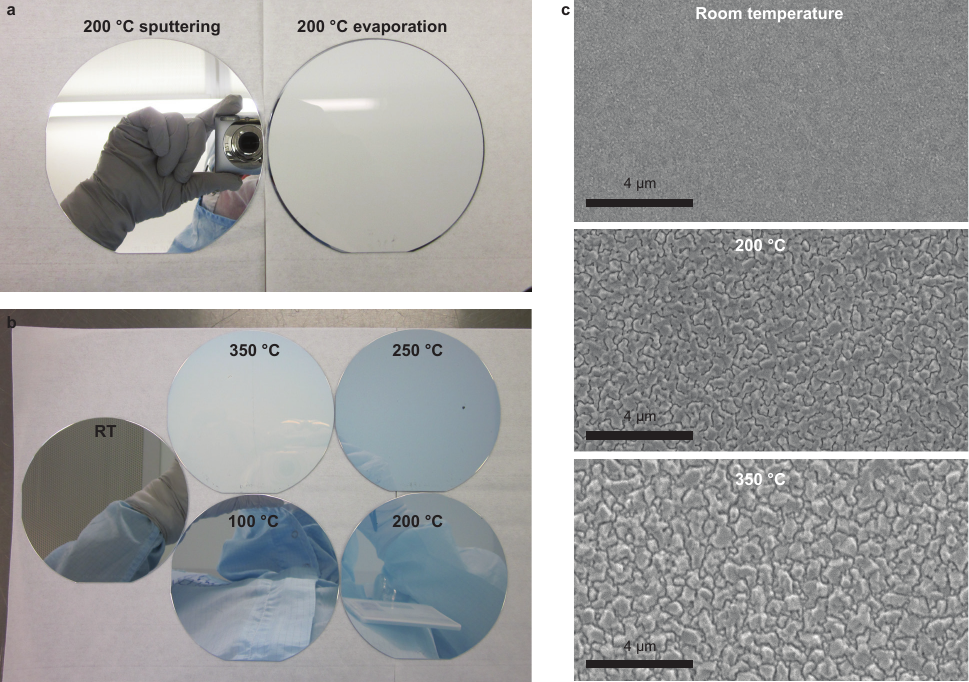} 
	\caption[High temperature aluminum deposition]{\textbf{High temperature aluminum deposition.} \textbf{a}, Comparison between 200°C sputtering and evaporation. \textbf{b}, Aluminum sputtering at high temperatures. \textbf{c}, SEM of high temperature aluminum sputtering showing significant increase of grain size by temperature.  }
	\label{fig:fab_high_T} 
\end{figure*}

We explored high temperature evaporation and sputtering of Al to reach to tensile stress at room temperature. Both evaporator and sputtering tools can deposit at higher temperatures up to 350 Celsius. We tested deposition of 250~nm Al on Si wafer at 200°C with both methods. Although both methods resulted in tensile stress of $\sim30$~MPa at room temperature, sputtering showed an acceptable film quality with $R_\mathrm{a}=15$~nm while the high temperature evaporation resulted in color change of the material to white and increasing of the grain size and roughness to $R_\mathrm{a}=50$~nm, which indicates compound formation between Al and Si (Fig.~\ref{fig:fab_high_T}).

\subsection{High temperature sputtering and surface roughness}
We deposited Al at different temperatures with the sputtering tool (Pfeiffer$^\text{®}$ SPIDER 600) and measured stress and roughness. Although the film stress increases by temperature, the roughness and grain size also increases which reduces the quality of the film.
\begin{table}[h!]
	\caption[High temperature aluminum sputtering]{High temperature aluminum sputtering.}
	\label{tab:high_T}
	\centering
	\begin{tabular}{cccccc}
		\toprule
		T (Celsius) & 20 & 100 & 200 & 250 & 350 \\ 
		\midrule
		$\sigma_\mathrm{Al}$ (MPa) & -53	& 35 	& 41 	& 47 & 61 \\
		$R_\mathrm{a}$ (nm)	& 2	& 10 	& 15 	& 17 & 20 \\
		\bottomrule 
	\end{tabular}
\end{table}

\subsection{Effect of evaporation rate on surface roughness}
We measured the film roughness in the three conditions of room temperature Al evaporation mentioned below and realized EVA 760 gives us the best film quality. We note that to remove the thin aluminum oxide layer for the galvanic connection, we have to deposit the top layer with Plassys$^\text{®}$ UHV evaporator with $10^{-8}$~mBar vacuum which also demonstrated high mechanical quality factors.
\begin{itemize}
	\item Alliance-Concept$^\text{®}$ EVA 760 with $10^{-6}$~mBar pressure, 45~cm working distance, and 5~\AA/s deposition rate. Film roughness: $R_\mathrm{a} = 2.1$~nm
	\item Leybold Optics$^\text{®}$ LAB 600H with $1.8\times10^{-6}$~mBar pressure, 100~cm working distance, and 4~\AA/s deposition rate. Film roughness: $R_\mathrm{a} = 3.5$~nm
	\item Leybold Optics$^\text{®}$ LAB 600H with $1.8\times10^{-6}$~mBar pressure, 100~cm working distance, and 1~\AA/s deposition rate. Film roughness: $R_\mathrm{a} = 8$~nm
\end{itemize}

\subsection{Effect of annealing cycle on the stress}
\begin{figure*}[h!] 
	\centering 
	\includegraphics[width=\textwidth]{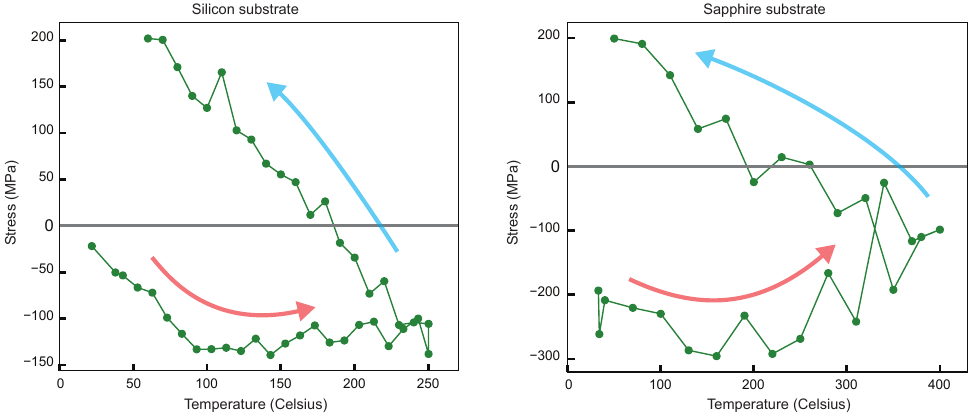} 
	\caption[Effect of annealing on the aluminum stress]{\textbf{Effect of annealing on the aluminum stress.} Annealing aluminum thin film and cooling it down adiabatically changes the stress to tensile. 150~nm Al is evaporated on silicon and sapphire substrates and annealing for about two hours cycle up to 250°C and 400°C respectively showed stress enhancement to 200~MPa at room temperature. One has to be careful to run annealing under vacuum to avoid oxidation of the film.}
	\label{fig:fab_annealing_stress} 
\end{figure*}
The effect of thermal annealing on the aluminum thin film stress has been studied in~\cite{gardner1990mechanical}. It has been shown that the slow annealing cycle of the Al-1\%Si film with 640~nm thickness after deposition by heating it up to $\sim 200$°C and cooling it down shows a hysteresis behavior in the stress resulting in higher tensile stress at the same initial temperature. We investigated this behavior on the pure Al with 150~nm thickness evaporated by 4~\AA/s rate. The result confirmed the same behavior where we could change the stress of the film deposited on Si or Sapphire wafers from initial compressive stress to $\sim200$~MPa tensile (Fig.~\ref{fig:fab_annealing_stress}). While this can be a useful technique to engineer the stress after deposition, we did not manage to use it for the final high-$Q_\mathrm{m}$ and reproducible devices because of concerns about the compound formation and oxidation on the Al film. We did not investigate further the change of roughness and grain size after annealing. However, we qualitatively did not observe any color change, severe roughness, or reduced transparency of the film after annealing. It is worth noting that annealing under a weak vacuum to 200°C is often used in the traditional fabrication process of drumhead capacitors to relax and uniforms stress in such drums before the release~\cite{toth2018dissipation}.

\subsection{Yield stress of aluminum films}

The stress-strain relation in materials generally has a linear behavior for small strains, which is called an elastic regime. By increasing the strain, in some cases, the stress does not scale linearly anymore, and the material goes to the plastic regime. Increasing the strain further results in buckling or cracking of the film. The stress which the elastic regime goes to plastic is called the \textit{yield stress}. Although the bulk yield stress can be theoretically calculated for some materials, the experimental values are often lower than the theoretical expectations~\cite{courtney2005mechanical}. The bulk yield stress of Al is calculated $\sim900$~MPa; however, the measured values for the bulk aluminum and its alloys are between $200-400$~MPa~\cite{courtney2005mechanical}. These values also depend on temperature. The low-temperature data on the mechanical properties of bulk Al and its alloys can also be found in~\cite{ekin2006experimental}. It is known that the yield stress in thin films or nano-structures can be higher than the bulk values approaching the theoretical limit depending on the thickness and grain size~\cite{richter2009ultrahigh,steinwall1990mechanical}. We did not find a systematic study on the yield stress of the sub-micron thin aluminum films at low temperatures. 

\begin{figure*}[h!] 
	\centering 
	\includegraphics[width=\textwidth]{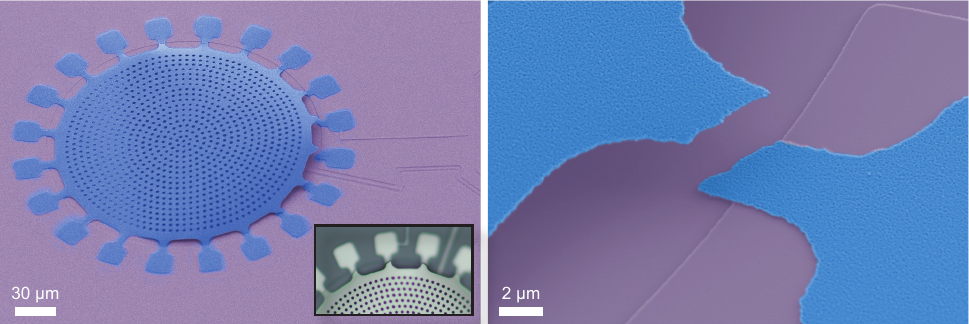} 
	\caption[Broken clamps and yield stress.]{\textbf{Broken clamps and yield stress.} SEM and microscope images show an example of devices after cool down with high clamp ratio (CR=7 for the device shown in the figure). Due to clamp tapering, the stress increases reaching to the yield stress of the thin film which results in the crack of the clamp. }
	\label{fig:fab_broken_legs} 
\end{figure*}

Tapering the clamps of the drumhead increases the local stress on the clamps. This enhanced stress should be below the yield stress to avoid breaking the legs or going to the plastic regime. To observe the ultimate limit of stress enhancement, we made a sweep over the clamping ratio (CR $\equiv$ the total perimeter of the trench divided by the total perimeter of the clamps) and cooled down these devices. After warming them up again, we observed drums with CR$>$4 are cracked (Fig.~{\ref{fig:fab_broken_legs}}), meaning that the maximum tolerable stress in our design at low temperatures should be below 1~GPa.

\section{Non-uniformity tolerance}
In many applications, we need to design arrays and lattices of coupled identical LC electro-mechanical circuits. Studying the disorder tolerance in such systems is crucial to understand the fabrication technique's limits and to improve it. In the LC circuits with the spiral inductor and vacuum gap capacitor, the dominant frequency disorder mechanism is the gap size imperfection which is directly defined by the total non-uniformity of the trench depth and bottom aluminum layer thickness at the end of the fabrication process. The frequency disorder between two identical circuits with a central frequency of $\omega_\mathrm{c}$ and target gap size of $d$ can be written as:
\begin{equation}
	\frac{\Delta\omega_\mathrm{c}}{\omega_\mathrm{c}} = \frac{\Delta d}{2d}.
\end{equation}
For example, this results in $15 \frac{\text{MHz}}{\text{nm}}$ shift of the cavity for a 6~GHz central cavity resonance frequency and 200~nm gap size. The total non-uniformity tolerance ($\epsilon \equiv$gap size variation divided by the lateral distance) is proportional to the maximum frequency tolerance of the circuit, $\Delta\omega_\mathrm{c}$, and the maximum lateral distance between two vacuum gap capacitors in the design, $l$, expressed by:
\begin{equation}
	\epsilon = \frac{2 d \Delta\omega_\mathrm{c}}{l \omega_\mathrm{c}}.
\end{equation}
The frequency tolerance is normally defined by the desired mutual coupling between the identical LCs winch is typically designed greater than 100~MHz in our designs. Considering optomechanical lattices~\cite{youssefi2022topological} as an example, we required 50~MHz minimum coupling rate for a 2~mm lattice which results in maximum non-uniformity tolerance of $\epsilon \simeq 2 \frac{\text{nm}}{\text{mm}}$.

Each step of the fabrication process can in principle induces non-uniformity. The total non-uniformity on the final device can be expressed based on the individual step's non-uniformity as $\epsilon_\mathrm{tot} = \sqrt{\sum \epsilon_i^2}$. Considering our process flow, the main non-uniformity origins are silicon plasma etching ($\epsilon < 0.5  \frac{\text{nm}}{\text{mm}}$), bottom layer aluminum evaporation ($\epsilon \simeq 0.1  \frac{\text{nm}}{\text{mm}}$, for sputtering it is higher value), LTO SiO$_2$ deposition ($\epsilon \simeq 0.2  \frac{\text{nm}}{\text{mm}}$), IBE etch-back ($\epsilon < 0.1  \frac{\text{nm}}{\text{mm}}$), and most importantly CMP, $\epsilon_\mathrm{CMP}$. Considering the above-mentioned example of the topological lattice, the maximum tolerated CMP non-uniformity should be $\epsilon_\mathrm{CMP} < \sqrt{\epsilon_\mathrm{tot}^2 - \sum_{i\neq \mathrm{CMP}} \epsilon_i^2} = 1.9 \frac{\text{nm}}{\text{mm}}$. This value can be easily achieved in CMP by optimizing the polishing parameters.

\section{LC circuits without the galvanic connection}
\begin{figure*}[h!] 
	\centering 
	\includegraphics[width=\textwidth]{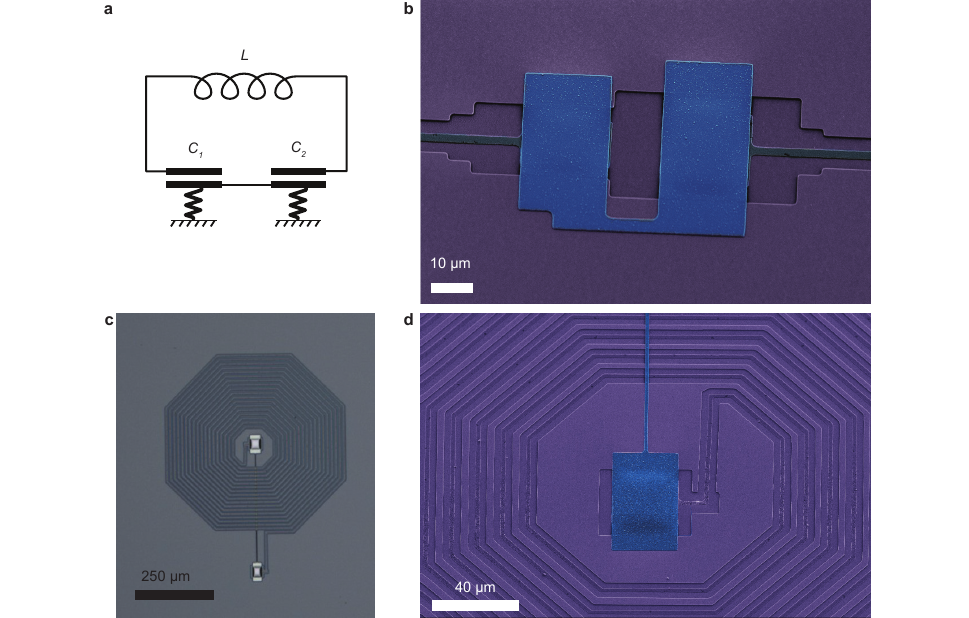} 
	\caption[Circuits without galvanic connection]{\textbf{Circuits without galvanic connection.} \textbf{a,} Circuit diagram of an electro-mechanical system with two mechanical oscillators which does not require galvanic connection of top to bottom layers. Since two capacitors are in series, the optomechanical coupling for each of them will be reduced proportional with their participation ratio. \textbf{b,} SEM of a double-capacitor circuit without galvanic connection. \textbf{c, d,} Micrograph and SEM of a circuit with spiral inductor and two capacitors inside and outside of the spiral connected through the spiral airbridges. }
	\label{fig:fab_nongalvanic} 
\end{figure*}

The galvanic connection can be evaded by making two parallel plate capacitors in series (as shown in Fig.~\ref{fig:fab_nongalvanic}). In this case, the optomechanical coupling rate will be diluted proportionally to the capacitors' participation ratio. However, we decided not to dilute the coupling and create a direct galvanic contact between the top and bottom layers.

\begin{figure*}[h!]
	\centering
	\includegraphics[width=\textwidth]{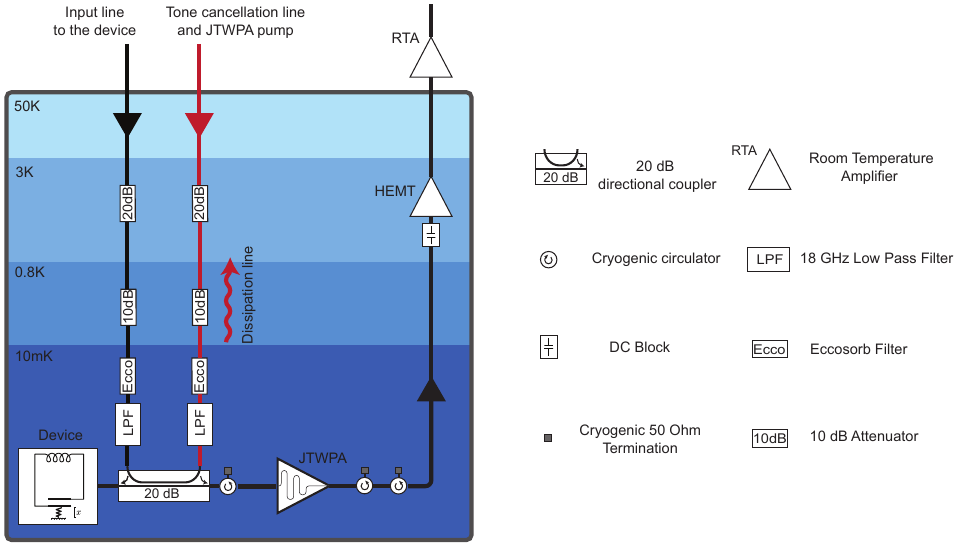}
	\caption[Microwave wiring in the fridge.]{\textbf{Microwave wiring in the fridge.} The standard microwave wiring used in circuit optomechanical experiments is shown. To reduce power dissipation in the base temperature flange we use directional couplers instead of cold attenuators and redirect the residual high-power optomechanical pump to higher stages to dissipate. The dissipation line simultaneously is used to combine tone cancellation signals to generate destructive interference before JTWPA. The dissipation line also carries the JTWPA pump.}
	\label{fig:exp_wiring}
\end{figure*}

\section{Sapphire substrate processing}\label{Sec:fab_sapphire}
\begin{figure}[h!] 
	\centering 
	\includegraphics[width=\textwidth]{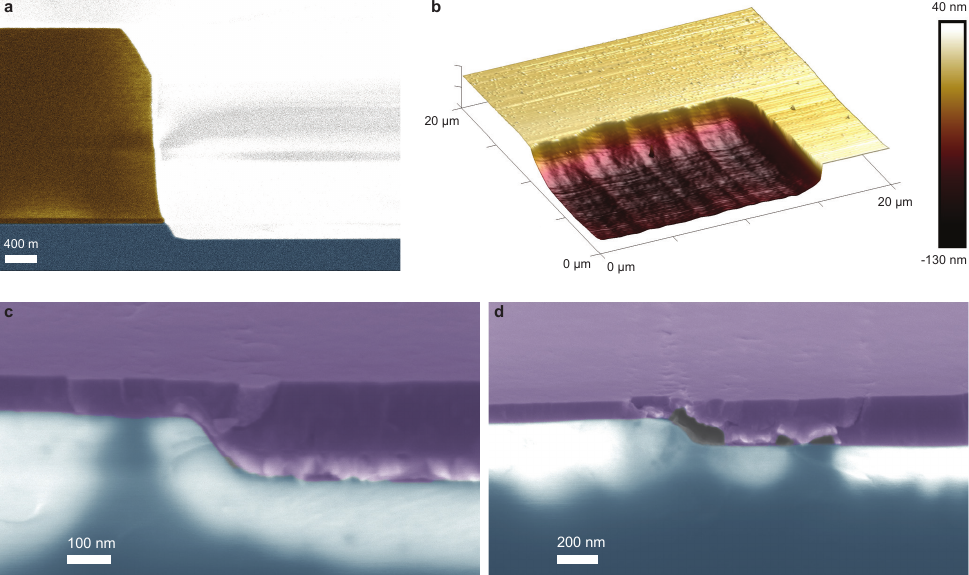} 
	\caption[Sapphire substrate processing]{\textbf{Sapphire substrate processing.} \textbf{a}, Cross section SEM of plasma etched trench in sapphire substrate. The orange color shows the photo resist. \textbf{b}, The AFM topography of a trench in sapphire. \textbf{c, d,}, Cross section SEM of a sapphire trench covered by amorphous silicon sacrificial layer after CMP planarization. Due to the low adhesion of the aSi to the substrate, the sacrificial layer delaminates in the CMP.}
	\label{fig:fab_sapphire} 
\end{figure}
In the early stage of our process development, we investigated implementing the idea of etching trenches on sapphire substrate, using amorphous Si as a sacrificial layer and planarizing it with CMP and releasing the drumhead with XeF$_2$ which is an isotropic gas etching, inspired by the traditional electro-mechanical platform were developed in LPQM-EPFL~\cite{toth2018dissipation}. The advantage of sapphire at fist sight was that it is a very resilient material to etchants, has high Young's modulus, low thermal expansion rate, and, most importantly, known to have less bulk dielectric loss for microwave circuits~\cite{read2022precision}. However, the other side of the coin was that micro-machining of sapphire is not trivial since it does not react with standard etchants. We realized that a few chlorine chemistries could be used to plasma-etch sapphire~\cite{jeong2002study}. Among the possible options we tried 20\%Cl$_2$-80\%BCl$_3$ argon plasma etching (STS$^\text{®}$ Multiplex ICP). Although the achieved etch rate was low (37 nm/min) and selectivity was below one (sapphire:PR $\sim$ 1:3.5), we managed to etch a few hundred nano-meter depth trenches with acceptable sidewall angle for our purpose and roughness of $R_\mathrm{a} = 1.2$~nm inside the trenches (Fig.~\ref{fig:fab_sapphire}).

After making the trench and deposition of the aSi sacrificial layer (sputtering with a good step coverage), we tried using CMP to planarize the topography. In this step, we realized two important challenges. First, because of the hardness of the sapphire wafer, the bow compensation with the back pressure was challenging, resulting in low uniformity after the CMP. The second problem was the low adhesion of aSi to the sapphire substrate, which resulted in the delamination of the sacrificial layer even when we stopped polishing above the wafer level. Considering such issues, we decided to switch to the high resistivity silicon substrate, which supports a wide range of standardized micro-machining processes.

\begin{figure*}[h!] 
	\centering 
	\includegraphics[width=\textwidth]{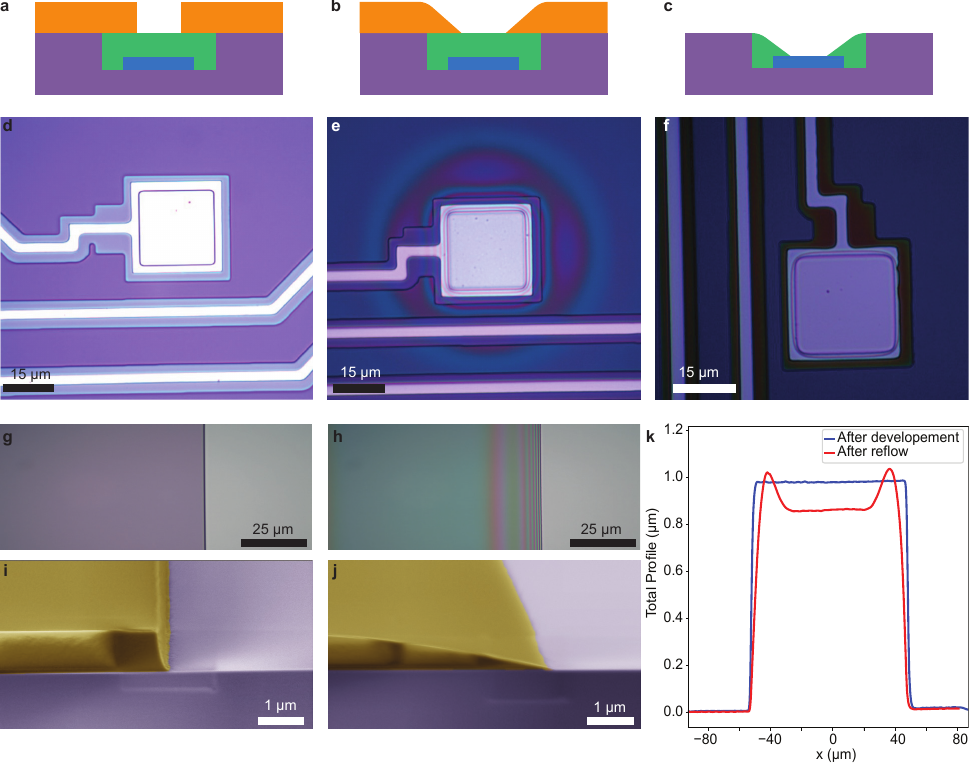}
	\caption[Reflow process for smooth SiO$_2$ opening.]{\textbf{Reflow process for smooth SiO$_2$ opening.} \textbf{a-c,} The schematic fabrication process showing the photoresist after patterning and development (a), Heating up the resist for reflow (b), and etching the SiO$_2$ with a low selectivity DRIE process (c).  \textbf{d-f,} Microscope images of the opening after the corresponding step shown above (a-c). The optical fringes in \textbf{e} indicates the smooth sidewall of the resist. \textbf{g-j,} Microscope top view images and SEM cross sections of normal resist and re-flowed resist respectively. \textbf{k,} Mechanical profilometry on normal and re-flowed resist.     }
	\label{fig:fab_opening} 
\end{figure*}

\begin{figure*}[h!] 
	\centering 
	\includegraphics[width=\textwidth]{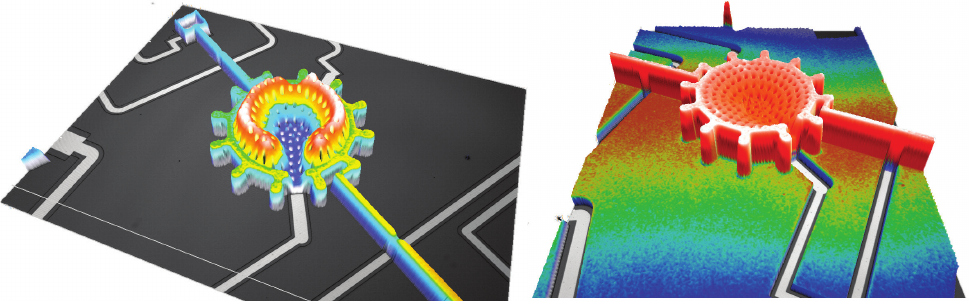} 
	\caption[Collapsed drums.]{\textbf{Collapsed drums.} Too small gap size ($<75$~nm), thin top layer($<50$~nm), and the big size of the drumhead ($R>300$~$\mu$m) can cause collapse of the structure after the release. Nevertheless, the collapse is rarely seen if the mentioned parameters are in the proper range, resulting in high yield fabrication i.e. $>$95$\%$ successful release.}
	\label{fig:fab_collaps} 
\end{figure*}

\begin{figure*}[h!] 
	\centering 
	\includegraphics[width=\textwidth]{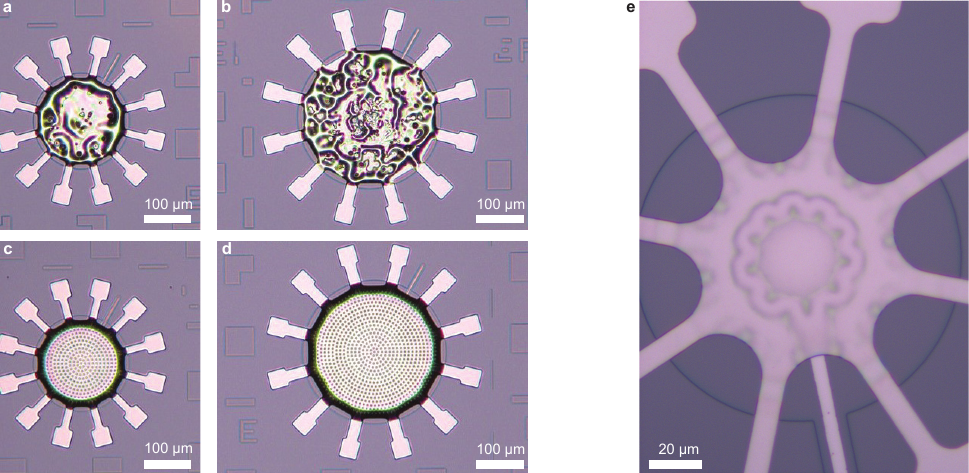} 
	\caption[Effect of the release holes.]{\textbf{Effect of the release holes.} \textbf{a-d}, Shows similar drums on a same chip after release. \textbf{a} and \textbf{b} shows wrinkled drums without release holes to assist HF vapor penetration under the structure. \textbf{c} and \textbf{d} shows the effect release holes which result in releasing drumheads in the fundamental symmetric buckling mode. \textbf{e,} Shows a drumhead without release holes after a long exposure to HF vapor. The central part which covers the bottom electrode is still not released because of smaller spacing between layers. HF vapor needs longer time to laterally penetrate under the top layer resulting in an incomplete release forming a wavy buckled shape on the released parts.   }
	\label{fig:fab_release_holes} 
\end{figure*}

\begin{figure*}[h!] 
	\centering 
	\includegraphics[width=\textwidth]{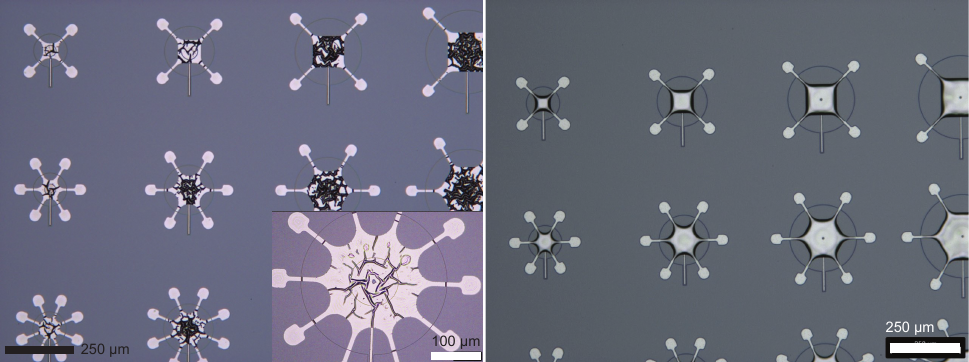} 
	\caption[Effect of top layer thickness on the release.]{\textbf{Effect of top layer thickness on the release.} Here we compare two identical designs after release with a same fabrication process but different top aluminum layer thicknesses of 50~nm (left image) and 150~nm (right image) with 200~nm gap size. The thin aluminum wrinkles instead of buckling up.  }
	\label{fig:fab_top_thickness} 
\end{figure*}

\begin{figure*}[h!] 
	\centering 
	\includegraphics[width=\textwidth]{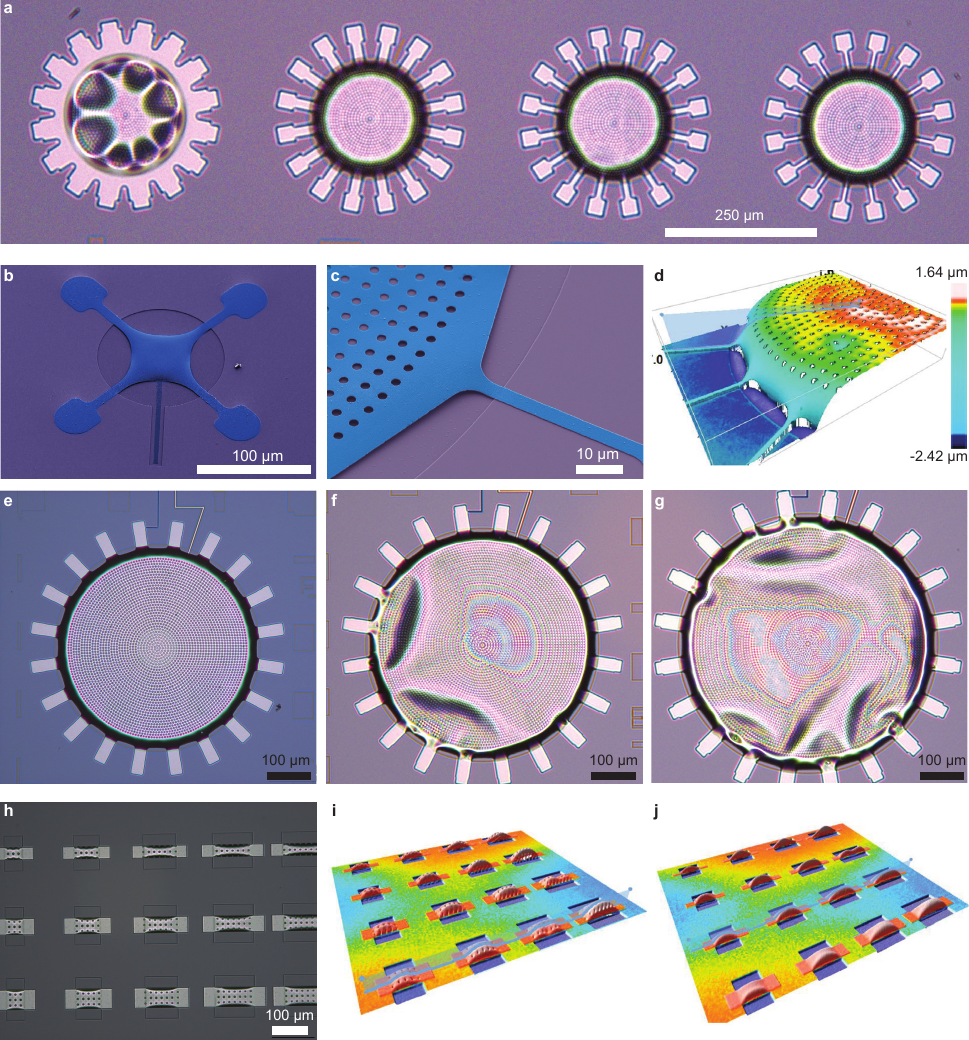} 
	\caption[Effect of clamps, size, and shape on the release.]{\textbf{Effect of clamps, size, and shape on the release.} \textbf{a}, Microscope image of drums with same parameters but sweeping the clamp ratio (The ration of the total trench perimeter over the clamps perimeter) from CR=1 (the left drum) to CR=4 (the right drum). Fully clamped drums (CR=1) in the presence of the compressive stress buckles up in a deformed shape, while the rest buckle in the fundamental mode. \textbf{b-d}, SEM and optical profilometry of released drums with high CR number. \textbf{e-g}, Show the maximum radius of successfully released drums with the gap size of 200~nm (\textbf{e}) and the collapse/deformation of bigger drums (\textbf{f, g}). \textbf{h-j}, Shows other possible geometry as rectangular beams released with and without using release holes.}
	\label{fig:fab_release_details} 
\end{figure*}

\begin{figure*}[h!] 
	\centering 
	\includegraphics[width=\textwidth]{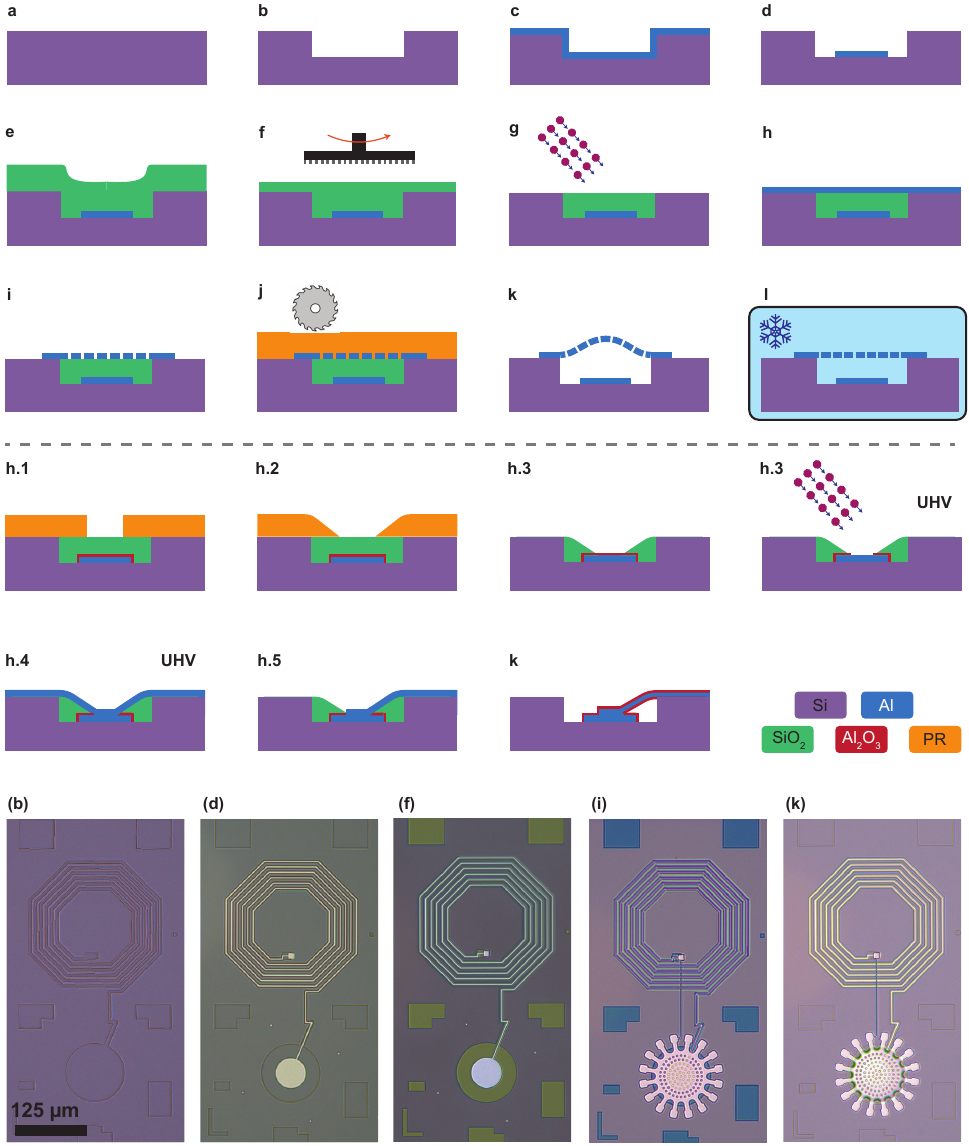} 
	\caption[Detail fabrication process flow]{\textbf{Detail fabrication process flow.} \textbf{a}, A high-resistivity silicon wafer is cleaned and used as substrate. \textbf{b}, Etching a trench in a silicon wafer (300 nm typical). \textbf{c}, Aluminum deposition of the bottom plate (100 nm typical). \textbf{d}, Patterning of the bottom Al. \textbf{e}, SiO$_2$ sacrificial layer deposition (2.5 $\mu$m typical). \textbf{f}, CMP planarization. \textbf{g}, Etching back and landing on the substrate using IBE. \textbf{h},Aluminum deposition of the top plate (200 nm typical). This step consists of opening the oxide for galvanic connections, removing native AlOx, and deposition, shown in h.1-5. \textbf{i}, Patterning the top Al layer. \textbf{j}, Dicing the wafer to chips. \textbf{k}, Releasing the structure using HF vapor. Depending on the compressive stress of Al top layer, the top plate may buckle up. \textbf{l}, At cryogenic temperatures, the drumhead shrinks and flattens. The optical micrographs show examples of selected steps of the process flow. Adapted from~\cite{youssefi2023squeezed}.}
	\label{fig:fab_full_PF} 
\end{figure*}

\begin{figure*}[h!] 
	\centering 
	\includegraphics[width=\textwidth]{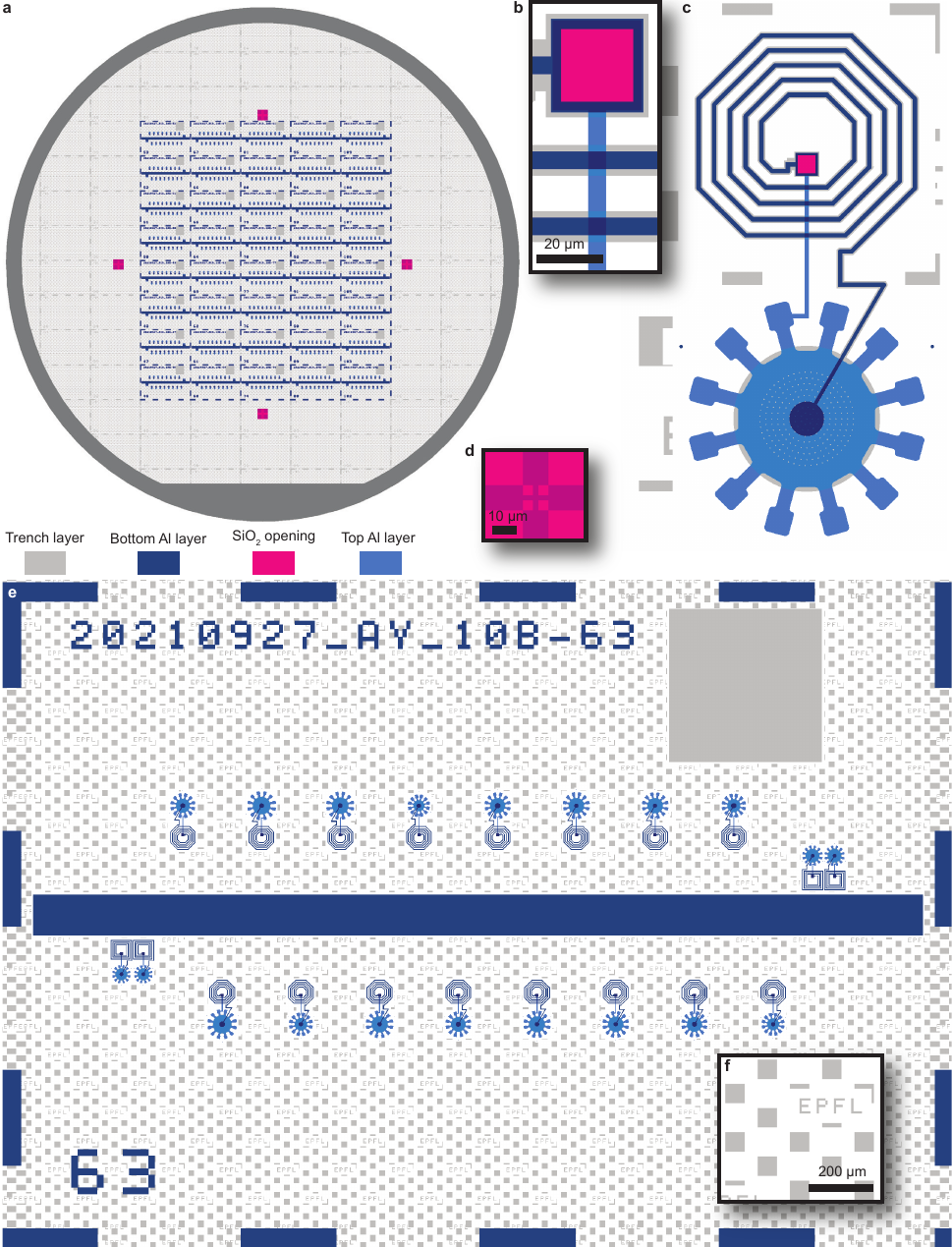} 
	\caption[Example of wafer and chip layout]{\textbf{Example of wafer and chip layout.} \textbf{a}, The wafer layout containing chips, alignment markers, and dummy trench patterns filling all the empty area to increase uniformity of the CMP planarization. \textbf{b}, the galvanic connection and air bridges of the spiral. \textbf{c}, Layout of a LC electro-mechanical resonator. \textbf{d,} The alignment markers for direct laser writer. \textbf{e,} The full layout of a chip containing several LC resonators (frequency multiplied) inductively coupled to a micro-strip waveguide. The top right rectangular big trench used for optical reflectometer measurement of the remaining SiO$_2$ after CMP. \textbf{f,} Shows the dummy patterns for CMP. }
	\label{fig:fab_chip_design} 
\end{figure*}

\end{document}